\documentclass[twocolumn,tighten]{aastex61}
\usepackage{apjfonts}
\usepackage{amsmath}
\usepackage{graphicx}
\usepackage{natbib}
\usepackage{longtable}
\usepackage{xcolor}
\def\pasa{\rm{PASA}}

\begin{document}

\title{Outcomes of Grazing Impacts between Sub-Neptunes in Kepler Multis}

\author{Jason Hwang}
\affil{Northwestern University, Department of Physics and Astronomy, Northwestern Univsersity, 2145
  Sheridan Road, Evanston, IL 60208, USA}
\affil{Center for Interdisciplinary Exploration and Research in Astrophysics (CIERA), Northwestern University, 2145
  Sheridan Road, Evanston, IL 60208, USA}
\author{Sourav Chatterjee}
\affil{Northwestern University, Department of Physics and Astronomy, Northwestern Univsersity, 2145
  Sheridan Road, Evanston, IL 60208, USA}
\affil{Center for Interdisciplinary Exploration and Research in Astrophysics (CIERA), Northwestern University, 2145
  Sheridan Road, Evanston, IL 60208, USA}
\author{James Lombardi Jr.}
\affil{Department of Physics, Allegheny College, Meadville, PA 16335, USA}
\author{Jason H. Steffen}
\affil{Department of Physics and Astronomy, University of Nevada, Las Vegas, 4505 S. Maryland Pkwy., Las Vegas, NV 89154-4002, USA}
\author{Frederic Rasio}
\affil{Northwestern University, Department of Physics and Astronomy, Northwestern Univsersity, 2145
  Sheridan Road, Evanston, IL 60208, USA}
\affil{Center for Interdisciplinary Exploration and Research in Astrophysics (CIERA), Northwestern University, 2145
  Sheridan Road, Evanston, IL 60208, USA}

\begin{abstract}
Studies of high-multiplicity, tightly-packed planetary systems suggest that dynamical instabilities are common and affect both the orbits and planet structures, where the compact orbits and typically low densities make physical collisions likely outcomes.
Since the structure of many of these planets is such that the mass is dominated by a rocky core, but the volume is dominated by a tenuous gas envelope, the sticky-sphere approximation, used in dynamical integrators, may be a poor model for these collisions.
We perform five sets of collision calculations, including detailed hydrodynamics, sampling mass ratios and core mass fractions typical in Kepler Multis.
In our primary set of calculations, we use Kepler-36 as a nominal remnant system, as the two planets have a small dynamical separation and an extreme density ratio.
We use an N-body code, {\it Mercury 6.2}, to integrate initially unstable systems and study the resultant collisions in detail.
We use these collisions, focusing on grazing collisions, in combination with realistic planet models created using gas profiles from {\it Modules for Experiments in Stellar Astrophysics} and core profiles using equations of state from \citet{2007ApJ...669.1279S}, to perform hydrodynamic calculations, finding scatterings, mergers, and even a potential planet-planet binary.
We dynamically integrate the remnant systems, examine the stability, and estimate the final densities, finding the remnant densities are sensitive to the core masses, and collisions result in generally more stable systems.
We provide prescriptions for predicting the outcomes and modeling the changes in mass and orbits following collisions for general use in dynamical integrators.
\end{abstract}


\section{Introduction}
\label{Sec:Intro}

Many studies of the dynamical evolution of high-multiplicity, tightly-packed planetary systems (known as Kepler Multis) have shown that systems similar to those observed in the Kepler sample may experience planet-planet instabilities that result in physical collisions (\citealt{2015ApJ...807...44P}; \citealt{2015ApJ...806L..26V}; \citealt{2017MNRAS.470.4145H}, hereafter referred to as {\it Paper 1}).
Studies of generic planetary systems show that post-disk dynamical interactions may be important in shaping the observed structure of Kepler Multis (\citealt{1996Icar..119..261C}; \citealt{2008ApJ...686..603J}; \citealt{2008ApJ...686..621F}; \citealt{2008ApJ...686..580C}), and additional studies (\citealt{2008Sci...321..814T}; \citealt{2010ApJ...725.1995M}) confirm the occurence of planet-planet interactions are consistent with orbits after the planetesimal disk dissipates.
\citet{2012ApJ...761...92F} find many Kepler Multis are dynamically packed, in that no additional planets may be added in between neighboring orbits without leading to planet-planet interactions, suggesting many observed systems are on the cusp of dynamical instability.
Studies of the structure of the planets in these systems have shown most of these planets are made up of a high-density core with a tenuous gas envelope dominating the volume (e.g., \citealt{2015ApJ...806..183W}; \citealt{2014ApJ...783L...6W}; \citealt{2015ApJ...801...41R}).
This highly differentiated structure suggests the outcomes of planet-planet collisions in these systems is very sensitive to the details of the collision, and in many cases the sticky-sphere approximation, ubiquitous in dynamical integrators, may not be valid for many of these collisions.
The sticky-sphere approximation assumes a collision occurs in the integration when the planets have a minimum separation less than the sum of the physical radii, and results in a single surviving planet with mass equal to the sum of the two colliding planets' masses, conserving center of mass, linear momentum, and angular momentum.
We perform five sets of detailed calculations, varying the mass ratio and core mass fractions, to better understand and predict the outcomes of planet-planet collisions that may be instrumental in shaping both the orbits and planet structures of these systems \citep{2016ApJ...817L..13I}.
For our primary suite of integrations, we use Kepler-36 as a nominal remnant of a previous planet-planet collision.
To examine how the mass ratio and planet structures affect the outcomes of these collisions, we also perform calculations between planets of mass ratio $q=1$ and $q=1/3$, where the less-massive planet has a mass of $4\ M_\mathrm{E}$, and core mass fractions of $m_\mathrm{c}/m=0.85$ and $m_\mathrm{c}/m=0.95.$

Kepler-36 is a 2-planet system discovered by the Kepler mission \citep{2012Sci...337..556C}, where Kepler-36b and Kepler-36c (henceforth referred to as {\it b} and {\it c}) have an orbital separation of less than $0.015$ AU, and a density ratio of $\rho_\mathrm{b}/\rho_\mathrm{c}=0.12$.
{\it b} has a density consistent with little or no gas envelope ($\rho_\mathrm{b}=7.5\ \mathrm{g}\ \mathrm{cm}^{-3}$, \citealt{2012Sci...337..556C}), while {\it c} has a density consistent with a gas mass fraction of $12\%$ and a gas radius fraction of $55\%$ ($\rho_\mathrm{c}=0.89\ \mathrm{g}\ \mathrm{cm}^{-3}$).
\citet{2012ApJ...755L..21D} show Kepler-36 likely undergoes short timescale dynamical chaos and provide best-fit densities of $\rho_b = 7.65\ \mathrm{g}\ \mathrm{cm}^{-3}$ and $\rho_c = 0.93\ \mathrm{g}\ \mathrm{cm}^{-3}$.
\citet{2013AN....334..992N} explore the possibility of additional planets in the system and use long term stability to contrain the semi-major axes of potential additional planets to $a<0.1\ \mathrm{AU}$ and $a>0.14\ \mathrm{AU}$.
The high density difference and small dynamical separation motivates many studies on the formation of Kepler-36.

Many formation channels have been explored to explain the observed properties of Kepler-36.
\citet{2013MNRAS.434.3018P} show migration in a turbulent disc can result in systems very similar to Kepler-36.
\citet{2013ApJ...776....2L} and \citet{2016ApJ...819L..10O} explore the possibility that the density difference can be explained by evaporation of the envelopes, and constrain the cooling timescale and initial structures of {\it b} and {\it c}.
Without migration, the planets' structure is unlikely to have formed in-situ due to the combination of small orbital separation and extreme density ratio, and may be a possible remnant of a previous planet-planet collision that depleted the gas envelope in the smaller planet.
We use Kepler-36 as a potential collision remnant to guide our choice of initial conditions for our primary set of collision calculations.
For this set of calculations we provide a generic analytic calculation of the orbits and amount of conservative mass-transfer required to create an initially unstable system that becomes stable after a collision resulting in two surviving planets (see \S\ref{A:Analytic}).

Many studies have been conducted to understand the details of planet-planet collisions and close encounters.
Much of the early work studies potential Moon-formation collisions, specifically between a proto-earth and an impactor, varying the planet structures (\citealt{1975Icar...24..504H}, \citealt{1986Icar...66..515B}; \citealt{2000orem.book..133C}; \citealt{2001Natur.412..708C}; \citealt{2004ARA&A..42..441C}; \citealt{2013Icar..222..200C}; \citealt{2014DPS....4650109C}) and using various tabulated equations of state to handle the abrupt changes in density and phase transitions of the impacted rock.
\citet{2015ApJ...812..164L} present calculations of direct collisions between a rocky, terrestrial planet and a gas giant, treating the rocky core with a multi-phase equation of state (Tillotson 1962) and the gas as a polytrope with $\gamma=5/3$.
\citet{2010arXiv1007.1418P} discuss the possibility of tidal dissipation of orbital energy leading to potentially stable gas-giant binaries.
\citet{2015MNRAS.448.1751I} and \citet{2016ApJ...817L..13I} combine 1-D hydrodynamic calculations and a thermal evolution model to examine the effect of collisions resulting in mergers during the giant-impact phase of planet formation, and find that these collisions result in a large range of planet densities, similar to the observed densities found in Kepler Multis.
In this work we explore the outcomes of grazing collisions between sub-Neptunes close to the host star, varying the mass ratio and gas mass fractions.
We present several suites of collision calculations from initial conditions sourced from dynamical integrations and find several distinct outcomes, including scatterings, mergers, and a potential planet-planet binary (formed from energy dissipation due to the physical collision).
We aggregate the results and develop a fit predicting the outcomes and changes in mass from such collisions for use in dynamical integrators.

The rest of the paper is organized as follows:
In \S\ref{Sec:Methods} we discuss the initial conditions and numerical methods used for our dynamical integrations of planetary systems, and our method for creating 3-D planet models and numerical methods used in the subsequent hydrodynamic calculations.
In \S\ref{Sec:Outcomes} we present the results of our hydrodynamic calculations, classifying the collision outcomes as bound planet-pairs, two surviving planets in stable or unstable orbits, or mergers, and explore the stability and potential observable indicators of collisions in both the orbits and planet structures.
In \S\ref{Sec:Predictions} we present a prescription, aggregating the results from all sets of collision calculations, for predicting the outcomes and modeling the mass loss of planet-planet collisions for use in dynamical integrators.
Finally, we summarize our conclusions in \S\ref{Sec:Conclusions}.

\section{Numerical Methods}
\label{Sec:Methods}
We perform five sets of collision calculations, varying the mass ratio and core mass fractions of the two-planet system.
We choose our combinations of mass ratios and core mass fractions to broadly cover the properties of neighboring planets seen in Kepler Multis.
Table~\ref{TBL:q_mc} shows the planet masses, radii, and core properties, and host star properties used in each set, where \S\ref{SSec:MassTransfer} describes how we assign the masses and core mass fractions for the Kepler-36 progenitor calculations.
Sub-Neptune structures are more dependent on the core mass fraction than the total mass; despite having three times more gas, the $12\ M_\mathrm{E}$ models have a physical size similar to the $4\ M_\mathrm{E}$ models of identical gas mass fraction.
For each set we generate a suite of dynamical integrations and use a subset of the resultant collisions as initial conditions for later detailed hydrodynamic calculations.

\begin{deluxetable*}{lccccccccccccccc}
\tabletypesize{\footnotesize}
\tablewidth{15.5cm}
\tablecolumns{11}
\tablecaption{System Properties\label{TBL:q_mc}}
\tablehead{
    \colhead{Set} & \colhead{$m_1$} & \colhead{$m_2$} & \colhead{$m_\mathrm{1,c}/m_1$} & \colhead{$m_\mathrm{2,c}/m_2$} & \colhead{$R_1$} & \colhead{$R_2$} & \colhead{$R_\mathrm{1,c}/R_1$} & \colhead{$R_\mathrm{2,c}/R_2$} & \colhead{$M_*$} & \colhead{$T_*$}}
\startdata
$ Kepler-36\ progenitor		$ & $	4.67	$ & $	7.87	$ & $	0.95	$ & $	0.91	$ & $	2.65	$ & $	3.37	$ & $	0.56	$ & $	0.49	$ & $	1.071	$ & $	5911	$ & \\
$ q=1;\ m_\mathrm{c}=0.85	$ & $	4.00	$ & $	4.00	$ & $	0.85	$ & $	0.85	$ & $	4.21	$ & $	3.96	$ & $	0.33	$ & $	0.35	$ & $	1.0	$ & $	5778	$ & \\
$ q=1;\ m_\mathrm{c}=0.95	$ & $	4.00	$ & $	4.00	$ & $	0.95	$ & $	0.95	$ & $	2.84	$ & $	2.72	$ & $	0.50	$ & $	0.52	$ & $	1.0	$ & $	5778	$ & \\
$ q=1/3;\ m_\mathrm{c}=0.85	$ & $	4.00	$ & $	12.00	$ & $	0.85	$ & $	0.85	$ & $	4.21	$ & $	4.01	$ & $	0.33	$ & $	0.45	$ & $	1.0	$ & $	5778	$ & \\
$ q=1/3;\ m_\mathrm{c}=0.95	$ & $	4.00	$ & $	12.00	$ & $	0.95	$ & $	0.95	$ & $	2.84	$ & $	2.94	$ & $	0.50	$ & $	0.63	$ & $	1.0	$ & $	5778	$ & \\
\enddata
\tablenotetext{}{$Set$ designates the set name, $m_1$ and $m_2$ are the masses of the inner and outer planet in Earth masses, $m_\mathrm{1,c}/m_1$ and $m_\mathrm{2,c}/m_2$ are the core mass fractions of the inner and outer planet, $R_1$ and $R_2$ are the radii of the inner and outer planet in Earth radii, $R_\mathrm{1,c}/R_1$ and $R_\mathrm{2,c}/R_2$ are the core radius fractions of the inner and outer planet, and $M_*$ and $T_*$ are the mass and temperature of the host star in solar masses and K.}
\end{deluxetable*}

\subsection{Dynamical Integrations}
\label{Sec:Nbody}
For each set of systems, we use a Monte Carlo method to generate 1000 realizations in the point-mass limit, where we set the planets' density to some arbitrary large value, using the Burlish-Stoer integrator within the N-body dynamics package, {\it Mercury} 6.2 \citep{1999MNRAS.304..793C}.
We later impose planet radii to generate collisions.
Table~\ref{TBL:a_e} summarizes the initial orbits used in each set of calculations, where we set the initial period ratio, $P_2/P_1=1.3$ and assign the eccentricities such that the total angular momentum is low enough to result in orbit crossings.
The results from \S\ref{A:Analytic} are used to generate the initial orbits for the Kepler-36 progenitor calculations, using the observed Kepler-36 properties to estimate the initial masses and semi-major axes.
In the more generic sets of calculations, the inner planet's semi-major axis is set to $a_1 = 0.1\ \mathrm{AU}$, and the initial eccentricities are randomly assigned, enforcing $a_1(1+e_1)=a_2(1-e_2)$ to ensure fast orbit crossing.
The inclinations are randomly drawn from a distribution uniform in $\cos(i)$, where we choose a small, but non-zero inclination, $-\theta\le i\le\theta;\ \theta=0.017$, ensuring that $a\sin(\theta)>R$, where $R$ is the radii of the planets, to simulate a nearly coplanar system.
Finally, we randomly sample the argument of periapsis, ascending node, and true anomaly from a uniform distribution from $0$ to $2\pi$.

\begin{deluxetable*}{lccccccccccc}
\tabletypesize{\footnotesize}
\tablewidth{10.0cm}
\tablecolumns{7}
\tablecaption{Initial Orbits\label{TBL:a_e}}
\tablehead{
    \colhead{Set} & \colhead{$a_1$} & \colhead{$a_2$} & \colhead{$e_{1,\mathrm{min}}$} & \colhead{$e_{1,\mathrm{max}}$} & \colhead{$e_{2,\mathrm{min}}$} & \colhead{$e_{2,\mathrm{max}}$}}
\startdata
$ Kepler-36\ progenitor		$ & $	0.111	$ & $	0.132	$ & $	0.11	$ & $	0.11	$ & $	0.07	$ & $	0.07	$ & \\
$ q=1				$ & $	0.100	$ & $	0.119	$ & $	0.00	$ & $	0.18	$ & $	0.01	$ & $	0.16	$ & \\
$ q=1/3				$ & $	0.100	$ & $	0.119	$ & $	0.00	$ & $	0.19	$ & $	0.00	$ & $	0.16	$ & \\
\enddata
\tablenotetext{}{$Set$ designates the set name, $a_1$ and $a_2$ are the semi-major axes of the inner and outer planet in AU, and $e_{1,\mathrm{min}}$, $e_{1,\mathrm{max}}$, $e_{2,\mathrm{min}}$, and $e_{2,\mathrm{max}}$ are the minimum and maximum eccentricities of the inner and outer planets. The initial orbits are identical for the sets of runs with a given mass ratio.}
\end{deluxetable*}

A collision occurs in the integration when the planets have a minimum separation less than the sum of the physical radii, calculated using {\it MESA} assuming an Earth-like core composition and a core-density calculated by interpolating core-mass core-density tables in \citet{2014ApJ...787..173H}.
For each calculation, the first planet-planet collision is characterized by the distance of closest approach, $d_\mathrm{min}$, in units of the sum of the planets' physical radii, $R_1$ and $R_2$, and the collision energy, $E_\mathrm{c}=E_\mathrm{k}+E_\mathrm{g}$ in units of the binding energy of both planets up to a coefficient, \begin{equation}\label{EQ:binding_energy}E_\mathrm{b}=\sum_{k=1,2}\frac{Gm_k^2}{R_k},\end{equation} where $E_\mathrm{k}$ and $E_\mathrm{g}$ are the kinetic and gravitational potential energy of the planets in the center of mass frame, ignoring the host star.
The degree of contact, \begin{equation}\label{EQ:Degree_of_Contact}\eta=\frac{d_\mathrm{min}}{R_1+R_2},\end{equation} characterizes the depth of the collision, where $\eta=1$ describes an extremely grazing collision and $\eta=0$ describes a head-on collision.

\begin{figure*}[htp]
\begin{center}
\begin{tabular}{cc}
\includegraphics[width=8.0cm]{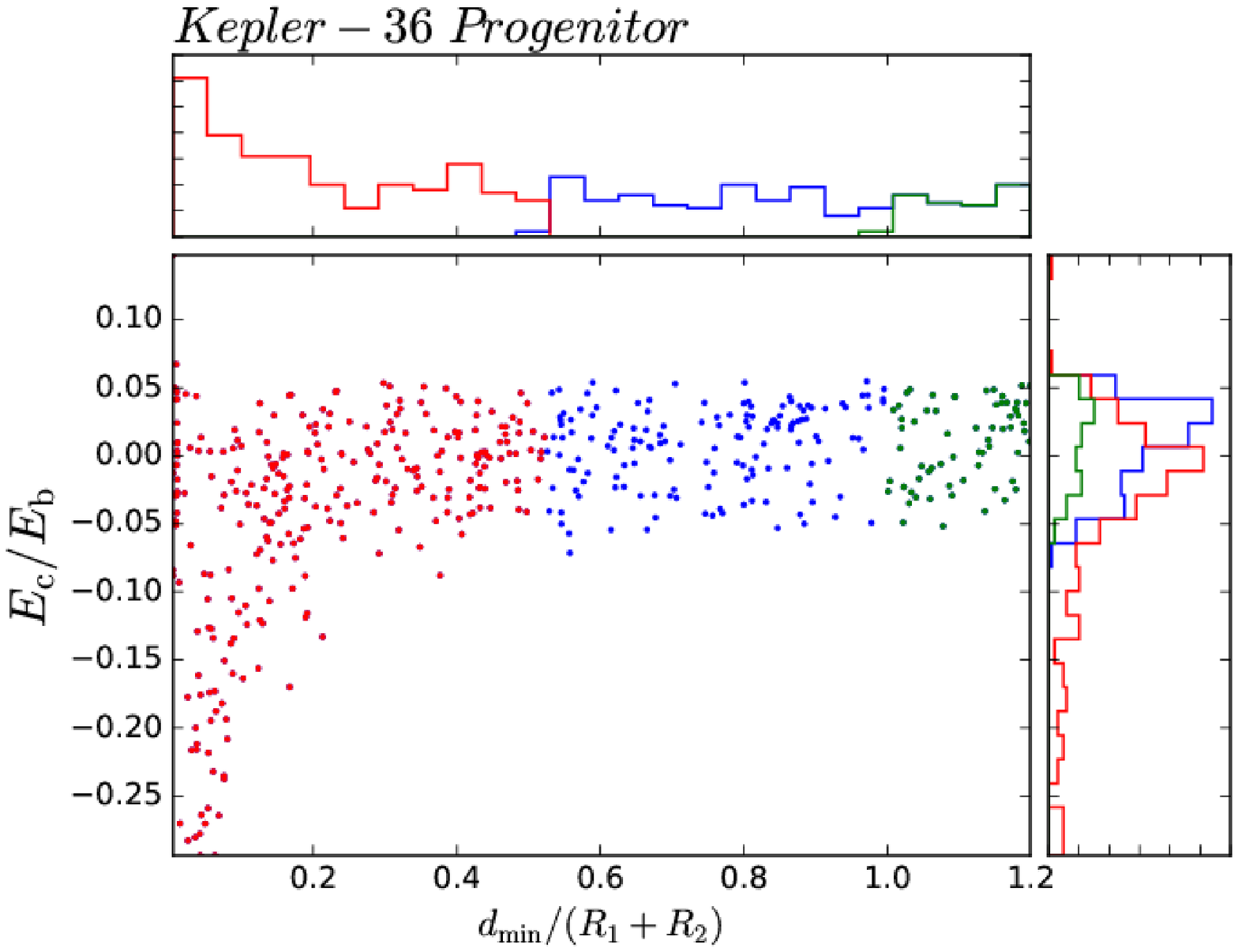}
\color{white}
\rule{8.0cm}{5.0cm}
\color{black} \\
\includegraphics[width=8.0cm]{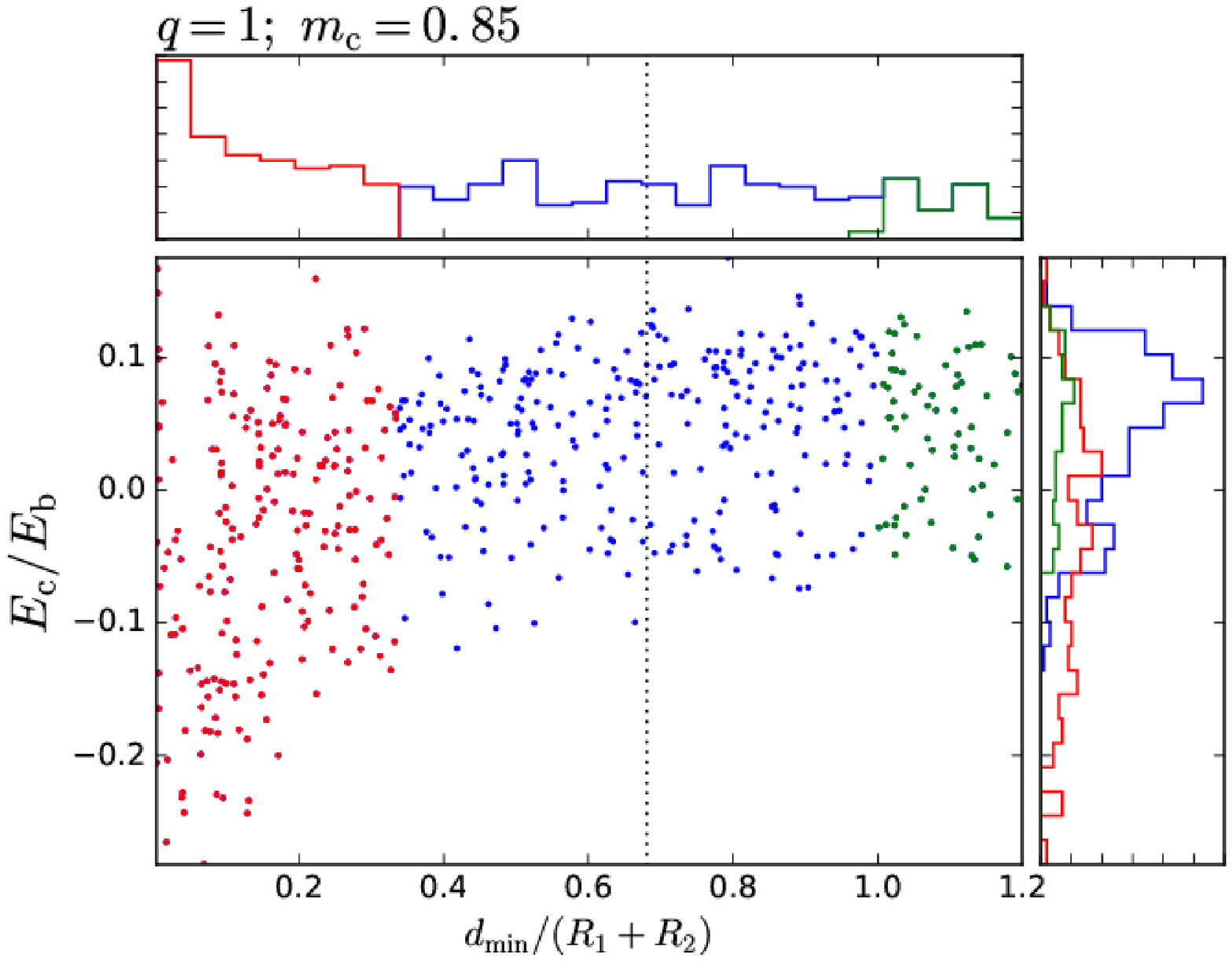}
\includegraphics[width=8.0cm]{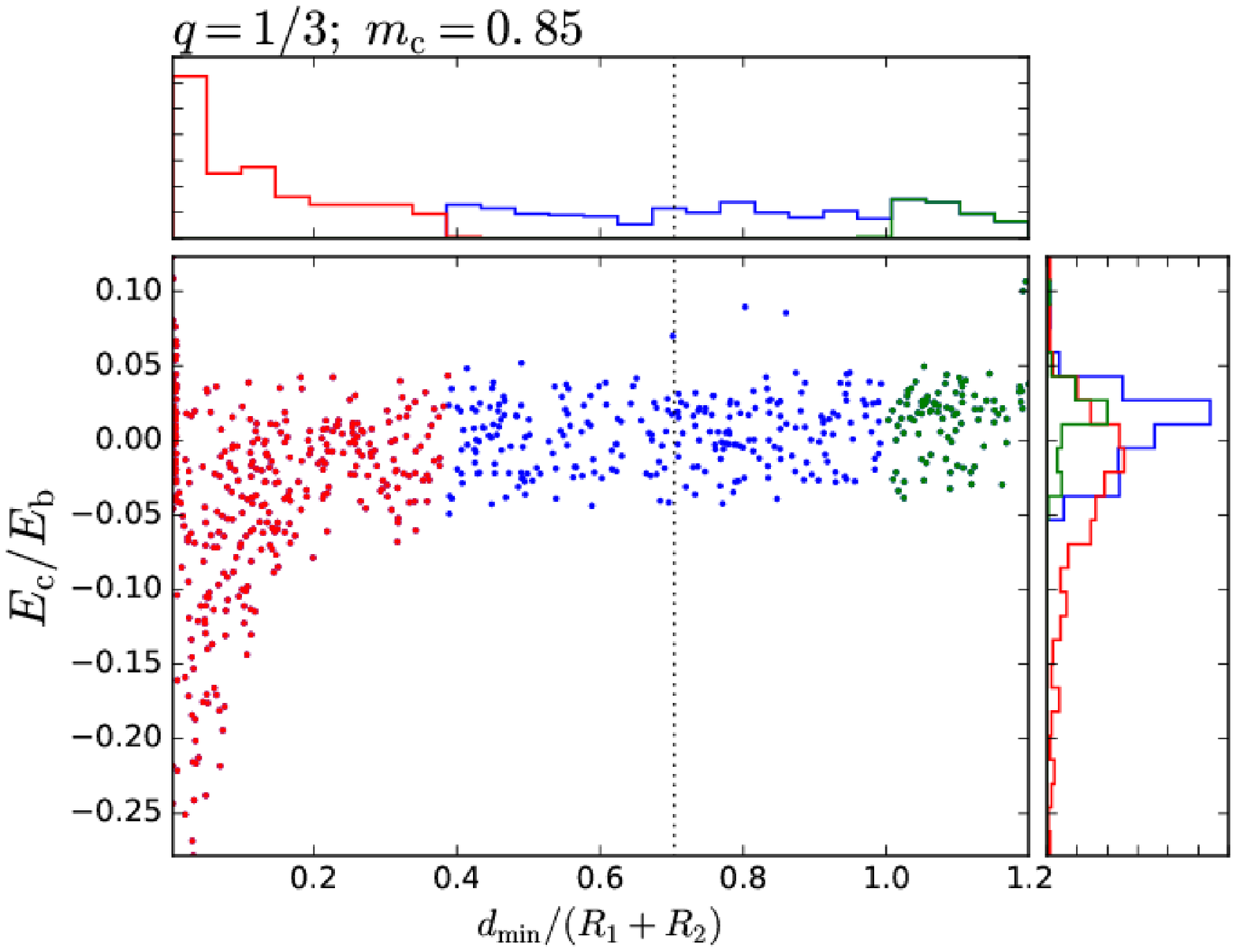}
\end{tabular}
\end{center}
\caption{Scatter plots showing the distance of closest approach, $d_\mathrm{min}$, in units of the sum of the planets' physical radii, and collision energy, $E_\mathrm{c}=E_\mathrm{k}+E_\mathrm{g}$ in units of the binding energy of both planets up to a coefficient, $E_\mathrm{b}=Gm_1^2/R_1+Gm_2^2/R_2$, where $E_\mathrm{k}$ and $E_\mathrm{g}$ are the kinetic and gravitational potential energy of the planets in the center of mass frame, ignoring the host star, for the first collision or near miss in each N-body integration. The histograms above the scatter plots show the distribution of distances of closest approach and the histograms to the right of the scatter plots show the distribution of collision energies for the first collision or near miss in each integration. We show the distribution of collisions with a small enough distance of closest approach to result in a direct impact between the two cores (red), using the nominal core radii given by \citet{2013ApJ...770..131L} for the Kepler-36 progenitor integrations (top left), and using the nominal core radii calculated using the assigned core mass fraction and a core density from \citet{2014ApJ...787..173H} for the $q=1$ (bottom left) and $q=1/3$ (bottom right) integrations. We also show the distribution of grazing collisions (blue), and near misses (green), defined as close encounters where $1<d_\mathrm{min}/(R_1+R_2)<1.2$. For the more generic calculations, collisions are calculated assuming a core mass fraction of $85\%$. Collisions in between planets with a higher core mass fraction require a smaller distance of closest approach, and we show the minimum distance for a core mass fraction of $95\%$ (dotted black line).}
\label{Fig:Collision_Parameters}
\end{figure*}

Table~\ref{TBL:NBody_Stats} shows the distribution of collisions for each set of calculations, categorizing the number of collisions that result in a direct impact of the two cores, grazing collisions, and near misses, defined as close encounters where $1<d_\mathrm{min}/(R_1+R_2)<1.2.$
We note that due to tidal bulging, many of the ``near misses'' would also result in physical contact between the planets.
We find that the number of collisions does not decrease significantly for systems with planets of the same mass but smaller physical size, and the fraction of collisions that result in a direct impact between the planets' cores increases with larger core mass fractions.
Figure~\ref{Fig:Collision_Parameters} shows the distribution of the first planet-planet collisions in each set of integrations.
In contrast to {\it Paper 1}, we find that the number of initial collisions decreases at higher distances of closest approach, likely due to the lower multiplicity of our systems; since higher-multiplicity systems can go unstable with wider orbital spacings (\citealt{1993Icar..106..247G}, \citealt{1996Icar..119..261C}), crossing orbits will extend to higher eccentricities---and hence, higher relative velocities.
In \S\ref{SSec:Hydro}, we describe how we perform detailed hydrodynamic calculations exploring how the distance of closest approach and collision energy affect the outcomes, specifically if the cores merge, if the planets survive, and how much gas is retained by the remnant planet(s) \citep{2008ApJ...686..580C}.

\begin{deluxetable*}{lccccccccc}
\tabletypesize{\footnotesize}
\tablewidth{9.0cm}
\tablecolumns{5}
\tablecaption{Statistics of N-body Collisions\label{TBL:NBody_Stats}}
\tablehead{
    \colhead{Set} & \colhead{$N_\mathrm{total}$} & \colhead{$N_\mathrm{direct\ impact}$} & \colhead{$N_\mathrm{grazing}$} & \colhead{$N_\mathrm{near\ miss}$}}
\startdata
$ Kepler-36\ progenitor		$ & $	502	$ & $	290	$ & $	148	$ & $	64	$ & \\
$ q=1;\ m_\mathrm{c} = 0.85	$ & $	578	$ & $	245	$ & $	266	$ & $	67	$ & \\
$ q=1;\ m_\mathrm{c} = 0.85	$ & $	572	$ & $	303	$ & $	181	$ & $	81	$ & \\
$ q=1/3;\ m_\mathrm{c} = 0.95	$ & $	708	$ & $	360	$ & $	256	$ & $	92	$ & \\
$ q=1/3;\ m_\mathrm{c} = 0.95	$ & $	701	$ & $	447	$ & $	171	$ & $	83	$ & \\
\enddata
\tablenotetext{}{$Set$ designates the set name, $N_\mathrm{total}$ is the number of collisions that occur within $1000$ years, $N_\mathrm{direct\ impact}$ is the number of collisions resulting in a direct impact between the planets' cores, $N_\mathrm{grazing}$ is the number of collisions that do not result in a direct impact between the planets' cores, and $N_\mathrm{near\ miss}$ is the number of near misses, defined as close encounters where $1<d_\mathrm{min}/(R_1+R_2)<1.2.$}
\end{deluxetable*}

\subsection{Hydrodynamics}
\label{SSec:Hydro}
We perform 102 hydrodynamic calculations sampled from each set of dynamical integrations described in \S\ref{Sec:Nbody}, using an SPH code, {\it StarSmasher}\footnote{{\it StarSmasher} is available at https://jalombar.github.io/starsmasher/.} (previously {\it StarCrash}, originally developed by \citealt{1991PhDT........11R} and later updated as described in \citealt{1999JCoPh.152..687L} and \citealt{2000PhRvD..62f4012F}), to treat the hydrodynamics.
{\it StarSmasher} implements variational equations of motion and libraries to calculate the gravitational forces between particles using direct summation on NVIDIA graphics cards as described in \citet{2010MNRAS.402..105G}.
Using direct summation instead of a tree-based algorithm for gravity increases the accuracy of the gravity calculations at the cost of speed \citep{2010ProCS...1.1119G}.
The code uses a cubic spline \citep{1985A&A...149..135M} for the smoothing kernel and an artificial viscosity prescription coupled with a Balsara Switch \citep{1995JCoPh.121..357B} to prevent unphysical inter-particle penetration, specifically described in \citet{2015ApJ...806..135H}.
We sample collisions at varying degrees of contact and collision energies, using the position and velocity coordinates of the planets many dynamical times defined as the free-fall time of a test-particle around the planet) prior to the close encounter, where we treat the host star as a point-mass particle that interacts only gravitationally.
We preferentially sample collisions to explore the boundary between merger and scattering results.

\subsection{Planet Models and Equations of State}
Following the method described in $\S A.3$ in {\it Paper 1}, we first use {\it MESA} to generate gas envelopes with a constant density core of mass, $m_\mathrm{c}$, where the core density is determined using tables from \citet{2014ApJ...787..173H} assuming an Earth-like core composition ($67.5\%$ silicate mantle and $32.5\%$ iron core).
We irradiate the planet at the chosen semi-major axis and host-star temperature (shown in Tables~\ref{TBL:q_mc} and \ref{TBL:a_e}) for $4.5\times10^9$ years.
We then replace the isothermal and constant-density cores {\it MESA} generates with a differentiated core, with equations of state from \citet{2007ApJ...669.1279S}, using the core mass and radius as boundary conditions to solve for the iron core and silicate-mantle mass fractions, recovering solutions very close to the Earth-like input values.

{\bf Planet structures are very time-sensitive.
\citet{2016ApJ...831..180C} and \citet{2013ApJ...776....2L} show the time-dependent structure of Kepler-36b and Kepler-36c (Figure 1 in both papers), where Kepler-36b contracts by a factor of 7 and Kepler-36c contracts by a factor of 3 between 100 Myr and 4.5 Gyr.
The age of planets when orbital instability first develops is not well understood, in part because we do not know what might trigger the instability in the system, just the internal secular evolution of the system (e.g. \citealt{2008ApJ...686..580C}; \citealt{2008ApJ...686..603J}; \citealt{2012ApJ...755L..21D}), or an external trigger (e.g., \citealt{2004AJ....128..869Z}).
In general, orbital instability timescales in chaotic systems have very broad distributions (e.g., \citealt{2008ApJ...686..580C}; \citealt{2007ApJ...666..423Z}; \citealt{2010A&A...516A..82F}), so here we take a conservative assumption, assuming the planets are old and therefore their radii have shrunk to near equilibrium.
With younger, more inflated planets, collisions happen even faster, but this would make our initial conditions very arbitrary (very sensitive to exact assumptions about the age of the planets at the time they collide).
Collisions occurring between younger planets would be more dramatic, due to lower density and hotter envelopes, and our calculations should be treated as a lower limit.}

{\it StarSmasher} uses the combined gas and core profiles, and because we are interested in the evolution of the low-density gas envelope, creates planet models with equal number-density, non-equal mass, particle-distributions \citep{2006MNRAS.365..991M} using $\sim10^5$ particles per planet.
Gas particles follow a semi-analytic equation of state fit to a grid of {\it MESA} models, \begin{equation}\label{EQ:gasEoS_SPH}p_i=\frac{\rho_iu_i}{\beta_i}+K_\mathrm{e}\rho_i^{\gamma_\mathrm{e}}\left(1-\frac{1}{\beta_i(\gamma_\mathrm{e}-1)}\right),\end{equation}\begin{equation}u_i=\frac{K_\mathrm{e}\rho_i^{\gamma_\mathrm{e}-1}}{\gamma_\mathrm{e}-1}+\frac{\beta_ik_\mathrm{B}T_i}{\mu_im_H},\end{equation} where we use $\gamma_\mathrm{e}=3$, while $K_e$ and $\beta_i$ are fitted parameters.
Core particles follow an equation of state from \citet{2007ApJ...669.1279S}, \begin{equation}p_i=\frac{u_i(\gamma_\mathrm{c}-1)\rho_i\left(1-\frac{\rho_{\rm c}^\prime}{\rho_i}\right)^{\gamma_\mathrm{c}}}{{}_2F_1\left(1-\gamma_\mathrm{c},-\gamma_\mathrm{c},2-\gamma_\mathrm{c},\frac{\rho_{\rm c}^\prime}{\rho_i}\right)},\end{equation} where the internal energy is initially set as \begin{equation}u_i=\frac{c\rho_i^{\gamma_\mathrm{c}-1}{}_2F_1\left(1-\gamma_\mathrm{c},-\gamma_\mathrm{c},2-\gamma_\mathrm{c},\frac{\rho_{\rm c}^\prime}{\rho}\right)}{\gamma_\mathrm{c}-1},\end{equation} $c$, $\gamma_\mathrm{c}$, and $\rho_{\rm c}^\prime$ are constants determined by the composition, and ${}_2F_1(1-\gamma_\mathrm{c},-\gamma_\mathrm{c},2-\gamma_\mathrm{c},\frac{\rho_{\rm c}^\prime}{\rho})$ is the ordinary hypergeometric function.
Expressions for the mantle are like those for the core but with coefficients applicable to a silicate composition employed.
Particles near the interfaces of the gas envelope, silicate mantle, and iron core are treated as mixed-composition in order to resolve the high density-gradient between components.
For more details of how we generate our planet models, including how we fit the equation of state used for the gas envelope, and the algorithms used to handle mixed-composition particles, see \S$A.3$ in {\it Paper 1}.

\begin{figure*}[htp]
\begin{center}
\begin{tabular}{cc}
\includegraphics[width=8.0cm]{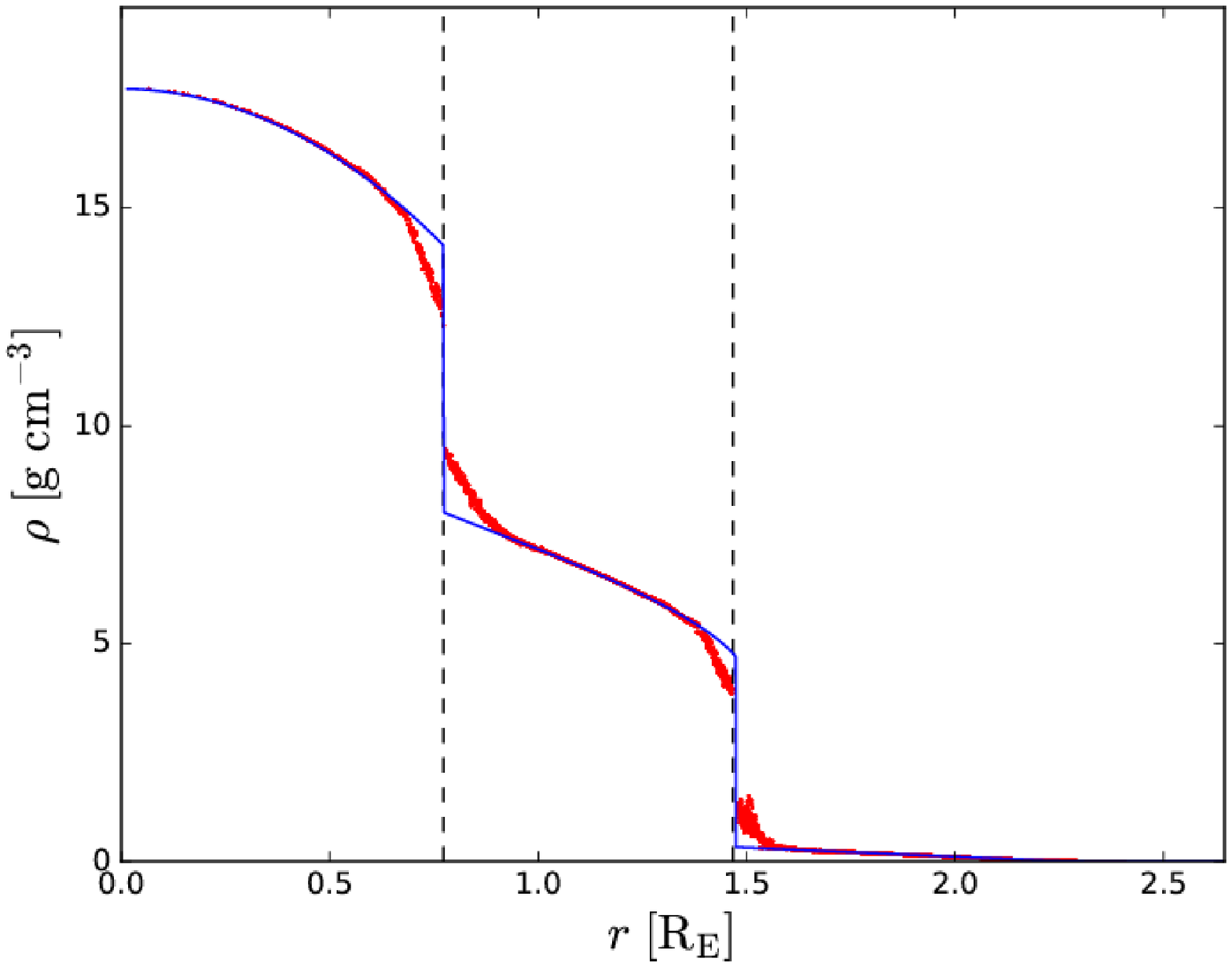}
\includegraphics[width=8.0cm]{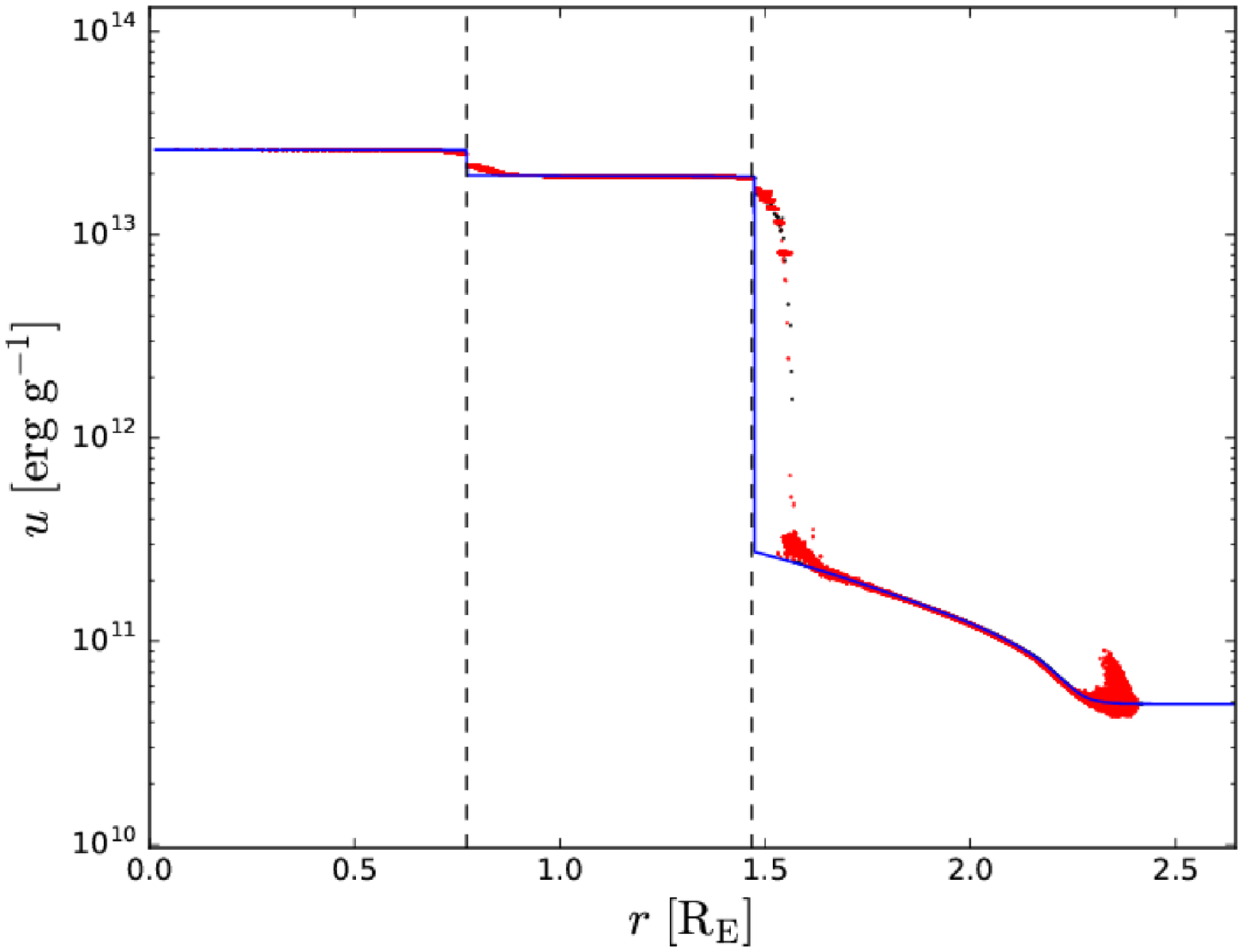}\\
\includegraphics[width=8.0cm]{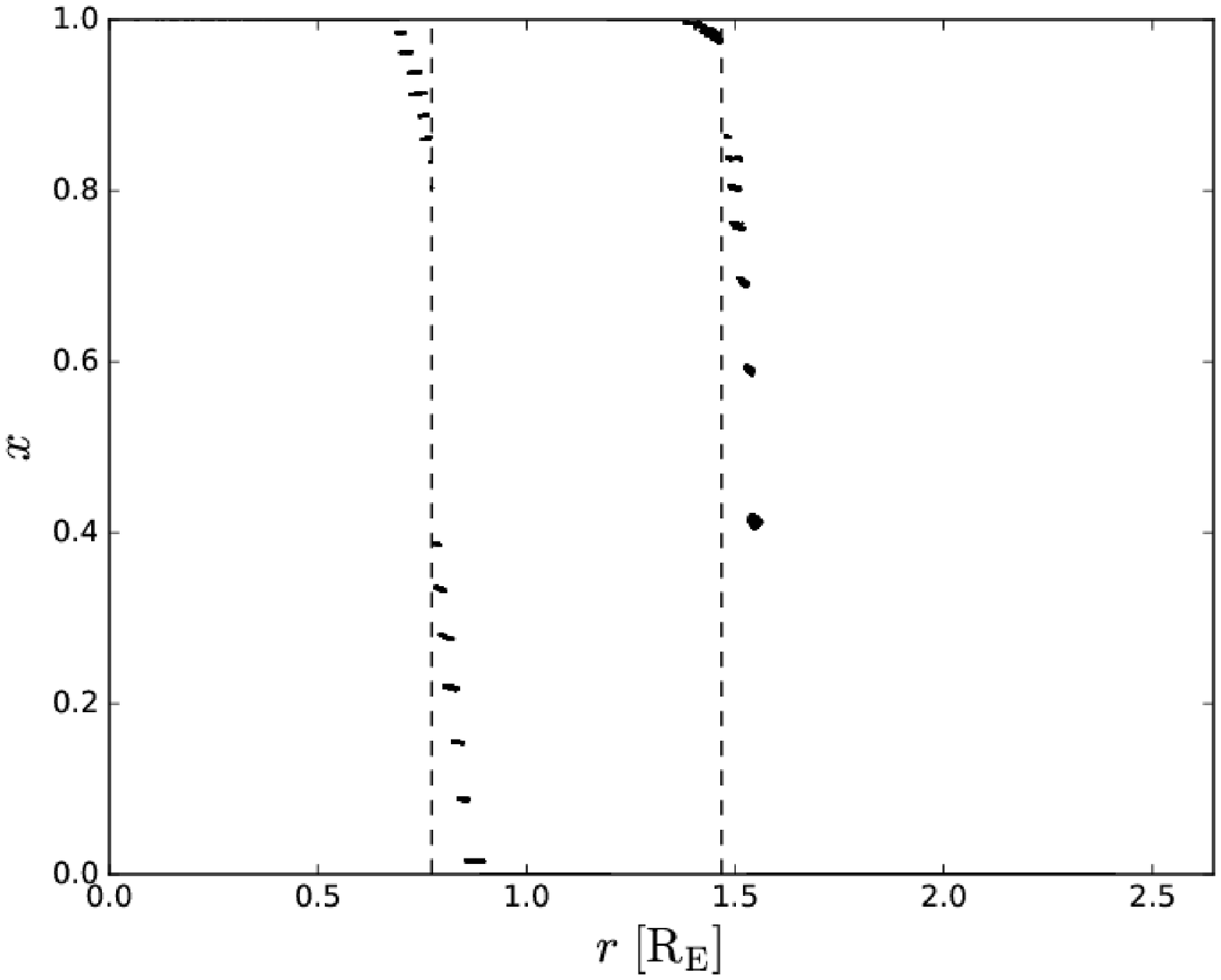}
\includegraphics[width=8.0cm]{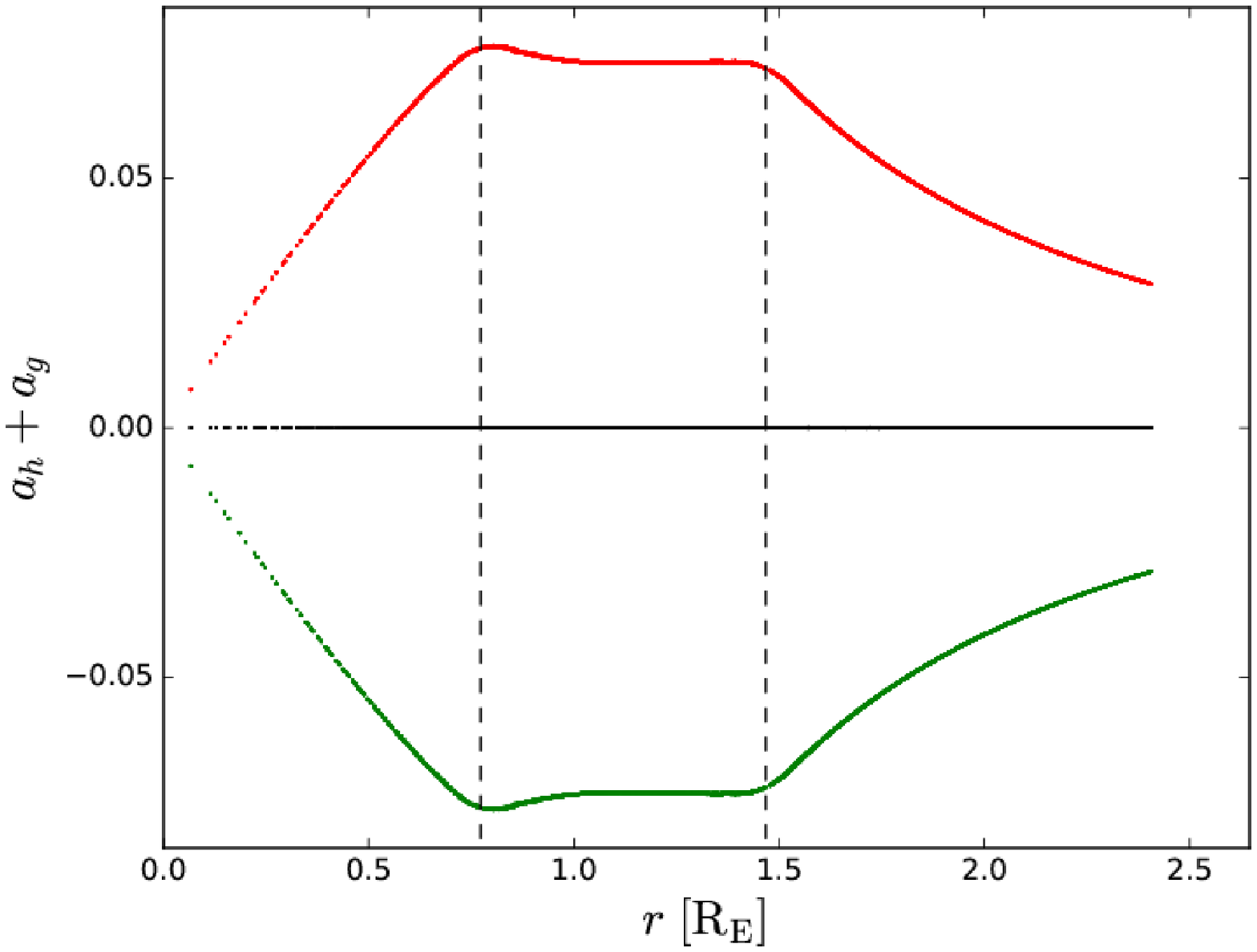}\\
\end{tabular}
\end{center}
\caption{Radial profiles at the end of 2000 code timescales for an isolated model of a Kepler-36b progenitor, where we enable a relaxation force (see \citealt{2015ApJ...806..135H} for details) for the first 100 code timescales.
In each plot we mark the transitions from core to mantle and from mantle to gas envelope (vertical dashed lines).
Density (top left) and internal energy (top right) profiles for the input profile (blue), generated using {\it MESA} and the equations of state from \citet{2007ApJ...669.1279S}, the initial values assigned to each particle (black), and the values after 2000 code timescales (red).
Particle composition, $x$, profile (bottom left) after 2000 code timescales, where $x$ represents the mass fraction of the particle belonging to the heavier composition near the interface.
Equilibrium profile (bottom right) after 2000 code timescales, showing the radial acceleration, {\bf in code units}, from the hydrostatic force (red), the gravitational force (green), and the total force (black).
These figures show how we assigned the particle composition and redistributed internal energy based on the smoothed densities.
The model relaxes into a very stable hydrostatic equilibrium, stable for at least the timescales used for the dynamical calculations.}
\label{Fig:Radial_Profiles_b}
\end{figure*}

\begin{figure*}[htp]
\begin{center}
\begin{tabular}{cc}
\includegraphics[width=8.0cm]{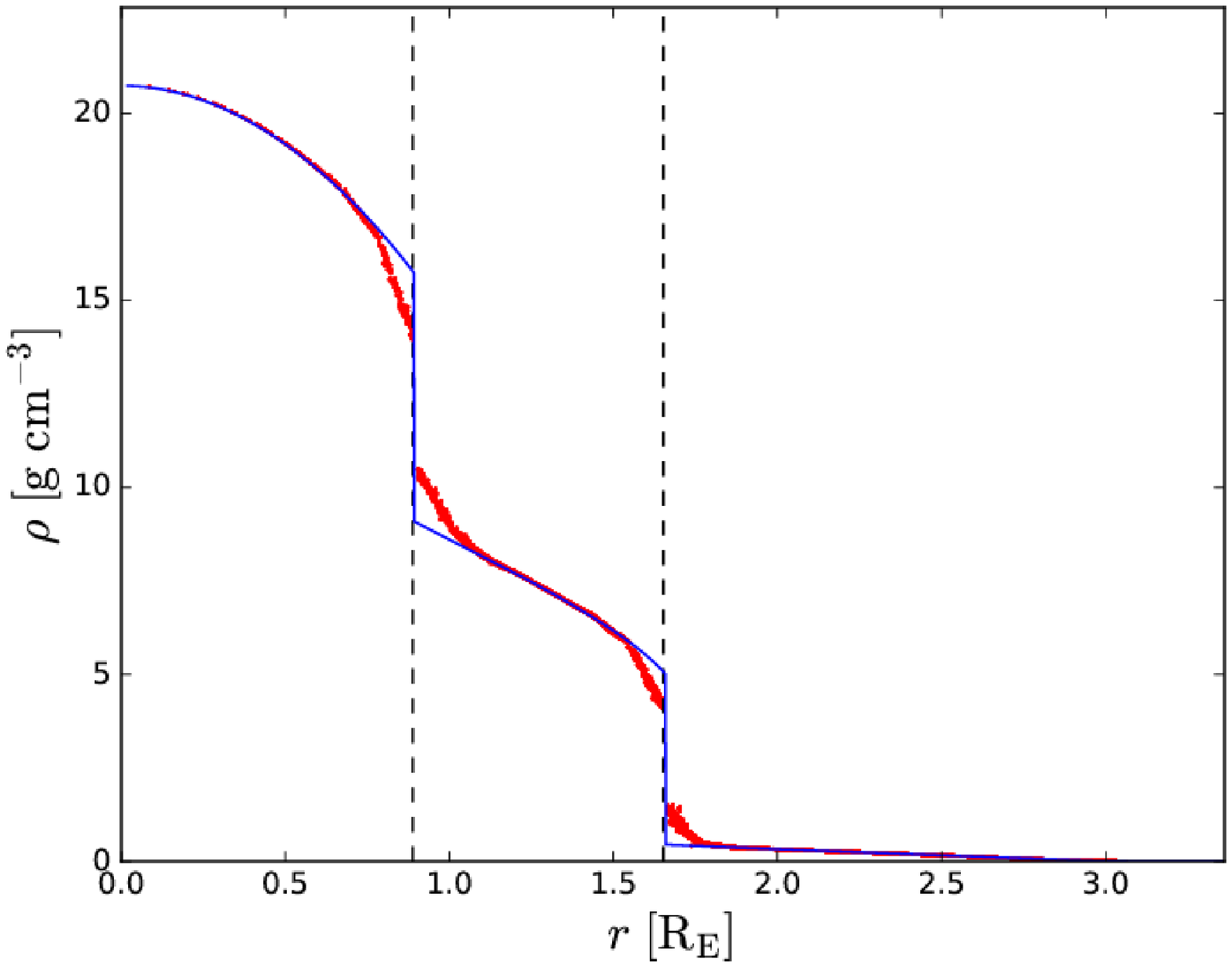}
\includegraphics[width=8.0cm]{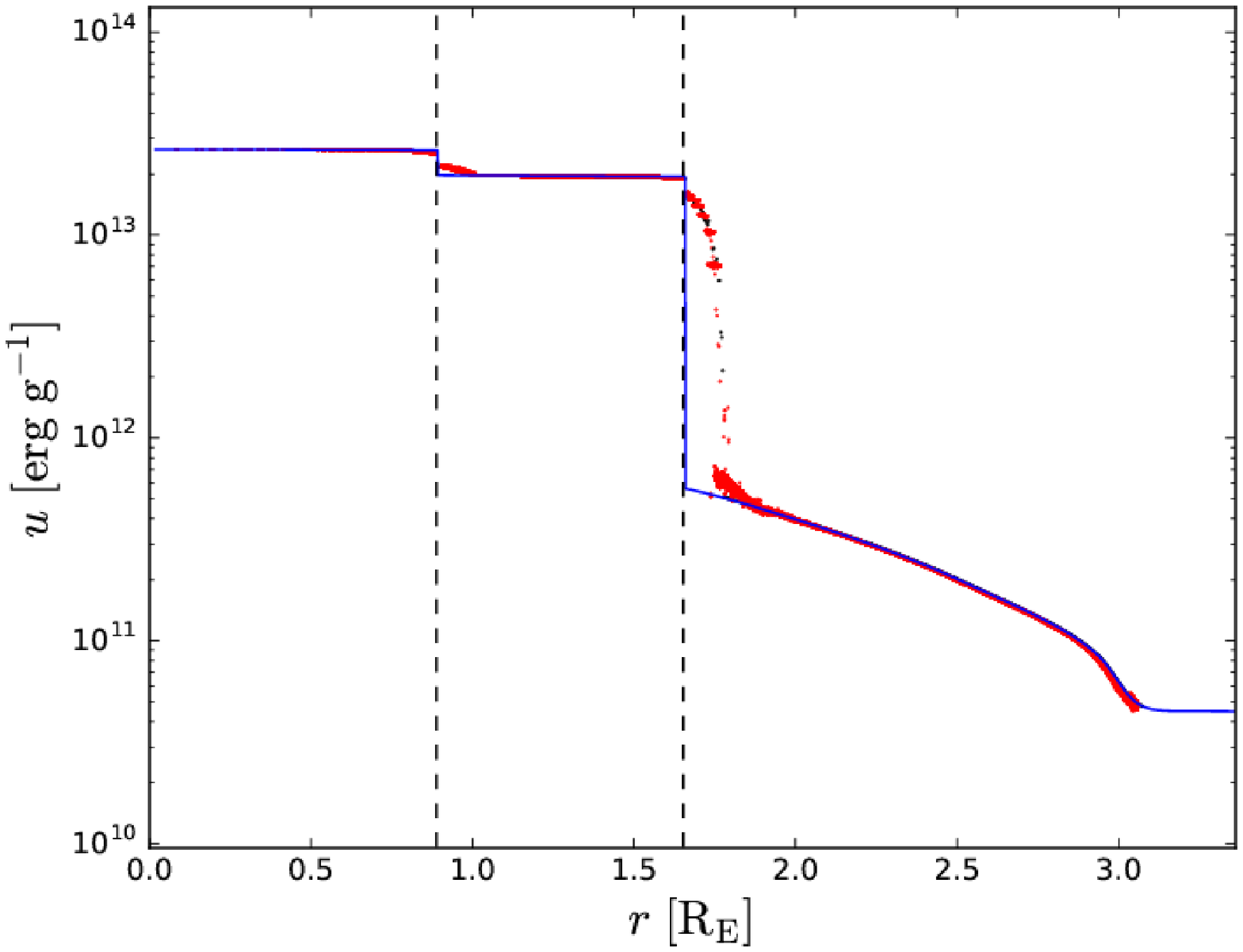}\\
\includegraphics[width=8.0cm]{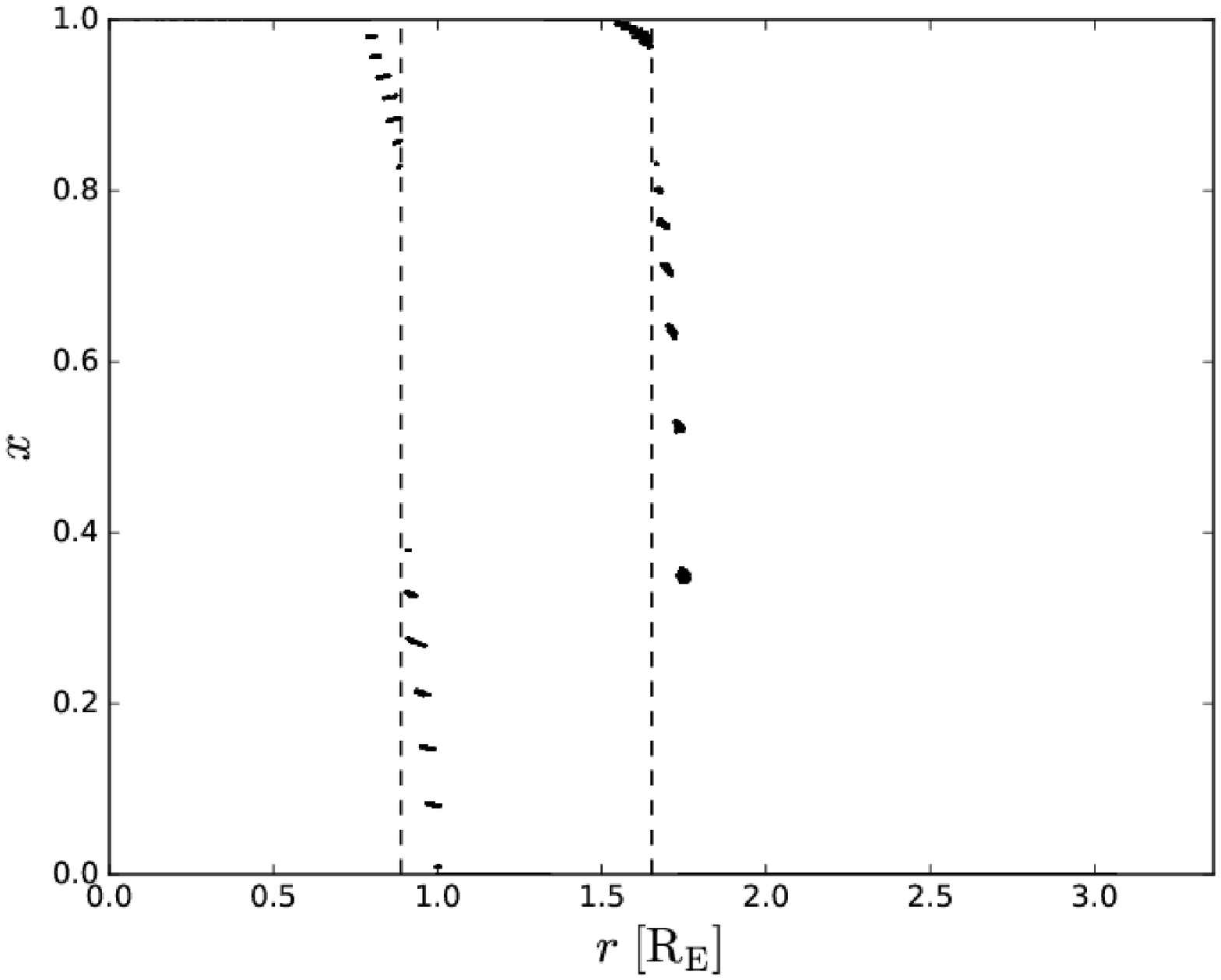}
\includegraphics[width=8.0cm]{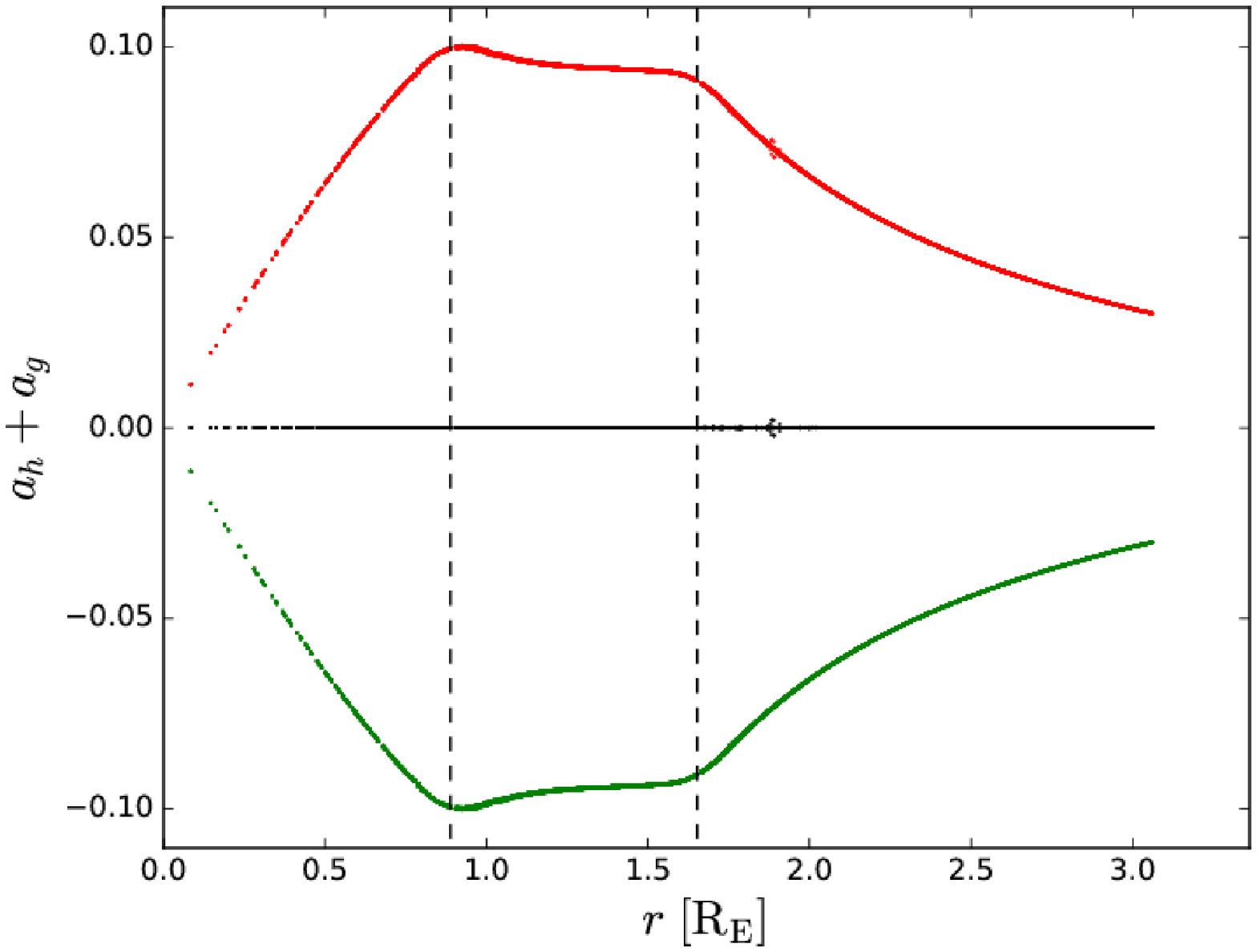}\\
\end{tabular}
\end{center}
\caption{Like Fig.~\ref{Fig:Radial_Profiles_b} but for an isolated model of a Kepler-36c progenitor.}
\label{Fig:Radial_Profiles_c}
\end{figure*}

Figures~\ref{Fig:Radial_Profiles_b} and \ref{Fig:Radial_Profiles_c} show two comparisons of the initial profiles from {\it MESA} and the semi-analytic core equations of state, to the planet profiles after 2000 code timescales in {\it StarSmasher}, relaxed in isolation to both test the stability of the models and minimize the initial spurious noise, where a code timescale is of order unity to the free-fall time of the planets, $t_\mathrm{free-fall}\sim\left(\frac{\rho_\odot}{\rho_\mathrm{planet}}\right)^{1/2}t_\mathrm{code}$.
The planet models are very stable at the end of the relaxation; the radial acceleration on the particles is near zero throughout the profile.
The models for {\it b} and {\it c} for the Kepler-36 progenitor calculations have 12114 and 8190 core particles and 87840 and 91764 gas particles, respectively.

We note that much of the previous literature exploring the mass-radius relationship of sub-Neptunes perform calculations with both a homogeneous and a differentiated equation of state (\citealt{2014ApJ...787..173H} and \citealt{2007ApJ...669.1279S}).
We emphasize that here we are focused on differentiated cores.
Homogeneous cores would be denser and smaller, which would impact the results of collisions simulations.

\subsection{Analysis of Hydro Calculations}
For each output snapshot from the hydrodynamic calculations, we build the planets starting from the dense iron core.
For each particle, in ascending order of distance from the center of mass of the planet, we calculate a Jacobi Constant, \begin{equation}C_\mathrm{J}=\frac{x^2+y^2}{a}+2a\left(\frac{k_1}{r_1}+\frac{k_2}{r_2}\right)-\frac{v_x^2+v_y^2+v_z^2}{n^2a^2},\end{equation} where $n$ is the orbital frequency, $k_1 = M_*/(m_\mathrm{p}+M_*)$ and $\mu_2 = m_\mathrm{p}/(m_\mathrm{p}+M_*)$ are the mass fractions of the star and planet, where $k_1+k_2=1$, $r_1$ and $r_2$ are the distances from the particle to the host star and planet, and the position and velocity vectors are defined in the corotating frame of the planet and the star, with $x^2+y^2+z^2\equiv 1$.
We calculate the Jacobi Constant at $L_1$, assuming a circular orbit \citep{1999ssd..book.....M}, \begin{equation}C_{L_1}\simeq3+3^{4/3}\mu_2^{2/3}-10\frac{\mu_2}{3},\end{equation} and assign the particle to the planet if $C_\mathrm{J} > C_{L_1}$, updating the planet's mass, position, and velocity.
For each particle initially assigned to both planets, we assign them to the planet with lower relative orbital energy with respect to the particle, \begin{equation}E=m\left(\frac{v^2}{2}-\frac{Gm_\mathrm{p}}{r}\right),\end{equation} where $m_\mathrm{p}$ is the mass of the planet, $m$ is the mass of the particle, and $r$ and $v$ are the scalar distance and relative velocity between the particle and the planet.

\begin{figure*}[htp]
\begin{center}
\begin{tabular}{cc}
\includegraphics[width=8.0cm]{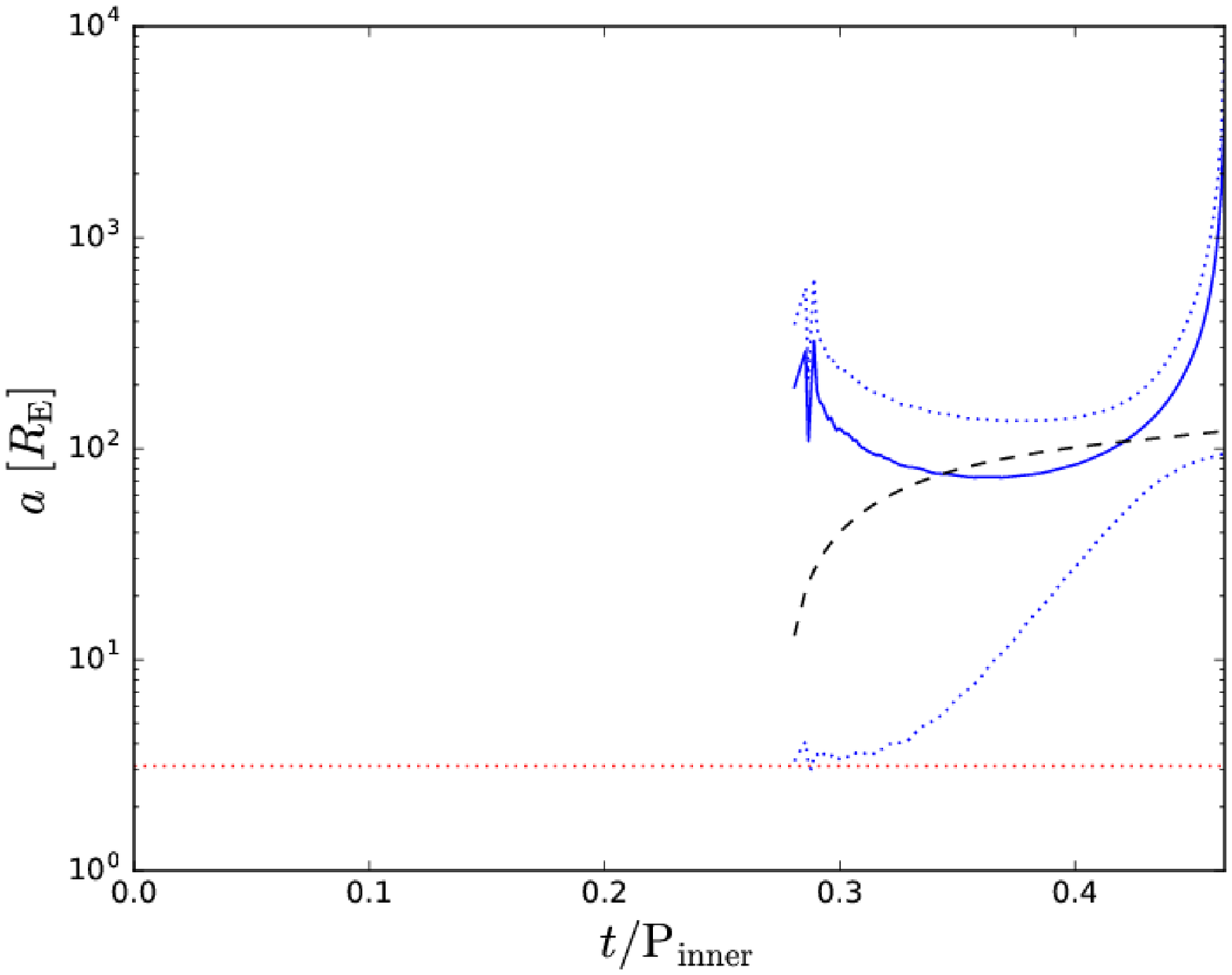}
\includegraphics[width=8.0cm]{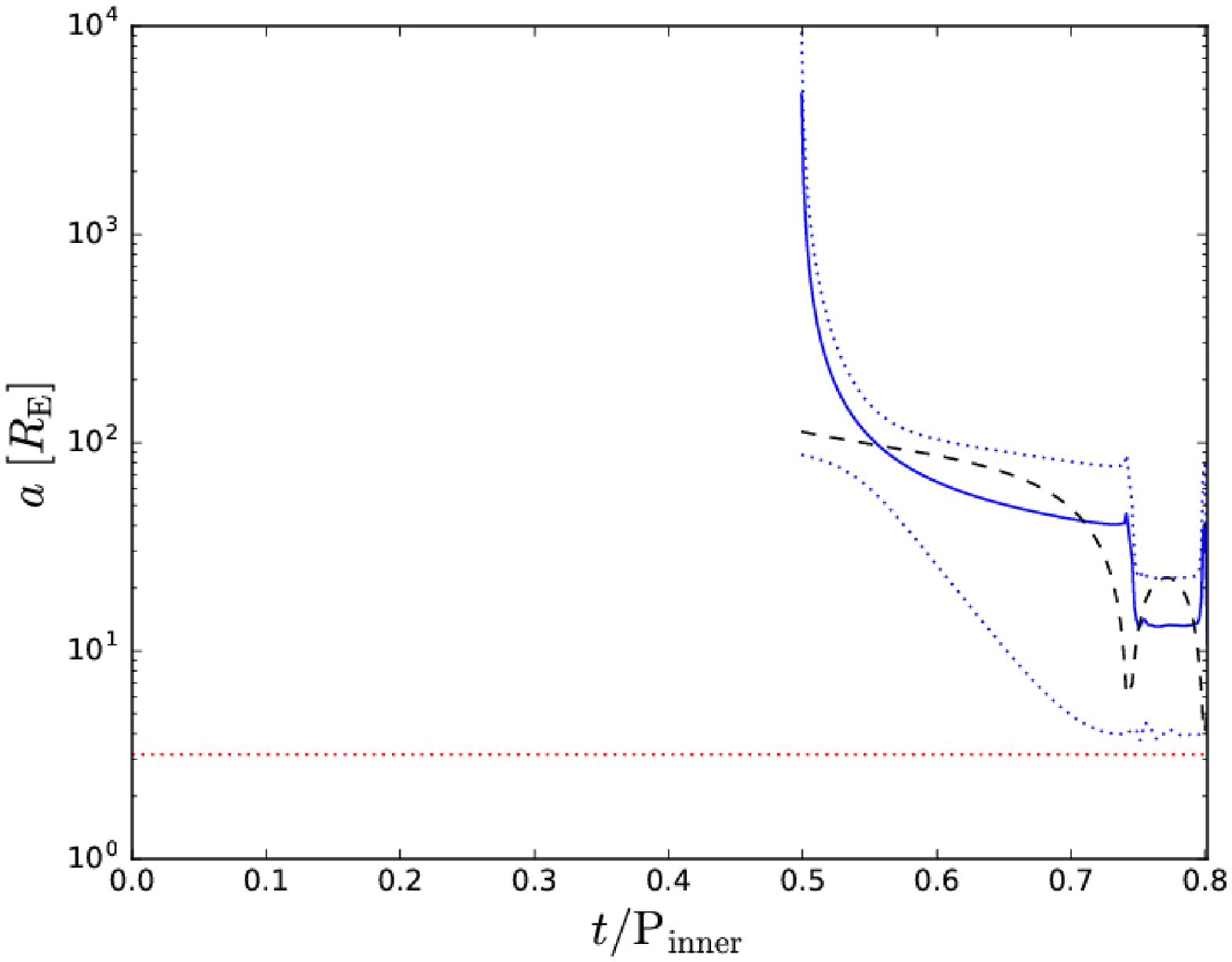} \\
\includegraphics[width=8.0cm]{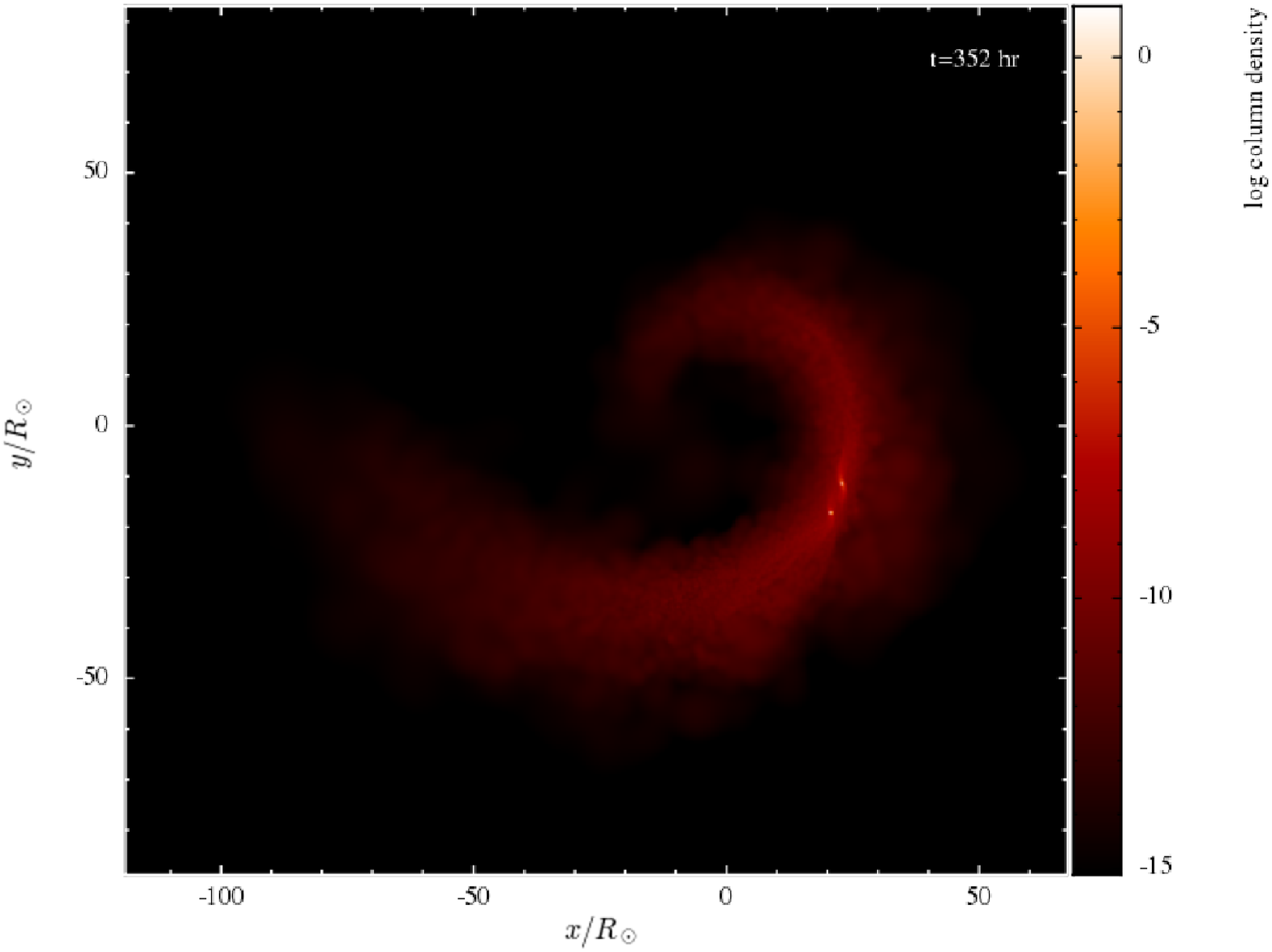}
\includegraphics[width=8.0cm]{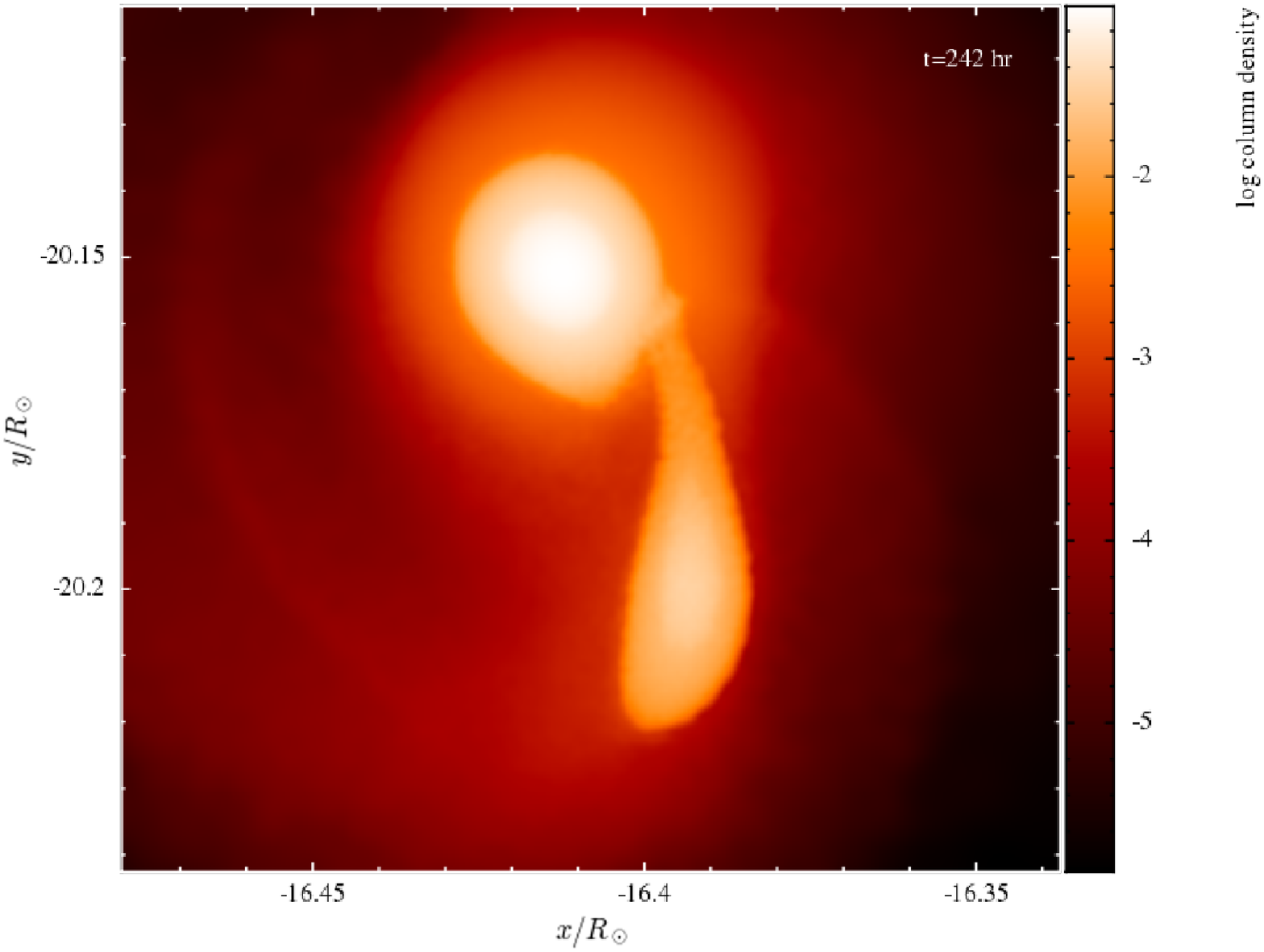}
\end{tabular}
\end{center}
\caption{Relative planet-planet orbits (top) and final snapshots (bottom) of collisions resulting in a scattering (left) and a merger (right).
The semi-major axes (solid blue line), and periastron and apoastron (dotted blue lines) are calculated using the relative separation (dashed black line) and velocity vectors of the two planets.
The calculation is categorized as a merger if the periapsis is less than twice the sum of the core radii (dotted red line), as a potential planet-planet binary if the apoapsis does not exceed the mutual Hill radius between the two planets, and as a scattering if the planets leave their mutual Hill sphere. The snapshots show the logarithm of column density plots in the orbital plane, centered on the host star (left) and the planet-planet merger (right). The axes are scaled to $R_\odot$, and the column density has units of $M_\odot R_\odot^{-2}$}
\label{Fig:Orbit_Examples}
\end{figure*}

We categorize the collision outcomes by examining the relative orbits of the two planets, ignoring the host star, where the semi-major axis, \begin{equation}a=\frac{-G(m_1+m_2)}{2E_\mathrm{orb}},\end{equation} and the eccentricity, \begin{equation}e=\left|\left(\frac{{\bf v}^2}{G(m_1+m_2)}-\frac{1}{|\bf{r}|}\right){\bf r}-\frac{\bf{r}\cdot\bf{v}}{G(m_1+m_2)}\bf{v}\right|,\end{equation} where \begin{equation}E_\mathrm{orb}=\frac{1}{2}\mu{\bf v}^2-\frac{Gm_1m_2}{|\bf{r}|}\end{equation} is the relative orbital energy, $\mu$ is the reduced mass, ${\bf v} = {\bf v}_1-{\bf v}_2,$ and ${\bf r}={\bf r}_1-{\bf r}_2$.
We categorize the outcome as a merger if the relative periapsis of the two planets is less than twice the sum of the core radii (to account for the observed tidal deformation of the cores), as a bound planet-pair if the relative orbital energy results in an apoapsis less than the mutual Hill radius of the planets, and a scattering if the two planets leave their mutual Hill spheres, where the mutual Hill radius, \begin{equation}\label{EQ:RHill}R_\mathrm{H} = a\left(\frac{\mu}{3M_*}\right)^{1/3}.\end{equation}
Figure~\ref{Fig:Orbit_Examples} shows the orbital evolution of two collisions after the initial contact, one resulting in a scattering and the other a merger.

\section{Collisional Outcomes}
\label{Sec:Outcomes}
\begin{figure*}[htp]
\begin{center}
\begin{tabular}{cc}
\includegraphics[width=8.0cm]{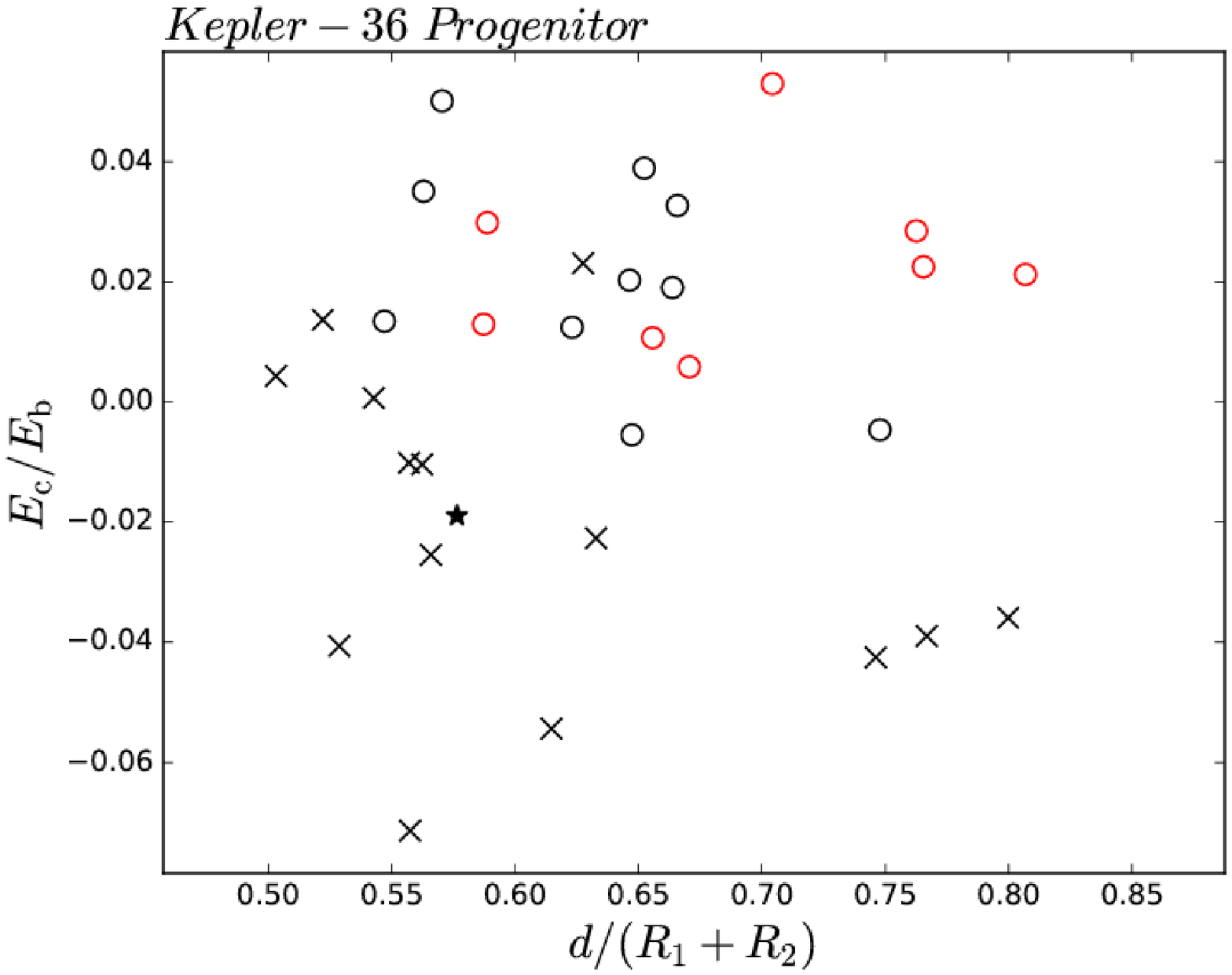} 
\color{white}
\rule{8.0cm}{5.0cm}
\color{black} \\
\includegraphics[width=8.0cm]{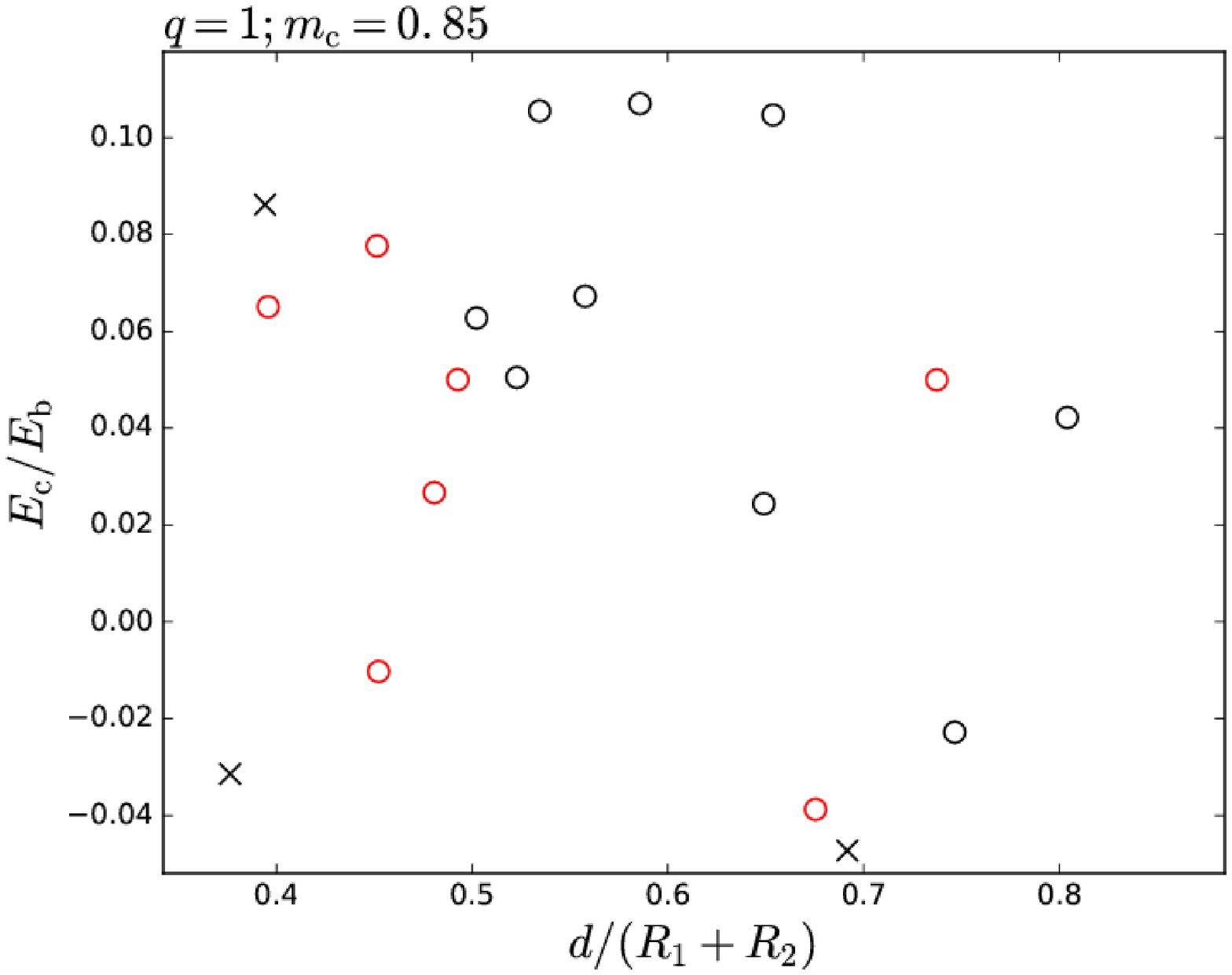}
\includegraphics[width=8.0cm]{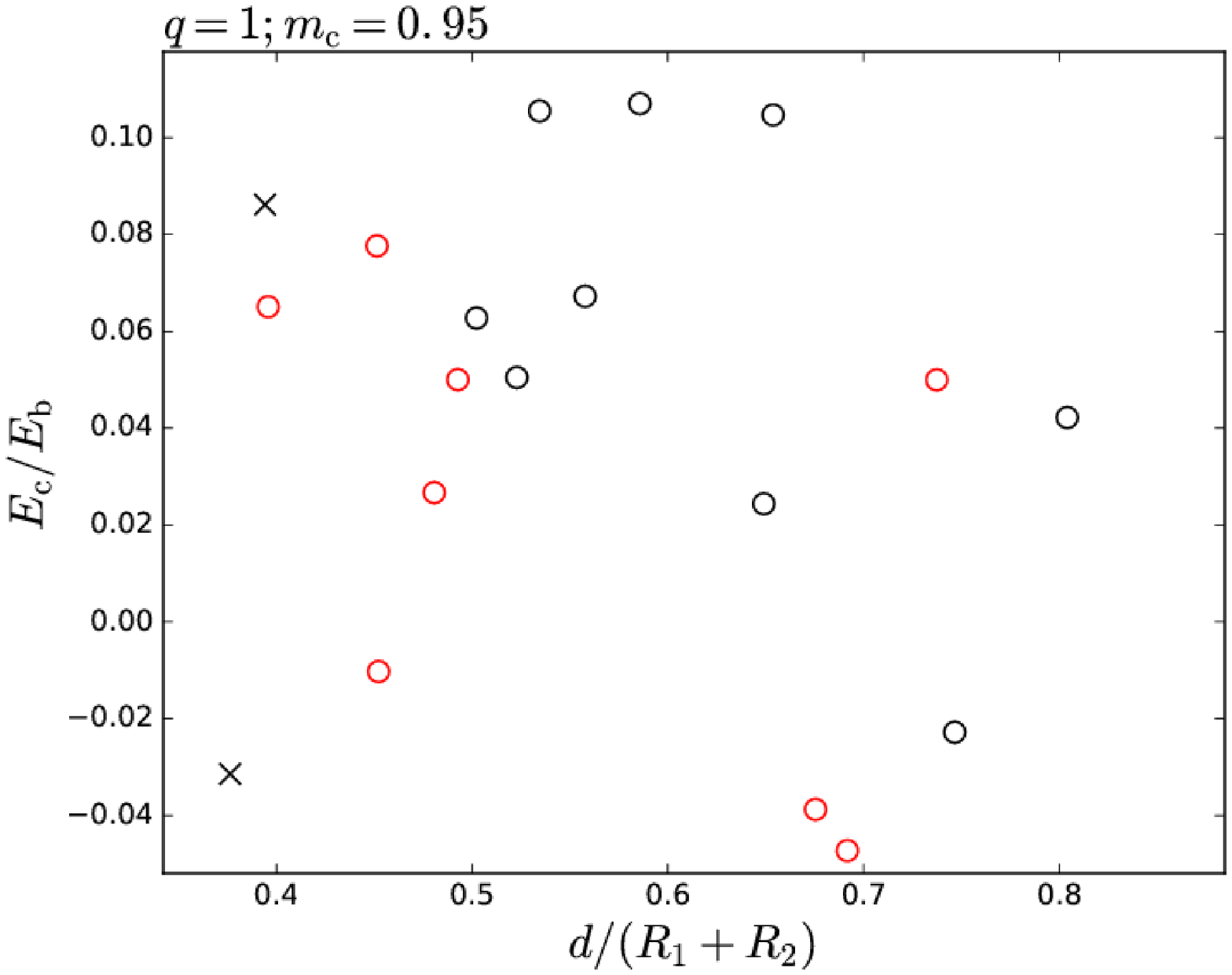} \\
\includegraphics[width=8.0cm]{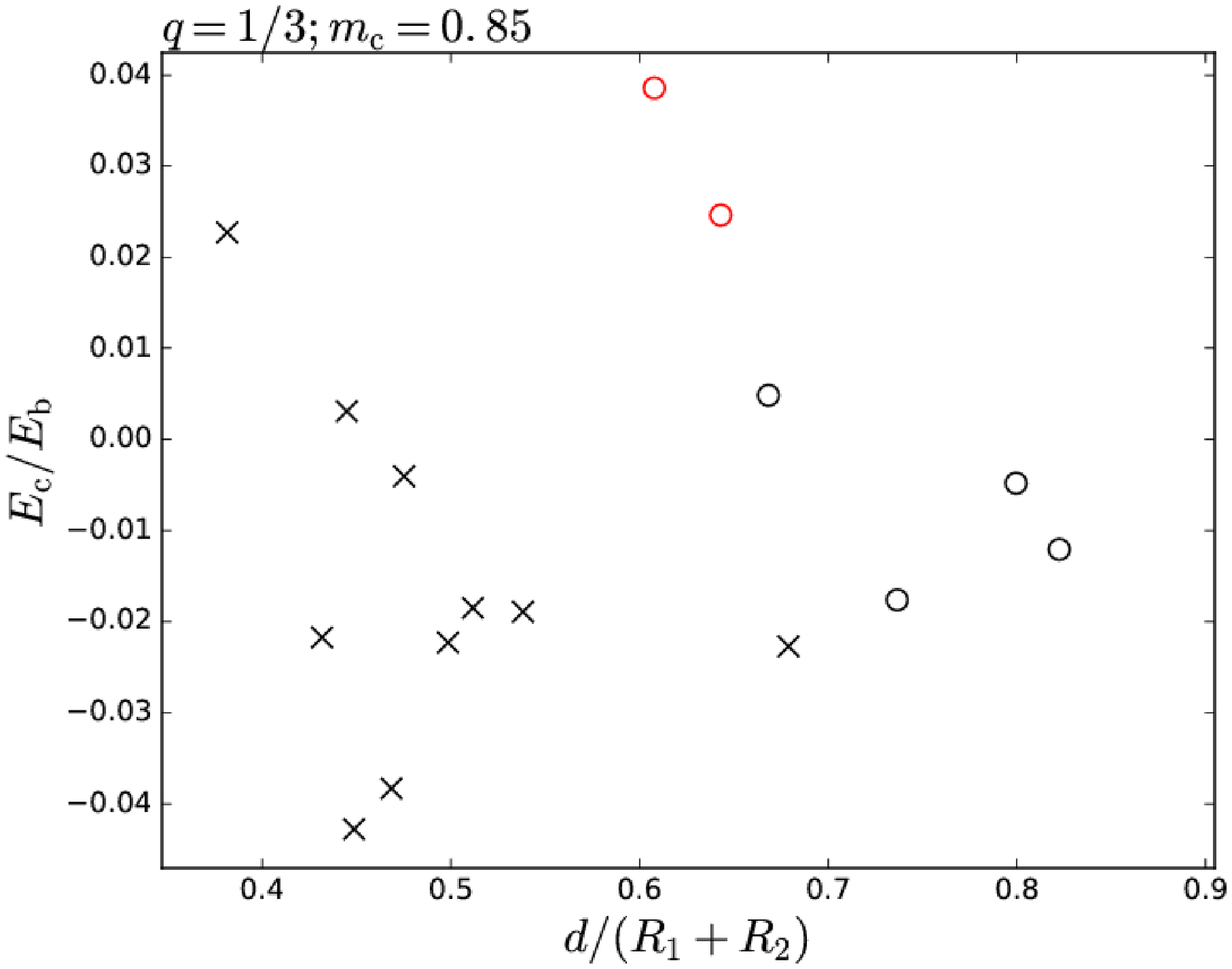}
\includegraphics[width=8.0cm]{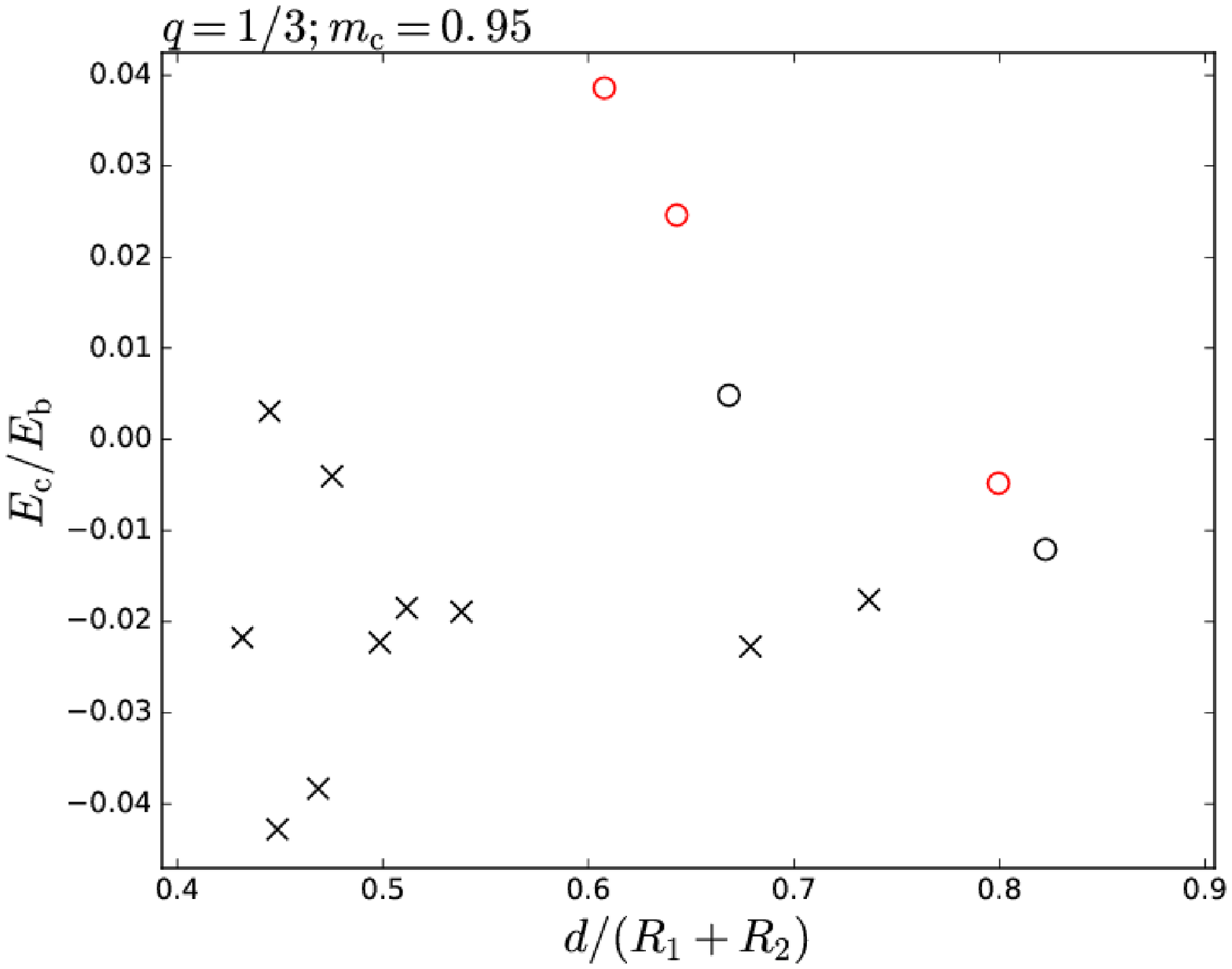}
\end{tabular}
\end{center}
\caption{Results of collisions as a function of the degree of contact, $\eta = d_\mathrm{min}/(R_1+R_2),$ and the collision energy in units of the binding energy, as described in \eqref{EQ:binding_energy}, from the N-body calculations.
In the Kepler-36 progenitor calculations (top left), we find 14 mergers ($\times$), 1 bound planet-pair ($\star$), and 18 scatterings (circle), in the $q=1;\ m_\mathrm{c}=0.85$ (middle left) calculations we find 3 mergers and 16 scatterings, in the $q=1;\ m_\mathrm{c}=0.95$ (middle right) calculations we find 2 mergers and 17 scatterings, in the $q=1/3;\ m_\mathrm{c}=0.85$ (bottom left) calculations we find 9 mergers and 7 scatterings, and in the $q=1/3;\ m_\mathrm{c}=0.95$ (bottom right) calculations we find 10 mergers and 5 scatterings.
Of the 63 collisions resulting in a scattering, 28 exhibit a flipped orbit (red edge-color), where the initially outer planet becomes the inner planet after the collision.}
\label{Fig:Collision_Results}
\end{figure*}

Tables~\ref{TBL:Results_Kepler36}-\ref{TBL:Results_q3} summarize the results of our collision calculations, showing the changes in mass and energy lost in the collision as a function of the N-body collision parameters.
We find that the distances of closest approach and relative collision energies are, in general, larger than the values from the N-body calculations, due to resolving the colliding atmospheres.
The calculations have adequate energy conservation, with the maximum fractional change in total energy $\Delta E_\mathrm{tot}/E_\mathrm{tot,i}<2\times10^{-5}$.
Figure~\ref{Fig:Collision_Results} shows the collision outcomes as a function of the distance of closest approach and collision energy, as calculated in {\it Mercury 6.2}, and in general, scatterings occur more often at larger distances of closest approach and collision energies, and planet-planet captures (bound planet-pairs or mergers) occur at smaller distances of closest approach and energies.
The same initial orbits are used for the two sets of calculations at a given mass ratio but different core mass fractions, and the results are very similar, with the only categorically different outcome occuring in the $q=1/3$ calculations; specifically, a merger in the $m_\mathrm{c}/m = 0.85$ calculations results instead in a scattering in the $m_\mathrm{c}/m = 0.95$ calculations due to less energy dissipated in the latter collision.
In total we find 62 scatterings, 39 mergers, and 1 potential bound-planet-pair.

\begin{deluxetable*}{ccc|cccccccc}
\tabletypesize{\footnotesize}
\tablewidth{14.5cm}
\tablecolumns{9}
\tablecaption{Results of Kepler-36 Progenitor SPH Calculations\label{TBL:Results_Kepler36}}
\tablehead{
    \colhead{} & \multicolumn{2}{c}{N-body} & \multicolumn{2}{c}{SPH} & \\
    \cline{2-3} \cline{4-5}
    \colhead{Run} & \colhead{$d_\mathrm{min}$} & \colhead{$E_\mathrm{c}$} & \colhead{$d_\mathrm{min}$} & \colhead{$E_\mathrm{c}$} & \colhead{$\Delta m_\mathrm{1}$} & \colhead{$\Delta m_\mathrm{2}$} & \colhead{$\Delta E_\mathrm{orb}$} & \colhead{$\epsilon_\mathrm{acc}$} & \colhead{outcome}}
\startdata
$ 1	$ & $	0.576	$ & $	-0.019	$ & $	0.641	$ & $	0.019	$ & $	-0.067	$ & $	0.009	$ & $	0.0404	$ & $	2.55\times10^{-6}	$ & $	bound	$ & \\
$ 2	$ & $	0.628	$ & $	0.023	$ & $	0.606	$ & $	0.023	$ & $	-1.310	$ & $	-0.082	$ & $	0.0491	$ & $	6.37\times10^{-6}	$ & $	merged	$ & \\
$ 3	$ & $	0.615	$ & $	-0.054	$ & $	0.669	$ & $	-0.025	$ & $	-0.604	$ & $	-0.030	$ & $	0.0275	$ & $	5.09\times10^{-7}	$ & $	merged	$ & \\
$ 4	$ & $	0.633	$ & $	-0.023	$ & $	0.668	$ & $	0.011	$ & $	-0.067	$ & $	0.003	$ & $	0.0356	$ & $	5.37\times10^{-5}	$ & $	merged	$ & \\
$ 5	$ & $	0.557	$ & $	-0.071	$ & $	0.647	$ & $	-0.025	$ & $	-1.104	$ & $	-0.293	$ & $	0.0439	$ & $	1.27\times10^{-6}	$ & $	merged	$ & \\
$ 6	$ & $	0.522	$ & $	0.014	$ & $	0.548	$ & $	0.044	$ & $	-0.851	$ & $	-0.704	$ & $	0.0635	$ & $	3.06\times10^{-6}	$ & $	merged	$ & \\
$ 7	$ & $	0.557	$ & $	-0.010	$ & $	0.553	$ & $	0.011	$ & $	-1.288	$ & $	-0.629	$ & $	0.0433	$ & $	7.64\times10^{-7}	$ & $	merged	$ & \\
$ 8	$ & $	0.562	$ & $	-0.010	$ & $	0.610	$ & $	0.005	$ & $	-1.012	$ & $	-0.503	$ & $	0.0455	$ & $	5.09\times10^{-7}	$ & $	merged	$ & \\
$ 9	$ & $	0.529	$ & $	-0.041	$ & $	0.613	$ & $	-0.012	$ & $	-1.029	$ & $	-0.077	$ & $	0.0267	$ & $	2.55\times10^{-7}	$ & $	merged	$ & \\
$ 10	$ & $	0.503	$ & $	0.004	$ & $	0.542	$ & $	0.035	$ & $	-0.968	$ & $	-0.047	$ & $	0.0592	$ & $	1.02\times10^{-6}	$ & $	merged	$ & \\
$ 11	$ & $	0.746	$ & $	-0.043	$ & $	0.768	$ & $	-0.017	$ & $	-0.779	$ & $	-0.216	$ & $	0.0157	$ & $	4.58\times10^{-6}	$ & $	merged	$ & \\
$ 12	$ & $	0.566	$ & $	-0.025	$ & $	0.586	$ & $	-0.023	$ & $	-1.023	$ & $	-0.041	$ & $	0.0193	$ & $	2.55\times10^{-7}	$ & $	merged	$ & \\
$ 13	$ & $	0.543	$ & $	0.001	$ & $	0.539	$ & $	0.017	$ & $	-1.158	$ & $	-0.742	$ & $	0.0512	$ & $	1.27\times10^{-6}	$ & $	merged	$ & \\
$ 14	$ & $	0.800	$ & $	-0.036	$ & $	0.791	$ & $	-0.017	$ & $	-0.008	$ & $	-0.006	$ & $	0.0086	$ & $	1.43\times10^{-5}	$ & $	merged	$ & \\
$ 15	$ & $	0.767	$ & $	-0.039	$ & $	0.772	$ & $	-0.022	$ & $	-0.021	$ & $	-0.043	$ & $	0.0103	$ & $	9.17\times10^{-6}	$ & $	merged	$ & \\
$ 16	$ & $	0.563	$ & $	0.035	$ & $	0.596	$ & $	0.052	$ & $	-0.104	$ & $	0.002	$ & $	0.0589	$ & $	2.09\times10^{-5}	$ & $	unstable	$ & \\
$ 17	$ & $	0.623	$ & $	0.012	$ & $	0.671	$ & $	0.041	$ & $	-0.049	$ & $	-0.001	$ & $	0.0379	$ & $	8.91\times10^{-6}	$ & $	unstable	$ & \\
$ 18	$ & $	0.664	$ & $	0.019	$ & $	0.682	$ & $	0.040	$ & $	-0.013	$ & $	-0.004	$ & $	0.0256	$ & $	2.52\times10^{-5}	$ & $	unstable	$ & \\
$ 19	$ & $	0.587	$ & $	0.013	$ & $	0.691	$ & $	0.029	$ & $	-0.017	$ & $	-0.004	$ & $	0.0257	$ & $	2.04\times10^{-6}	$ & $	unstable	$ & \\
$ 20	$ & $	0.547	$ & $	0.013	$ & $	0.664	$ & $	0.045	$ & $	-0.112	$ & $	-0.007	$ & $	0.0537	$ & $	9.42\times10^{-6}	$ & $	unstable	$ & \\
$ 21	$ & $	0.652	$ & $	0.039	$ & $	0.737	$ & $	0.044	$ & $	-0.001	$ & $	-0.002	$ & $	0.0129	$ & $	3.06\times10^{-6}	$ & $	unstable	$ & \\
$ 22	$ & $	0.666	$ & $	0.033	$ & $	0.771	$ & $	0.047	$ & $	-0.001	$ & $	-0.002	$ & $	0.0111	$ & $	4.58\times10^{-6}	$ & $	unstable	$ & \\
$ 23	$ & $	0.807	$ & $	0.021	$ & $	0.771	$ & $	0.047	$ & $	-0.001	$ & $	-0.002	$ & $	0.0111	$ & $	7.89\times10^{-6}	$ & $	unstable	$ & \\
$ 24	$ & $	0.570	$ & $	0.050	$ & $	0.667	$ & $	0.055	$ & $	-0.095	$ & $	0.002	$ & $	0.0565	$ & $	9.42\times10^{-6}	$ & $	unstable	$ & \\
$ 25	$ & $	0.589	$ & $	0.030	$ & $	0.736	$ & $	0.054	$ & $	-0.008	$ & $	-0.004	$ & $	0.0238	$ & $	2.47\times10^{-5}	$ & $	unstable	$ & \\
$ 26	$ & $	0.671	$ & $	0.006	$ & $	0.737	$ & $	0.015	$ & $	-0.005	$ & $	-0.003	$ & $	0.0182	$ & $	2.55\times10^{-6}	$ & $	unstable	$ & \\
$ 27	$ & $	0.748	$ & $	-0.005	$ & $	0.751	$ & $	0.015	$ & $	-0.001	$ & $	-0.003	$ & $	0.0129	$ & $	2.04\times10^{-6}	$ & $	unstable	$ & \\
$ 28	$ & $	0.656	$ & $	0.011	$ & $	0.564	$ & $	0.057	$ & $	-0.187	$ & $	-0.013	$ & $	0.0595	$ & $	1.27\times10^{-6}	$ & $	unstable	$ & \\
$ 29	$ & $	0.647	$ & $	-0.005	$ & $	0.693	$ & $	0.029	$ & $	-0.043	$ & $	0.000	$ & $	0.0337	$ & $	1.53\times10^{-6}	$ & $	unstable	$ & \\
$ 30	$ & $	0.765	$ & $	0.023	$ & $	0.754	$ & $	0.048	$ & $	-0.001	$ & $	-0.002	$ & $	0.0130	$ & $	3.31\times10^{-6}	$ & $	unstable	$ & \\
$ 31	$ & $	0.763	$ & $	0.028	$ & $	0.776	$ & $	0.045	$ & $	-0.001	$ & $	-0.002	$ & $	0.0101	$ & $	1.53\times10^{-6}	$ & $	unstable	$ & \\
$ 32	$ & $	0.646	$ & $	0.020	$ & $	0.828	$ & $	0.053	$ & $	-0.001	$ & $	-0.001	$ & $	0.0074	$ & $	2.80\times10^{-6}	$ & $	unstable	$ & \\
$ 33	$ & $	0.704	$ & $	0.053	$ & $	0.740	$ & $	0.059	$ & $	-0.003	$ & $	-0.003	$ & $	0.0180	$ & $	1.53\times10^{-6}	$ & $	unstable	$ & \\
\enddata

\tablenotetext{}{$Run$ designates the run number, $d_\mathrm{min}$ and $E_\mathrm{c}$ are the distance of closest approach and energy of each collision from the N-body and SPH calculations in units of the combined physical radii, $R_1+R_2$, and binding energy, $E_\mathrm{b}$, described in \eqref{EQ:binding_energy}, $\Delta m_1$ and $\Delta m_2$ are the changes in mass for the inner and outer planet in Earth masses, $\Delta E_\mathrm{orb}$ is the change in relative planet-planet orbital energy during the collision in units of the binding energy, $\epsilon_\mathrm{acc}$ is the fractional change in the total energy, and outcome designates the result of the collision.}
\end{deluxetable*}

\begin{deluxetable*}{ccc|cccccccc}
\tabletypesize{\footnotesize}
\tablewidth{14.5cm}
\tablecolumns{9}
\tablecaption{Results of $q=1$ SPH Calculations\label{TBL:Results_q1}}
\tablehead{
    \colhead{} & \multicolumn{2}{c}{N-body} & \multicolumn{2}{c}{SPH} & \\
    \cline{2-3} \cline{4-5}
    \colhead{Run} & \colhead{$d_\mathrm{min}$} & \colhead{$E_\mathrm{c}$} & \colhead{$d_\mathrm{min}$} & \colhead{$E_\mathrm{c}$} & \colhead{$\Delta m_\mathrm{1}$} & \colhead{$\Delta m_\mathrm{2}$} & \colhead{$\Delta E_\mathrm{orb}$} & \colhead{$\epsilon_\mathrm{acc}$} & \colhead{outcome}}
\startdata
$ m_\mathrm{c}/m = 0.85$ & \\
$ 1	$ & $	0.394	$ & $	0.086	$ & $	0.408	$ & $	0.087	$ & $	-0.142	$ & $	-0.167	$ & $	0.1333	$ & $	3.36\times10^{-7}	$ & $	merged	$ & \\
$ 2	$ & $	0.376	$ & $	-0.031	$ & $	0.382	$ & $	-0.030	$ & $	-1.213	$ & $	-1.100	$ & $	0.0744	$ & $	6.71\times10^{-7}	$ & $	merged	$ & \\
$ 3	$ & $	0.692	$ & $	-0.047	$ & $	0.690	$ & $	-0.007	$ & $	-0.013	$ & $	-0.011	$ & $	0.0102	$ & $	6.72\times10^{-7}	$ & $	merged	$ & \\
$ 4	$ & $	0.452	$ & $	-0.010	$ & $	0.451	$ & $	0.032	$ & $	-0.020	$ & $	-0.021	$ & $	0.0502	$ & $	2.35\times10^{-6}	$ & $	unstable	$ & \\
$ 5	$ & $	0.737	$ & $	0.050	$ & $	0.680	$ & $	0.067	$ & $	-0.001	$ & $	-0.001	$ & $	0.0047	$ & $	6.71\times10^{-7}	$ & $	unstable	$ & \\
$ 6	$ & $	0.649	$ & $	0.024	$ & $	0.608	$ & $	0.042	$ & $	-0.001	$ & $	-0.001	$ & $	0.0126	$ & $	3.36\times10^{-7}	$ & $	unstable	$ & \\
$ 7	$ & $	0.746	$ & $	-0.023	$ & $	0.755	$ & $	-0.011	$ & $	-0.000	$ & $	-0.000	$ & $	0.0040	$ & $	1.34\times10^{-6}	$ & $	unstable	$ & \\
$ 8	$ & $	0.451	$ & $	0.078	$ & $	0.470	$ & $	0.082	$ & $	-0.021	$ & $	-0.024	$ & $	0.0539	$ & $	3.36\times10^{-7}	$ & $	unstable	$ & \\
$ 9	$ & $	0.396	$ & $	0.065	$ & $	0.384	$ & $	0.093	$ & $	-0.063	$ & $	-0.063	$ & $	0.1045	$ & $	2.01\times10^{-6}	$ & $	unstable	$ & \\
$ 10	$ & $	0.534	$ & $	0.106	$ & $	0.561	$ & $	0.136	$ & $	-0.003	$ & $	-0.003	$ & $	0.0185	$ & $	2.35\times10^{-6}	$ & $	unstable	$ & \\
$ 11	$ & $	0.654	$ & $	0.105	$ & $	0.742	$ & $	0.127	$ & $	-0.000	$ & $	-0.000	$ & $	0.0034	$ & $	1.68\times10^{-6}	$ & $	unstable	$ & \\
$ 12	$ & $	0.558	$ & $	0.067	$ & $	0.554	$ & $	0.117	$ & $	-0.003	$ & $	-0.003	$ & $	0.0187	$ & $	3.36\times10^{-7}	$ & $	unstable	$ & \\
$ 13	$ & $	0.804	$ & $	0.042	$ & $	0.795	$ & $	0.081	$ & $	-0.000	$ & $	-0.000	$ & $	0.0027	$ & $	3.36\times10^{-6}	$ & $	unstable	$ & \\
$ 14	$ & $	0.481	$ & $	0.027	$ & $	0.538	$ & $	0.103	$ & $	-0.003	$ & $	-0.004	$ & $	0.0210	$ & $	1.01\times10^{-6}	$ & $	unstable	$ & \\
$ 15	$ & $	0.502	$ & $	0.063	$ & $	0.531	$ & $	0.082	$ & $	-0.004	$ & $	-0.003	$ & $	0.0234	$ & $	<10^{-7}		$ & $	unstable	$ & \\
$ 16	$ & $	0.675	$ & $	-0.039	$ & $	0.676	$ & $	-0.010	$ & $	-0.001	$ & $	-0.000	$ & $	0.0055	$ & $	1.01\times10^{-6}	$ & $	unstable	$ & \\
$ 17	$ & $	0.586	$ & $	0.107	$ & $	0.571	$ & $	0.115	$ & $	-0.029	$ & $	-0.036	$ & $	0.0017	$ & $	2.69\times10^{-6}	$ & $	unstable	$ & \\
$ 18	$ & $	0.523	$ & $	0.050	$ & $	0.564	$ & $	0.064	$ & $	-0.003	$ & $	-0.002	$ & $	0.0122	$ & $	1.01\times10^{-6}	$ & $	unstable	$ & \\
$ 19	$ & $	0.493	$ & $	0.050	$ & $	0.474	$ & $	0.099	$ & $	-0.017	$ & $	-0.016	$ & $	0.0485	$ & $	1.68\times10^{-6}	$ & $	unstable	$ & \\
\hline \\
$ m_\mathrm{c}/m = 0.95$ & \\
$ 1	$ & $	0.394	$ & $	0.086	$ & $	0.578	$ & $	0.060	$ & $	-0.046	$ & $	-0.036	$ & $	0.0876	$ & $	1.11\times10^{-6}	$ & $	merged	$ & \\
$ 2	$ & $	0.376	$ & $	-0.031	$ & $	0.540	$ & $	-0.021	$ & $	-1.112	$ & $	-1.152	$ & $	0.0417	$ & $	7.39\times10^{-7}	$ & $	merged	$ & \\
$ 3	$ & $	0.452	$ & $	-0.010	$ & $	0.640	$ & $	0.022	$ & $	-0.005	$ & $	-0.005	$ & $	0.0375	$ & $	7.39\times10^{-7}	$ & $	unstable	$ & \\
$ 4	$ & $	0.737	$ & $	0.050	$ & $	0.999	$ & $	0.046	$ & $	-0.000	$ & $	-0.000	$ & $	0.0009	$ & $	5.91\times10^{-6}	$ & $	unstable	$ & \\
$ 5	$ & $	0.649	$ & $	0.024	$ & $	0.888	$ & $	0.029	$ & $	-0.000	$ & $	-0.000	$ & $	0.0042	$ & $	4.43\times10^{-6}	$ & $	unstable	$ & \\
$ 6	$ & $	0.746	$ & $	-0.023	$ & $	1.107	$ & $	-0.007	$ & $	-0.000	$ & $	0.000	$ & $	0.0008	$ & $	4.80\times10^{-6}	$ & $	unstable	$ & \\
$ 7	$ & $	0.451	$ & $	0.078	$ & $	0.690	$ & $	0.056	$ & $	-0.005	$ & $	-0.006	$ & $	0.0390	$ & $	3.69\times10^{-7}	$ & $	unstable	$ & \\
$ 8	$ & $	0.396	$ & $	0.065	$ & $	0.551	$ & $	0.063	$ & $	-0.032	$ & $	-0.031	$ & $	0.0721	$ & $	1.11\times10^{-6}	$ & $	unstable	$ & \\
$ 9	$ & $	0.534	$ & $	0.106	$ & $	0.816	$ & $	0.093	$ & $	-0.000	$ & $	-0.000	$ & $	0.0079	$ & $	4.06\times10^{-6}	$ & $	unstable	$ & \\
$ 10	$ & $	0.654	$ & $	0.105	$ & $	1.090	$ & $	0.087	$ & $	-0.000	$ & $	-0.000	$ & $	0.0007	$ & $	5.17\times10^{-6}	$ & $	unstable	$ & \\
$ 11	$ & $	0.558	$ & $	0.067	$ & $	0.803	$ & $	0.080	$ & $	-0.000	$ & $	-0.000	$ & $	0.0082	$ & $	2.22\times10^{-6}	$ & $	unstable	$ & \\
$ 12	$ & $	0.804	$ & $	0.042	$ & $	1.168	$ & $	0.055	$ & $	-0.000	$ & $	0.000	$ & $	0.0005	$ & $	6.65\times10^{-6}	$ & $	unstable	$ & \\
$ 13	$ & $	0.481	$ & $	0.027	$ & $	0.779	$ & $	0.070	$ & $	-0.000	$ & $	-0.000	$ & $	0.0094	$ & $	7.39\times10^{-7}	$ & $	unstable	$ & \\
$ 14	$ & $	0.502	$ & $	0.063	$ & $	0.767	$ & $	0.056	$ & $	-0.001	$ & $	-0.001	$ & $	0.0112	$ & $	3.69\times10^{-6}	$ & $	unstable	$ & \\
$ 15	$ & $	0.675	$ & $	-0.039	$ & $	0.992	$ & $	-0.007	$ & $	-0.000	$ & $	0.000	$ & $	0.0006	$ & $	4.06\times10^{-6}	$ & $	unstable	$ & \\
$ 16	$ & $	0.586	$ & $	0.107	$ & $	1.090	$ & $	0.087	$ & $	-0.000	$ & $	-0.000	$ & $	0.0007	$ & $	5.17\times10^{-6}	$ & $	unstable	$ & \\
$ 17	$ & $	0.523	$ & $	0.050	$ & $	0.823	$ & $	0.043	$ & $	-0.000	$ & $	-0.000	$ & $	0.0049	$ & $	2.59\times10^{-6}	$ & $	unstable	$ & \\
$ 18	$ & $	0.493	$ & $	0.050	$ & $	0.693	$ & $	0.067	$ & $	-0.003	$ & $	-0.003	$ & $	0.0325	$ & $	3.32\times10^{-6}	$ & $	unstable	$ & \\
$ 19	$ & $	0.692	$ & $	-0.047	$ & $	1.013	$ & $	-0.004	$ & $	-0.000	$ & $	0.000	$ & $	0.0036	$ & $	3.69\times10^{-6}	$ & $	unstable	$ & \\
\enddata

\tablenotetext{}{$Run$ designates the run number, $d_\mathrm{min}$ and $E_\mathrm{c}$ are the distance of closest approach and energy of each collision from the N-body and SPH calculations in units of the combined physical radii, $R_1+R_2$, and binding energy, $E_\mathrm{b}$, described in \eqref{EQ:binding_energy}, $\Delta m_1$ and $\Delta m_2$ are the changes in mass for the inner and outer planet in Earth masses, $\Delta E_\mathrm{orb}$ is the change in relative planet-planet orbital energy during the collision in units of the binding energy, $\epsilon_\mathrm{acc}$ is the fractional change in the total energy, and outcome designates the result of the collision.}
\end{deluxetable*}

\begin{deluxetable*}{ccc|cccccccc}
\tabletypesize{\footnotesize}
\tablewidth{14.5cm}
\tablecolumns{9}
\tablecaption{Results of $q=1/3$ SPH Calculations\label{TBL:Results_q3}}
\tablehead{
    \colhead{} & \multicolumn{2}{c}{N-body} & \multicolumn{2}{c}{SPH} & \\
    \cline{2-3} \cline{4-5}
    \colhead{Run} & \colhead{$d_\mathrm{min}$} & \colhead{$E_\mathrm{c}$} & \colhead{$d_\mathrm{min}$} & \colhead{$E_\mathrm{c}$} & \colhead{$\Delta m_\mathrm{1}$} & \colhead{$\Delta m_\mathrm{2}$} & \colhead{$\Delta E_\mathrm{orb}$} & \colhead{$\epsilon_\mathrm{acc}$} & \colhead{outcome}}
\startdata
$ m_\mathrm{c}/m = 0.85$ & \\
$ 1	$ & $	0.512	$ & $	-0.018	$ & $	0.530	$ & $	-0.018	$ & $	-1.135	$ & $	-0.016	$ & $	0.0185	$ & $	6.02\times10^{-6}	$ & $	merged	$ & \\
$ 2	$ & $	0.445	$ & $	0.003	$ & $	0.530	$ & $	0.020	$ & $	-1.066	$ & $	-0.264	$ & $	0.0000	$ & $	3.19\times10^{-6}	$ & $	merged	$ & \\
$ 3	$ & $	0.449	$ & $	-0.043	$ & $	0.485	$ & $	-0.014	$ & $	-1.177	$ & $	-0.172	$ & $	0.0000	$ & $	4.07\times10^{-6}	$ & $	merged	$ & \\
$ 4	$ & $	0.475	$ & $	-0.004	$ & $	0.532	$ & $	-0.002	$ & $	-0.814	$ & $	-0.102	$ & $	0.0102	$ & $	4.78\times10^{-6}	$ & $	merged	$ & \\
$ 5	$ & $	0.498	$ & $	-0.022	$ & $	0.530	$ & $	-0.018	$ & $	-0.644	$ & $	-0.038	$ & $	0.0092	$ & $	6.37\times10^{-6}	$ & $	merged	$ & \\
$ 6	$ & $	0.432	$ & $	-0.022	$ & $	0.519	$ & $	-0.015	$ & $	-1.238	$ & $	-0.139	$ & $	0.0235	$ & $	6.72\times10^{-6}	$ & $	merged	$ & \\
$ 7	$ & $	0.679	$ & $	-0.023	$ & $	0.683	$ & $	-0.021	$ & $	-1.171	$ & $	0.088	$ & $	0.0052	$ & $	5.13\times10^{-6}	$ & $	merged	$ & \\
$ 8	$ & $	0.469	$ & $	-0.038	$ & $	0.489	$ & $	-0.018	$ & $	-1.104	$ & $	-0.151	$ & $	0.0000	$ & $	4.78\times10^{-6}	$ & $	merged	$ & \\
$ 9	$ & $	0.538	$ & $	-0.019	$ & $	0.566	$ & $	-0.019	$ & $	-1.061	$ & $	-0.088	$ & $	0.0160	$ & $	5.84\times10^{-6}	$ & $	merged	$ & \\
$ 10	$ & $	0.382	$ & $	0.023	$ & $	0.440	$ & $	0.041	$ & $	-1.324	$ & $	-0.286	$ & $	0.0105	$ & $	1.24\times10^{-6}	$ & $	merged	$ & \\
$ 11	$ & $	0.800	$ & $	-0.005	$ & $	0.887	$ & $	-0.005	$ & $	-0.014	$ & $	-0.002	$ & $	0.0002	$ & $	3.01\times10^{-6}	$ & $	unstable	$ & \\
$ 12	$ & $	0.822	$ & $	-0.012	$ & $	0.850	$ & $	0.002	$ & $	-0.019	$ & $	0.015	$ & $	0.0017	$ & $	6.37\times10^{-6}	$ & $	unstable	$ & \\
$ 13	$ & $	0.737	$ & $	-0.018	$ & $	0.737	$ & $	-0.005	$ & $	-0.032	$ & $	0.019	$ & $	0.0048	$ & $	4.96\times10^{-6}	$ & $	unstable	$ & \\
$ 14	$ & $	0.643	$ & $	0.025	$ & $	0.678	$ & $	0.027	$ & $	-0.039	$ & $	0.013	$ & $	0.0088	$ & $	4.60\times10^{-6}	$ & $	unstable	$ & \\
$ 15	$ & $	0.608	$ & $	0.039	$ & $	0.631	$ & $	0.039	$ & $	-0.057	$ & $	0.014	$ & $	0.0125	$ & $	5.13\times10^{-6}	$ & $	unstable	$ & \\
$ 16	$ & $	0.668	$ & $	0.005	$ & $	0.668	$ & $	0.020	$ & $	-0.036	$ & $	0.013	$ & $	0.0052	$ & $	1.42\times10^{-6}	$ & $	unstable	$ & \\
\hline \\
$ m_\mathrm{c}/m = 0.95$ & \\
$ 1	$ & $	0.512	$ & $	-0.018	$ & $	0.744	$ & $	-0.013	$ & $	-0.938	$ & $	0.006	$ & $	0.0000	$ & $	1.77\times10^{-6}	$ & $	merged	$ & \\
$ 2	$ & $	0.445	$ & $	0.003	$ & $	0.741	$ & $	0.014	$ & $	-1.133	$ & $	-0.036	$ & $	0.0075	$ & $	9.03\times10^{-6}	$ & $	merged	$ & \\
$ 3	$ & $	0.449	$ & $	-0.043	$ & $	0.666	$ & $	-0.010	$ & $	-1.186	$ & $	-0.074	$ & $	0.0000	$ & $	1.18\times10^{-6}	$ & $	merged	$ & \\
$ 4	$ & $	0.498	$ & $	-0.022	$ & $	0.738	$ & $	-0.013	$ & $	-0.693	$ & $	-0.022	$ & $	0.0000	$ & $	3.93\times10^{-7}	$ & $	merged	$ & \\
$ 5	$ & $	0.737	$ & $	-0.018	$ & $	1.045	$ & $	-0.003	$ & $	-0.018	$ & $	-0.015	$ & $	0.0018	$ & $	1.81\times10^{-5}	$ & $	merged	$ & \\
$ 6	$ & $	0.432	$ & $	-0.022	$ & $	0.739	$ & $	-0.011	$ & $	-0.852	$ & $	-0.016	$ & $	0.0001	$ & $	1.77\times10^{-6}	$ & $	merged	$ & \\
$ 7	$ & $	0.679	$ & $	-0.023	$ & $	0.971	$ & $	-0.016	$ & $	-0.018	$ & $	-0.002	$ & $	0.0017	$ & $	1.37\times10^{-6}	$ & $	merged	$ & \\
$ 8	$ & $	0.469	$ & $	-0.038	$ & $	0.678	$ & $	-0.013	$ & $	-0.987	$ & $	-0.058	$ & $	0.0000	$ & $	9.03\times10^{-6}	$ & $	merged	$ & \\
$ 9	$ & $	0.538	$ & $	-0.019	$ & $	0.796	$ & $	-0.013	$ & $	-1.017	$ & $	-0.019	$ & $	0.0102	$ & $	4.91\times10^{-6}	$ & $	merged	$ & \\
$ 10	$ & $	0.475	$ & $	-0.004	$ & $	0.766	$ & $	-0.002	$ & $	-0.845	$ & $	-0.036	$ & $	0.0000	$ & $	2.94\times10^{-6}	$ & $	merged	$ & \\
$ 11	$ & $	0.800	$ & $	-0.005	$ & $	1.046	$ & $	-0.003	$ & $	-0.004	$ & $	0.003	$ & $	0.0007	$ & $	2.16\times10^{-6}	$ & $	unstable	$ & \\
$ 12	$ & $	0.822	$ & $	-0.012	$ & $	1.204	$ & $	0.002	$ & $	-0.000	$ & $	0.000	$ & $	0.0003	$ & $	2.75\times10^{-6}	$ & $	unstable	$ & \\
$ 13	$ & $	0.643	$ & $	0.025	$ & $	0.962	$ & $	0.020	$ & $	-0.006	$ & $	0.005	$ & $	0.0042	$ & $	2.16\times10^{-6}	$ & $	unstable	$ & \\
$ 14	$ & $	0.608	$ & $	0.039	$ & $	0.893	$ & $	0.028	$ & $	-0.012	$ & $	0.006	$ & $	0.0090	$ & $	3.93\times10^{-6}	$ & $	unstable	$ & \\
$ 15	$ & $	0.668	$ & $	0.005	$ & $	0.951	$ & $	0.014	$ & $	-0.005	$ & $	0.004	$ & $	0.0018	$ & $	1.77\times10^{-6}	$ & $	unstable	$ & \\
\enddata

\tablenotetext{}{$Run$ designates the run number, $d_\mathrm{min}$ and $E_\mathrm{c}$ are the distance of closest approach and energy of each collision from the N-body and SPH calculations in units of the combined physical radii, $R_1+R_2$, and binding energy, $E_\mathrm{b}$, described in \eqref{EQ:binding_energy}, $\Delta m_1$ and $\Delta m_2$ are the changes in mass for the inner and outer planet in Earth masses, $\Delta E_\mathrm{orb}$ is the change in relative planet-planet orbital energy during the collision in units of the binding energy, $\epsilon_\mathrm{acc}$ is the fractional change in the total energy, and outcome designates the result of the collision.}
\end{deluxetable*}

Tables~\ref{TBL:Observable_Kepler36}-\ref{TBL:Observable_q3} summarize the orbits and planet properties of the remnant planets, where the angular momentum deficit is used to quantify the stability of the systems, discussed in more detail in \S\ref{SSec:Scatterings}, and the densities are estimated using the same method presented in {\it Paper 1}, where we match the total mass and core mass fraction of the remnant planets to a grid of {\it MESA} models.
The initial gas mass fractions and densities are listed on the first row for each set of calculations.
We examine each set of outcomes in more detail and discuss the orbits and planet structures of the remnants.
In \S\ref{Sec:Predictions} we use these calculations to develop prescriptions for predicting the outcomes and modeling the changes in mass during these collisions.

\begin{deluxetable*}{c|ccccccc|cccccc}
\tabletypesize{\footnotesize}
\tablewidth{14.cm}
\tablecolumns{11}
\tablecaption{Post-Collision Orbits and Planet Properties of Kepler-36 Progenitor Calculations\label{TBL:Observable_Kepler36}}
\tablehead{
    \colhead{} & \multicolumn{7}{c}{Orbit Properties} & \multicolumn{5}{c}{Planet Properties} & \\
    \cline{2-8} \cline{9-13}
    \colhead{Run} & \colhead{$a_1$} & \colhead{$e_1$} & \colhead{$a_2$} & \colhead{$e_2$} & \colhead{$P_2/P_1$} & \colhead{$C$} & \colhead{$C_\mathrm{min}$} & \colhead{$m_\mathrm{gas,1}$} & \colhead{$m_\mathrm{gas,2}$} & \colhead{$\rho_1$} & \colhead{$\rho_2$} & \colhead{$\rho_1/\rho_2$}}
\startdata
$ 	$ & $	$ & $	$ & $	$ & $	$ & $	$  & \multicolumn{2}{l}{Initial Properties}	 & $	0.233	$ & $	0.708	$ & $	1.12	$ & $	1.29	$ & $	1.15	$ & \\
\hline
$ 1	$ & $	0.121	$ & $	0.028	$ & $	0.125	$ & $	0.042	$ & $	1.055	$ & $	0.003	$ & $	0.003	$ & $	0.024	$ & $	0.086	$ & $	1.875	$ & $	1.376	$ & $	0.734	$ & \\
$ 2	$ & $	0.137	$ & $	0.115	$ & $	0.118	$ & $	0.061	$ & $	1.251	$ & $	0.010	$ & $	0.005	$ & $	0.003	$ & $	0.051	$ & $	2.263	$ & $	1.672	$ & $	0.739	$ & \\
$ 3	$ & $	0.120	$ & $	0.045	$ & $	0.127	$ & $	0.081	$ & $	1.083	$ & $	0.011	$ & $	0.010	$ & $	0.015	$ & $	0.065	$ & $	2.041	$ & $	1.571	$ & $	0.770	$ & \\
$ 4	$ & $	0.127	$ & $	0.148	$ & $	0.123	$ & $	0.023	$ & $	1.046	$ & $	0.019	$ & $	0.019	$ & $	0.021	$ & $	0.086	$ & $	1.970	$ & $	1.372	$ & $	0.696	$ & \\
$ 5	$ & $	0.138	$ & $	0.109	$ & $	0.119	$ & $	0.097	$ & $	1.000	$ & $	0.018	$ & $	0.013	$ & $	0.008	$ & $	0.049	$ & $	2.216	$ & $	1.744	$ & $	0.787	$ & \\
$ 6	$ & $	0.104	$ & $	0.199	$ & $	0.140	$ & $	0.114	$ & $	1.000	$ & $	0.036	$ & $	0.010	$ & $	0.008	$ & $	0.036	$ & $	2.484	$ & $	2.182	$ & $	0.879	$ & \\
$ 7	$ & $	0.141	$ & $	0.143	$ & $	0.116	$ & $	0.069	$ & $	1.000	$ & $	0.014	$ & $	0.006	$ & $	0.002	$ & $	0.033	$ & $	2.897	$ & $	2.270	$ & $	0.784	$ & \\
$ 8	$ & $	0.116	$ & $	0.193	$ & $	0.132	$ & $	0.059	$ & $	1.000	$ & $	0.022	$ & $	0.018	$ & $	0.004	$ & $	0.031	$ & $	2.681	$ & $	2.355	$ & $	0.878	$ & \\
$ 9	$ & $	0.132	$ & $	0.115	$ & $	0.119	$ & $	0.022	$ & $	1.167	$ & $	0.007	$ & $	0.004	$ & $	0.008	$ & $	0.054	$ & $	2.290	$ & $	1.645	$ & $	0.718	$ & \\
$ 10	$ & $	0.134	$ & $	0.133	$ & $	0.120	$ & $	0.057	$ & $	1.000	$ & $	0.014	$ & $	0.011	$ & $	0.004	$ & $	0.068	$ & $	2.585	$ & $	1.553	$ & $	0.601	$ & \\
$ 11	$ & $	0.118	$ & $	0.135	$ & $	0.127	$ & $	0.053	$ & $	1.122	$ & $	0.014	$ & $	0.012	$ & $	0.012	$ & $	0.050	$ & $	2.053	$ & $	1.639	$ & $	0.798	$ & \\
$ 12	$ & $	0.136	$ & $	0.121	$ & $	0.118	$ & $	0.056	$ & $	1.230	$ & $	0.012	$ & $	0.007	$ & $	0.007	$ & $	0.056	$ & $	2.327	$ & $	1.635	$ & $	0.702	$ & \\
$ 13	$ & $	0.124	$ & $	0.134	$ & $	0.126	$ & $	0.088	$ & $	1.000	$ & $	0.017	$ & $	0.017	$ & $	0.002	$ & $	0.023	$ & $	3.274	$ & $	2.421	$ & $	0.739	$ & \\
$ 14	$ & $	0.167	$ & $	0.249	$ & $	0.109	$ & $	0.167	$ & $	1.000	$ & $	0.099	$ & $	0.026	$ & $	0.040	$ & $	0.088	$ & $	1.633	$ & $	1.340	$ & $	0.821	$ & \\
$ 15	$ & $	0.099	$ & $	0.265	$ & $	0.149	$ & $	0.171	$ & $	1.000	$ & $	0.099	$ & $	0.032	$ & $	0.034	$ & $	0.079	$ & $	1.729	$ & $	1.464	$ & $	0.847	$ & \\
$ 16	$ & $	0.120	$ & $	0.008	$ & $	0.126	$ & $	0.004	$ & $	1.071	$ & $	0.000	$ & $	-0.001	$ & $	0.013	$ & $	0.081	$ & $	2.451	$ & $	1.451	$ & $	0.592	$ & \\
$ 17	$ & $	0.119	$ & $	0.057	$ & $	0.126	$ & $	0.006	$ & $	1.092	$ & $	0.003	$ & $	0.001	$ & $	0.026	$ & $	0.087	$ & $	1.843	$ & $	1.357	$ & $	0.736	$ & \\
$ 18	$ & $	0.121	$ & $	0.036	$ & $	0.125	$ & $	0.030	$ & $	1.047	$ & $	0.002	$ & $	0.002	$ & $	0.038	$ & $	0.088	$ & $	1.682	$ & $	1.332	$ & $	0.792	$ & \\
$ 19	$ & $	0.128	$ & $	0.017	$ & $	0.121	$ & $	0.045	$ & $	1.093	$ & $	0.003	$ & $	0.002	$ & $	0.036	$ & $	0.088	$ & $	1.706	$ & $	1.331	$ & $	0.780	$ & \\
$ 20	$ & $	0.122	$ & $	0.034	$ & $	0.124	$ & $	0.004	$ & $	1.025	$ & $	0.001	$ & $	0.001	$ & $	0.010	$ & $	0.080	$ & $	2.957	$ & $	1.464	$ & $	0.495	$ & \\
$ 21	$ & $	0.121	$ & $	0.040	$ & $	0.125	$ & $	0.042	$ & $	1.057	$ & $	0.004	$ & $	0.003	$ & $	0.042	$ & $	0.089	$ & $	1.554	$ & $	1.321	$ & $	0.850	$ & \\
$ 22	$ & $	0.121	$ & $	0.055	$ & $	0.125	$ & $	0.027	$ & $	1.045	$ & $	0.004	$ & $	0.003	$ & $	0.043	$ & $	0.089	$ & $	1.549	$ & $	1.320	$ & $	0.852	$ & \\
$ 23	$ & $	0.132	$ & $	0.083	$ & $	0.119	$ & $	0.027	$ & $	1.157	$ & $	0.007	$ & $	0.003	$ & $	0.043	$ & $	0.089	$ & $	1.545	$ & $	1.320	$ & $	0.855	$ & \\
$ 24	$ & $	0.120	$ & $	0.003	$ & $	0.125	$ & $	0.006	$ & $	1.061	$ & $	0.000	$ & $	-0.001	$ & $	0.015	$ & $	0.083	$ & $	2.325	$ & $	1.432	$ & $	0.616	$ & \\
$ 25	$ & $	0.134	$ & $	0.081	$ & $	0.118	$ & $	0.043	$ & $	1.202	$ & $	0.008	$ & $	0.002	$ & $	0.040	$ & $	0.088	$ & $	1.634	$ & $	1.330	$ & $	0.814	$ & \\
$ 26	$ & $	0.128	$ & $	0.032	$ & $	0.121	$ & $	0.048	$ & $	1.091	$ & $	0.004	$ & $	0.003	$ & $	0.041	$ & $	0.088	$ & $	1.603	$ & $	1.327	$ & $	0.828	$ & \\
$ 27	$ & $	0.121	$ & $	0.041	$ & $	0.125	$ & $	0.036	$ & $	1.049	$ & $	0.003	$ & $	0.003	$ & $	0.043	$ & $	0.089	$ & $	1.546	$ & $	1.324	$ & $	0.857	$ & \\
$ 28	$ & $	0.130	$ & $	0.039	$ & $	0.120	$ & $	0.024	$ & $	1.127	$ & $	0.002	$ & $	-0.000	$ & $	0.008	$ & $	0.079	$ & $	3.680	$ & $	1.472	$ & $	0.400	$ & \\
$ 29	$ & $	0.122	$ & $	0.027	$ & $	0.124	$ & $	0.030	$ & $	1.034	$ & $	0.002	$ & $	0.002	$ & $	0.029	$ & $	0.087	$ & $	1.795	$ & $	1.353	$ & $	0.753	$ & \\
$ 30	$ & $	0.135	$ & $	0.086	$ & $	0.117	$ & $	0.058	$ & $	1.236	$ & $	0.011	$ & $	0.003	$ & $	0.043	$ & $	0.089	$ & $	1.547	$ & $	1.320	$ & $	0.853	$ & \\
$ 31	$ & $	0.130	$ & $	0.076	$ & $	0.120	$ & $	0.030	$ & $	1.129	$ & $	0.006	$ & $	0.004	$ & $	0.043	$ & $	0.089	$ & $	1.547	$ & $	1.319	$ & $	0.852	$ & \\
$ 32	$ & $	0.120	$ & $	0.044	$ & $	0.126	$ & $	0.048	$ & $	1.084	$ & $	0.005	$ & $	0.004	$ & $	0.043	$ & $	0.089	$ & $	1.544	$ & $	1.315	$ & $	0.852	$ & \\
$ 33	$ & $	0.135	$ & $	0.092	$ & $	0.118	$ & $	0.048	$ & $	1.232	$ & $	0.011	$ & $	0.003	$ & $	0.042	$ & $	0.089	$ & $	1.580	$ & $	1.324	$ & $	0.838	$ & \\
\enddata

\tablenotetext{}{$Run$ designates the run number, $a$, $e$, $m_\mathrm{gas}$, and $\rho$ are the semi-major axis (in AU), eccentricity, gas mass (in Earth masses), and density (in CGS) of the inner out and outer planets, $P_2/P_1$ is the period ratio, $C$ and $C_\mathrm{min}$ are the angular momentum deficit and minimum possible angular momentum deficit, normalized as described in \S\ref{SSec:Scatterings}, and $\rho_1/\rho_2$ is the density ratio.
The initial gas mass fractions and densities are listed on the first row of each set of calculations.}
\end{deluxetable*}

\begin{deluxetable*}{c|ccccccc|cccccc}
\tabletypesize{\footnotesize}
\tablewidth{14.cm}
\tablecolumns{11}
\tablecaption{Post-Collision Orbits and Planet Properties of $q=1$ Calculations\label{TBL:Observable_q1}}
\tablehead{
    \colhead{} & \multicolumn{7}{c}{Orbit Properties} & \multicolumn{5}{c}{Planet Properties} & \\
    \cline{2-8} \cline{9-13}
    \colhead{Run} & \colhead{$a_1$} & \colhead{$e_1$} & \colhead{$a_2$} & \colhead{$e_2$} & \colhead{$P_2/P_1$} & \colhead{$C$} & \colhead{$C_\mathrm{min}$} & \colhead{$m_\mathrm{gas,1}$} & \colhead{$m_\mathrm{gas,2}$} & \colhead{$\rho_1$} & \colhead{$\rho_2$} & \colhead{$\rho_1/\rho_2$}}
\startdata
$ m_\mathrm{c}/m = 0.85 $ & $	$ & $	$ & $	$ & $	$ & $	$ & \multicolumn{2}{l}{Initial Properties}	 & $	0.600	$ & $	0.600	$ & $	0.509	$ & $	0.509	$ & $	1.00	$ & \\
\hline
$ 1	$ & $	0.102	$ & $	0.060	$ & $	0.118	$ & $	0.148	$ & $	1.000	$ & $	0.016	$ & $	0.011	$ & $	0.079	$ & $	0.067	$ & $	0.829	$ & $	0.856	$ & $	1.033	$ & \\
$ 2	$ & $	0.117	$ & $	0.145	$ & $	0.102	$ & $	0.048	$ & $	1.000	$ & $	0.006	$ & $	0.005	$ & $	0.006	$ & $	0.009	$ & $	1.705	$ & $	1.962	$ & $	1.150	$ & \\
$ 3	$ & $	0.127	$ & $	0.109	$ & $	0.096	$ & $	0.194	$ & $	1.000	$ & $	0.032	$ & $	0.013	$ & $	0.139	$ & $	0.139	$ & $	0.503	$ & $	0.504	$ & $	1.001	$ & \\
$ 4	$ & $	0.112	$ & $	0.033	$ & $	0.105	$ & $	0.078	$ & $	1.107	$ & $	0.005	$ & $	0.004	$ & $	0.136	$ & $	0.135	$ & $	0.499	$ & $	0.498	$ & $	0.998	$ & \\
$ 5	$ & $	0.113	$ & $	0.042	$ & $	0.105	$ & $	0.062	$ & $	1.110	$ & $	0.004	$ & $	0.002	$ & $	0.144	$ & $	0.144	$ & $	0.508	$ & $	0.508	$ & $	1.000	$ & \\
$ 6	$ & $	0.104	$ & $	0.095	$ & $	0.114	$ & $	0.022	$ & $	1.141	$ & $	0.006	$ & $	0.004	$ & $	0.144	$ & $	0.143	$ & $	0.508	$ & $	0.508	$ & $	1.000	$ & \\
$ 7	$ & $	0.106	$ & $	0.066	$ & $	0.112	$ & $	0.082	$ & $	1.078	$ & $	0.007	$ & $	0.007	$ & $	0.144	$ & $	0.144	$ & $	0.509	$ & $	0.508	$ & $	1.000	$ & \\
$ 8	$ & $	0.113	$ & $	0.107	$ & $	0.105	$ & $	0.053	$ & $	1.117	$ & $	0.009	$ & $	0.008	$ & $	0.135	$ & $	0.134	$ & $	0.498	$ & $	0.496	$ & $	0.996	$ & \\
$ 9	$ & $	0.111	$ & $	0.041	$ & $	0.106	$ & $	0.042	$ & $	1.075	$ & $	0.002	$ & $	0.002	$ & $	0.117	$ & $	0.116	$ & $	0.474	$ & $	0.473	$ & $	0.999	$ & \\
$ 10	$ & $	0.105	$ & $	0.067	$ & $	0.112	$ & $	0.092	$ & $	1.103	$ & $	0.009	$ & $	0.007	$ & $	0.143	$ & $	0.143	$ & $	0.508	$ & $	0.508	$ & $	1.000	$ & \\
$ 11	$ & $	0.106	$ & $	0.096	$ & $	0.112	$ & $	0.054	$ & $	1.083	$ & $	0.008	$ & $	0.007	$ & $	0.144	$ & $	0.144	$ & $	0.509	$ & $	0.508	$ & $	1.000	$ & \\
$ 12	$ & $	0.101	$ & $	0.127	$ & $	0.118	$ & $	0.076	$ & $	1.271	$ & $	0.014	$ & $	0.008	$ & $	0.143	$ & $	0.143	$ & $	0.508	$ & $	0.508	$ & $	0.999	$ & \\
$ 13	$ & $	0.105	$ & $	0.040	$ & $	0.113	$ & $	0.091	$ & $	1.122	$ & $	0.007	$ & $	0.005	$ & $	0.144	$ & $	0.144	$ & $	0.509	$ & $	0.508	$ & $	1.000	$ & \\
$ 14	$ & $	0.117	$ & $	0.103	$ & $	0.101	$ & $	0.059	$ & $	1.251	$ & $	0.009	$ & $	0.004	$ & $	0.143	$ & $	0.142	$ & $	0.507	$ & $	0.507	$ & $	0.999	$ & \\
$ 15	$ & $	0.102	$ & $	0.017	$ & $	0.117	$ & $	0.144	$ & $	1.232	$ & $	0.014	$ & $	0.010	$ & $	0.143	$ & $	0.142	$ & $	0.507	$ & $	0.507	$ & $	1.000	$ & \\
$ 16	$ & $	0.111	$ & $	0.092	$ & $	0.106	$ & $	0.042	$ & $	1.065	$ & $	0.007	$ & $	0.006	$ & $	0.144	$ & $	0.144	$ & $	0.508	$ & $	0.508	$ & $	1.000	$ & \\
$ 17	$ & $	0.106	$ & $	0.082	$ & $	0.113	$ & $	0.082	$ & $	1.099	$ & $	0.009	$ & $	0.008	$ & $	0.132	$ & $	0.128	$ & $	0.493	$ & $	0.488	$ & $	0.990	$ & \\
$ 18	$ & $	0.107	$ & $	0.054	$ & $	0.111	$ & $	0.104	$ & $	1.058	$ & $	0.009	$ & $	0.009	$ & $	0.143	$ & $	0.143	$ & $	0.508	$ & $	0.508	$ & $	1.000	$ & \\
$ 19	$ & $	0.110	$ & $	0.056	$ & $	0.107	$ & $	0.057	$ & $	1.049	$ & $	0.004	$ & $	0.004	$ & $	0.137	$ & $	0.137	$ & $	0.501	$ & $	0.501	$ & $	1.000	$ & \\
\hline
$ m_\mathrm{c}/m = 0.95 $ & $	$ & $	$ & $	$ & $	$ & $	$ & \multicolumn{2}{l}{Initial Properties}	 & $	0.200	$ & $	0.200	$ & $	0.977	$ & $	0.977	$ & $	1.00	$ & \\
\hline
$ 1	$ & $	0.114	$ & $	0.034	$ & $	0.104	$ & $	0.136	$ & $	1.000	$ & $	0.013	$ & $	0.010	$ & $	0.025	$ & $	0.028	$ & $	1.648	$ & $	1.605	$ & $	0.974	$ & \\
$ 2	$ & $	0.106	$ & $	0.064	$ & $	0.112	$ & $	0.105	$ & $	1.000	$ & $	0.004	$ & $	0.004	$ & $	0.003	$ & $	0.002	$ & $	2.119	$ & $	2.225	$ & $	1.050	$ & \\
$ 3	$ & $	0.112	$ & $	0.039	$ & $	0.106	$ & $	0.072	$ & $	1.084	$ & $	0.004	$ & $	0.004	$ & $	0.045	$ & $	0.044	$ & $	1.305	$ & $	1.314	$ & $	1.007	$ & \\
$ 4	$ & $	0.112	$ & $	0.041	$ & $	0.105	$ & $	0.062	$ & $	1.107	$ & $	0.004	$ & $	0.003	$ & $	0.047	$ & $	0.047	$ & $	1.198	$ & $	1.208	$ & $	1.008	$ & \\
$ 5	$ & $	0.104	$ & $	0.098	$ & $	0.114	$ & $	0.022	$ & $	1.152	$ & $	0.007	$ & $	0.004	$ & $	0.047	$ & $	0.047	$ & $	1.200	$ & $	1.209	$ & $	1.008	$ & \\
$ 6	$ & $	0.106	$ & $	0.069	$ & $	0.111	$ & $	0.079	$ & $	1.069	$ & $	0.007	$ & $	0.007	$ & $	0.047	$ & $	0.047	$ & $	1.198	$ & $	1.208	$ & $	1.009	$ & \\
$ 7	$ & $	0.114	$ & $	0.113	$ & $	0.104	$ & $	0.046	$ & $	1.142	$ & $	0.010	$ & $	0.008	$ & $	0.044	$ & $	0.044	$ & $	1.322	$ & $	1.345	$ & $	1.018	$ & \\
$ 8	$ & $	0.111	$ & $	0.040	$ & $	0.106	$ & $	0.042	$ & $	1.071	$ & $	0.002	$ & $	0.002	$ & $	0.032	$ & $	0.031	$ & $	1.568	$ & $	1.579	$ & $	1.007	$ & \\
$ 9	$ & $	0.105	$ & $	0.066	$ & $	0.112	$ & $	0.091	$ & $	1.104	$ & $	0.008	$ & $	0.007	$ & $	0.046	$ & $	0.046	$ & $	1.210	$ & $	1.219	$ & $	1.007	$ & \\
$ 10	$ & $	0.106	$ & $	0.097	$ & $	0.112	$ & $	0.053	$ & $	1.086	$ & $	0.008	$ & $	0.007	$ & $	0.047	$ & $	0.047	$ & $	1.198	$ & $	1.208	$ & $	1.009	$ & \\
$ 11	$ & $	0.101	$ & $	0.128	$ & $	0.118	$ & $	0.076	$ & $	1.277	$ & $	0.014	$ & $	0.008	$ & $	0.047	$ & $	0.046	$ & $	1.206	$ & $	1.216	$ & $	1.008	$ & \\
$ 12	$ & $	0.105	$ & $	0.041	$ & $	0.113	$ & $	0.092	$ & $	1.123	$ & $	0.007	$ & $	0.005	$ & $	0.047	$ & $	0.047	$ & $	1.198	$ & $	1.208	$ & $	1.009	$ & \\
$ 13	$ & $	0.118	$ & $	0.104	$ & $	0.101	$ & $	0.061	$ & $	1.260	$ & $	0.010	$ & $	0.004	$ & $	0.046	$ & $	0.046	$ & $	1.211	$ & $	1.219	$ & $	1.007	$ & \\
$ 14	$ & $	0.101	$ & $	0.015	$ & $	0.117	$ & $	0.147	$ & $	1.245	$ & $	0.015	$ & $	0.010	$ & $	0.046	$ & $	0.046	$ & $	1.213	$ & $	1.222	$ & $	1.007	$ & \\
$ 15	$ & $	0.112	$ & $	0.073	$ & $	0.105	$ & $	0.072	$ & $	1.103	$ & $	0.007	$ & $	0.006	$ & $	0.047	$ & $	0.047	$ & $	1.198	$ & $	1.208	$ & $	1.008	$ & \\
$ 16	$ & $	0.106	$ & $	0.097	$ & $	0.112	$ & $	0.053	$ & $	1.086	$ & $	0.008	$ & $	0.007	$ & $	0.047	$ & $	0.047	$ & $	1.198	$ & $	1.208	$ & $	1.009	$ & \\
$ 17	$ & $	0.107	$ & $	0.051	$ & $	0.111	$ & $	0.107	$ & $	1.063	$ & $	0.009	$ & $	0.009	$ & $	0.047	$ & $	0.046	$ & $	1.206	$ & $	1.215	$ & $	1.008	$ & \\
$ 18	$ & $	0.111	$ & $	0.064	$ & $	0.106	$ & $	0.052	$ & $	1.074	$ & $	0.005	$ & $	0.004	$ & $	0.045	$ & $	0.045	$ & $	1.271	$ & $	1.281	$ & $	1.008	$ & \\
$ 19	$ & $	0.111	$ & $	0.051	$ & $	0.106	$ & $	0.077	$ & $	1.075	$ & $	0.006	$ & $	0.005	$ & $	0.047	$ & $	0.047	$ & $	1.198	$ & $	1.208	$ & $	1.008	$ & \\
\enddata

\tablenotetext{}{$Run$ designates the run number, $a$, $e$, $m_\mathrm{gas}$, and $\rho$ are the semi-major axis (in AU), eccentricity, gas mass (in Earth masses), and density (in CGS) of the inner out and outer planets, $P_2/P_1$ is the period ratio, $C$ and $C_\mathrm{min}$ are the angular momentum deficit and minimum possible angular momentum deficit, normalized as described in \S\ref{SSec:Scatterings}, and $\rho_1/\rho_2$ is the density ratio.
The initial gas mass fractions and densities are listed on the first row of each set of calculations.}
\end{deluxetable*}

\begin{deluxetable*}{c|ccccccc|cccccc}
\tabletypesize{\footnotesize}
\tablewidth{14.cm}
\tablecolumns{11}
\tablecaption{Post-Collision Orbits and Planet Properties of $q=1/3$ Calculations\label{TBL:Observable_q3}}
\tablehead{
    \colhead{} & \multicolumn{7}{c}{Orbit Properties} & \multicolumn{5}{c}{Planet Properties} & \\
    \cline{2-8} \cline{9-13}
    \colhead{Run} & \colhead{$a_1$} & \colhead{$e_1$} & \colhead{$a_2$} & \colhead{$e_2$} & \colhead{$P_2/P_1$} & \colhead{$C$} & \colhead{$C_\mathrm{min}$} & \colhead{$m_\mathrm{gas,1}$} & \colhead{$m_\mathrm{gas,2}$} & \colhead{$\rho_1$} & \colhead{$\rho_2$} & \colhead{$\rho_1/\rho_2$}}
\startdata
$ m_\mathrm{c}/m = 0.85 $ & $	$ & $	$ & $	$ & $	$ & $	$ & \multicolumn{2}{l}{Initial Properties}	 & $	0.200	$ & $	0.600	$ & $	0.509	$ & $	1.204	$ & $	2.37	$ & \\
\hline
$ 1	$ & $	0.153	$ & $	0.288	$ & $	0.106	$ & $	0.095	$ & $	1.000	$ & $	0.046	$ & $	0.022	$ & $	0.006	$ & $	0.140	$ & $	1.849	$ & $	1.224	$ & $	0.662	$ & \\
$ 2	$ & $	0.117	$ & $	0.211	$ & $	0.114	$ & $	0.112	$ & $	1.000	$ & $	0.039	$ & $	0.039	$ & $	0.005	$ & $	0.108	$ & $	1.893	$ & $	1.252	$ & $	0.661	$ & \\
$ 3	$ & $	0.142	$ & $	0.246	$ & $	0.108	$ & $	0.009	$ & $	1.509	$ & $	0.019	$ & $	0.006	$ & $	0.003	$ & $	0.112	$ & $	1.821	$ & $	1.254	$ & $	0.688	$ & \\
$ 4	$ & $	0.109	$ & $	0.152	$ & $	0.115	$ & $	0.052	$ & $	1.083	$ & $	0.015	$ & $	0.014	$ & $	0.012	$ & $	0.127	$ & $	1.667	$ & $	1.228	$ & $	0.737	$ & \\
$ 5	$ & $	0.110	$ & $	0.068	$ & $	0.115	$ & $	0.088	$ & $	1.075	$ & $	0.018	$ & $	0.017	$ & $	0.021	$ & $	0.136	$ & $	1.307	$ & $	1.226	$ & $	0.938	$ & \\
$ 6	$ & $	0.148	$ & $	0.234	$ & $	0.107	$ & $	0.141	$ & $	1.000	$ & $	0.055	$ & $	0.038	$ & $	0.002	$ & $	0.119	$ & $	2.101	$ & $	1.237	$ & $	0.589	$ & \\
$ 7	$ & $	0.125	$ & $	0.139	$ & $	0.111	$ & $	0.113	$ & $	1.200	$ & $	0.032	$ & $	0.029	$ & $	0.006	$ & $	0.148	$ & $	1.813	$ & $	1.223	$ & $	0.674	$ & \\
$ 8	$ & $	0.133	$ & $	0.196	$ & $	0.109	$ & $	0.074	$ & $	1.000	$ & $	0.023	$ & $	0.016	$ & $	0.008	$ & $	0.124	$ & $	2.050	$ & $	1.226	$ & $	0.598	$ & \\
$ 9	$ & $	0.133	$ & $	0.116	$ & $	0.110	$ & $	0.147	$ & $	1.000	$ & $	0.047	$ & $	0.039	$ & $	0.008	$ & $	0.132	$ & $	2.122	$ & $	1.224	$ & $	0.577	$ & \\
$ 10	$ & $	0.165	$ & $	0.339	$ & $	0.104	$ & $	0.063	$ & $	2.010	$ & $	0.040	$ & $	0.009	$ & $	0.001	$ & $	0.109	$ & $	2.492	$ & $	1.246	$ & $	0.500	$ & \\
$ 11	$ & $	0.095	$ & $	0.339	$ & $	0.123	$ & $	0.065	$ & $	1.469	$ & $	0.082	$ & $	0.056	$ & $	0.138	$ & $	0.146	$ & $	0.502	$ & $	1.215	$ & $	2.419	$ & \\
$ 12	$ & $	0.111	$ & $	0.103	$ & $	0.115	$ & $	0.113	$ & $	1.043	$ & $	0.033	$ & $	0.033	$ & $	0.136	$ & $	0.148	$ & $	0.500	$ & $	1.212	$ & $	2.426	$ & \\
$ 13	$ & $	0.101	$ & $	0.035	$ & $	0.119	$ & $	0.123	$ & $	1.283	$ & $	0.032	$ & $	0.022	$ & $	0.131	$ & $	0.149	$ & $	0.491	$ & $	1.211	$ & $	2.464	$ & \\
$ 14	$ & $	0.127	$ & $	0.088	$ & $	0.110	$ & $	0.062	$ & $	1.237	$ & $	0.013	$ & $	0.006	$ & $	0.127	$ & $	0.148	$ & $	0.487	$ & $	1.213	$ & $	2.492	$ & \\
$ 15	$ & $	0.123	$ & $	0.111	$ & $	0.111	$ & $	0.006	$ & $	1.170	$ & $	0.009	$ & $	0.004	$ & $	0.119	$ & $	0.148	$ & $	0.476	$ & $	1.212	$ & $	2.545	$ & \\
$ 16	$ & $	0.112	$ & $	0.127	$ & $	0.114	$ & $	0.084	$ & $	1.035	$ & $	0.025	$ & $	0.025	$ & $	0.129	$ & $	0.148	$ & $	0.489	$ & $	1.212	$ & $	2.480	$ & \\
\hline
$ m_\mathrm{c}/m = 0.95 $ & $	$ & $	$ & $	$ & $	$ & $	$ & \multicolumn{2}{l}{Initial Properties}	 & $	0.600	$ & $	1.800	$ & $	1.204	$ & $	2.314	$ & $	2.37	$ & \\
\hline
$ 1	$ & $	0.143	$ & $	0.244	$ & $	0.107	$ & $	0.075	$ & $	1.536	$ & $	0.035	$ & $	0.018	$ & $	0.010	$ & $	0.046	$ & $	1.854	$ & $	2.729	$ & $	1.472	$ & \\
$ 2	$ & $	0.136	$ & $	0.262	$ & $	0.108	$ & $	0.054	$ & $	1.415	$ & $	0.028	$ & $	0.018	$ & $	0.003	$ & $	0.042	$ & $	2.064	$ & $	2.922	$ & $	1.415	$ & \\
$ 3	$ & $	0.150	$ & $	0.288	$ & $	0.106	$ & $	0.004	$ & $	1.669	$ & $	0.027	$ & $	0.007	$ & $	0.002	$ & $	0.040	$ & $	2.598	$ & $	3.000	$ & $	1.155	$ & \\
$ 4	$ & $	0.104	$ & $	0.140	$ & $	0.118	$ & $	0.093	$ & $	1.206	$ & $	0.026	$ & $	0.022	$ & $	0.018	$ & $	0.044	$ & $	1.551	$ & $	2.805	$ & $	1.808	$ & \\
$ 5	$ & $	0.083	$ & $	0.277	$ & $	0.132	$ & $	0.210	$ & $	1.000	$ & $	0.141	$ & $	0.053	$ & $	0.038	$ & $	0.045	$ & $	1.484	$ & $	2.769	$ & $	1.866	$ & \\
$ 6	$ & $	0.121	$ & $	0.174	$ & $	0.113	$ & $	0.100	$ & $	1.113	$ & $	0.032	$ & $	0.031	$ & $	0.012	$ & $	0.045	$ & $	1.610	$ & $	2.787	$ & $	1.732	$ & \\
$ 7	$ & $	0.089	$ & $	0.352	$ & $	0.127	$ & $	0.129	$ & $	1.000	$ & $	0.111	$ & $	0.061	$ & $	0.038	$ & $	0.047	$ & $	1.488	$ & $	2.615	$ & $	1.758	$ & \\
$ 8	$ & $	0.115	$ & $	0.095	$ & $	0.115	$ & $	0.083	$ & $	1.010	$ & $	0.017	$ & $	0.017	$ & $	0.008	$ & $	0.042	$ & $	2.195	$ & $	2.954	$ & $	1.346	$ & \\
$ 9	$ & $	0.127	$ & $	0.105	$ & $	0.112	$ & $	0.140	$ & $	1.000	$ & $	0.043	$ & $	0.040	$ & $	0.005	$ & $	0.041	$ & $	1.956	$ & $	2.987	$ & $	1.527	$ & \\
$ 10	$ & $	0.110	$ & $	0.175	$ & $	0.116	$ & $	0.052	$ & $	1.086	$ & $	0.017	$ & $	0.017	$ & $	0.012	$ & $	0.044	$ & $	1.637	$ & $	2.842	$ & $	1.736	$ & \\
$ 11	$ & $	0.118	$ & $	0.142	$ & $	0.112	$ & $	0.104	$ & $	1.077	$ & $	0.036	$ & $	0.035	$ & $	0.045	$ & $	0.048	$ & $	1.289	$ & $	2.545	$ & $	1.975	$ & \\
$ 12	$ & $	0.106	$ & $	0.138	$ & $	0.116	$ & $	0.106	$ & $	1.146	$ & $	0.036	$ & $	0.033	$ & $	0.047	$ & $	0.047	$ & $	1.207	$ & $	2.585	$ & $	2.142	$ & \\
$ 13	$ & $	0.128	$ & $	0.094	$ & $	0.110	$ & $	0.064	$ & $	1.253	$ & $	0.014	$ & $	0.006	$ & $	0.044	$ & $	0.048	$ & $	1.330	$ & $	2.524	$ & $	1.897	$ & \\
$ 14	$ & $	0.124	$ & $	0.116	$ & $	0.111	$ & $	0.008	$ & $	1.187	$ & $	0.010	$ & $	0.005	$ & $	0.041	$ & $	0.048	$ & $	1.432	$ & $	2.506	$ & $	1.751	$ & \\
$ 15	$ & $	0.112	$ & $	0.131	$ & $	0.114	$ & $	0.083	$ & $	1.034	$ & $	0.026	$ & $	0.025	$ & $	0.044	$ & $	0.048	$ & $	1.319	$ & $	2.529	$ & $	1.917	$ & \\
\enddata

\tablenotetext{}{$Run$ designates the run number, $a$, $e$, $m_\mathrm{gas}$, and $\rho$ are the semi-major axis (in AU), eccentricity, gas mass (in Earth masses), and density (in CGS) of the inner out and outer planets, $P_2/P_1$ is the period ratio, $C$ and $C_\mathrm{min}$ are the angular momentum deficit and minimum possible angular momentum deficit, normalized as described in \S\ref{SSec:Scatterings}, and $\rho_1/\rho_2$ is the density ratio.
The initial gas mass fractions and densities are listed on the first row of each set of calculations.}
\end{deluxetable*}

\subsection{Scatterings}
\label{SSec:Scatterings}
\begin{figure*}[htp]
\begin{center}
\begin{tabular}{cc}
\includegraphics[width=8.0cm]{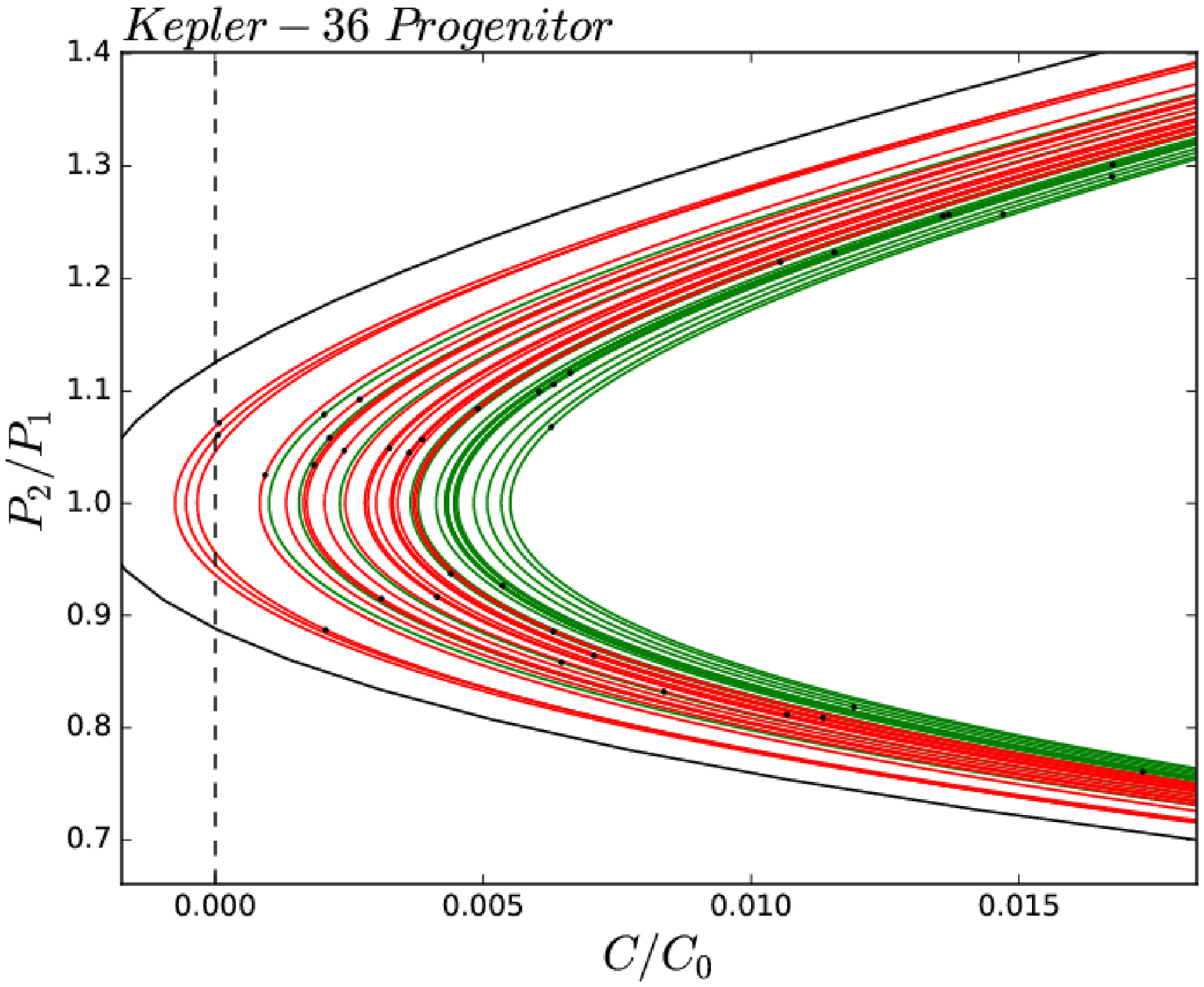} 
\color{white}
\rule{8.0cm}{5.0cm}
\color{black} \\
\includegraphics[width=8.0cm]{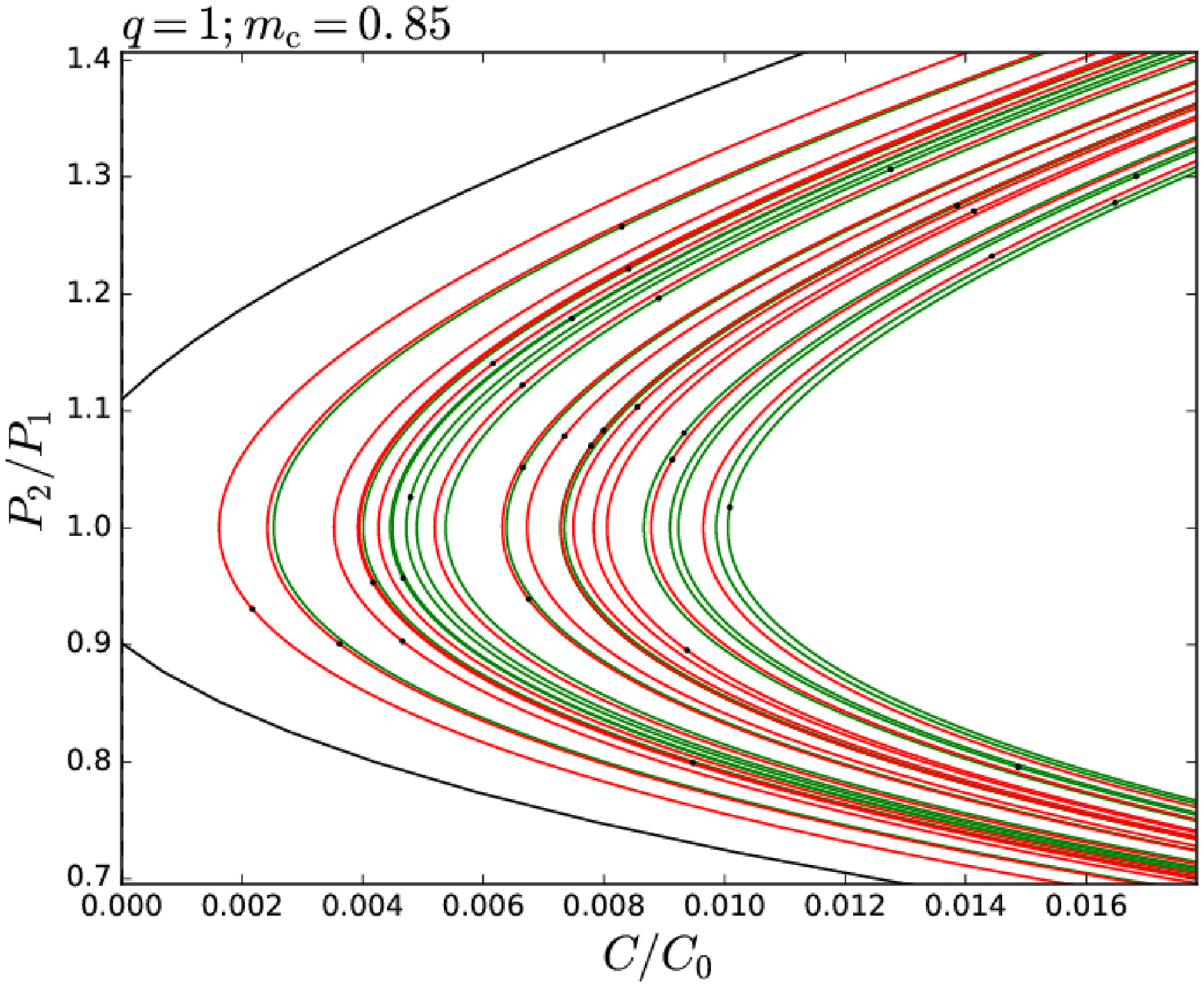}
\includegraphics[width=8.0cm]{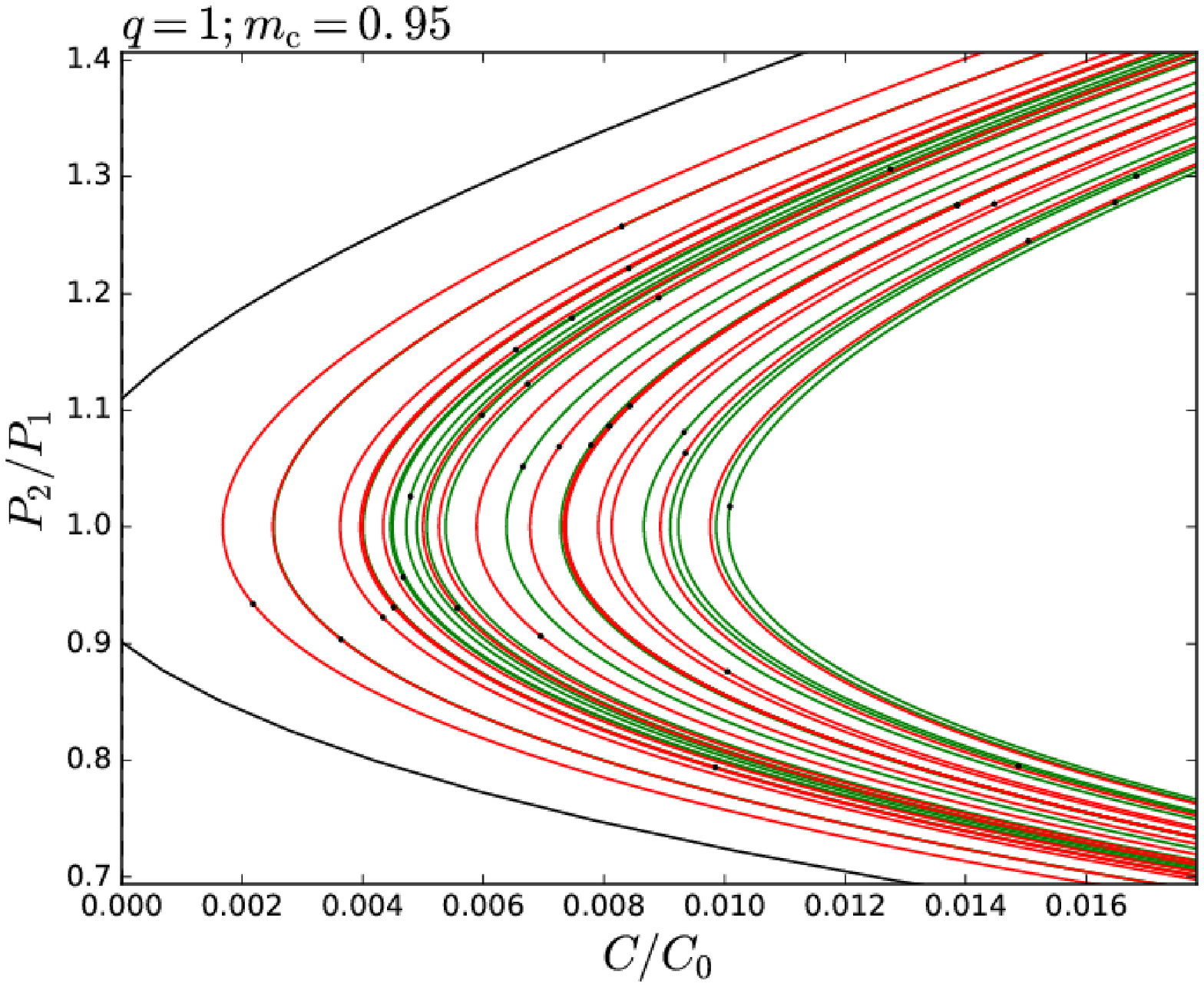} \\
\includegraphics[width=8.0cm]{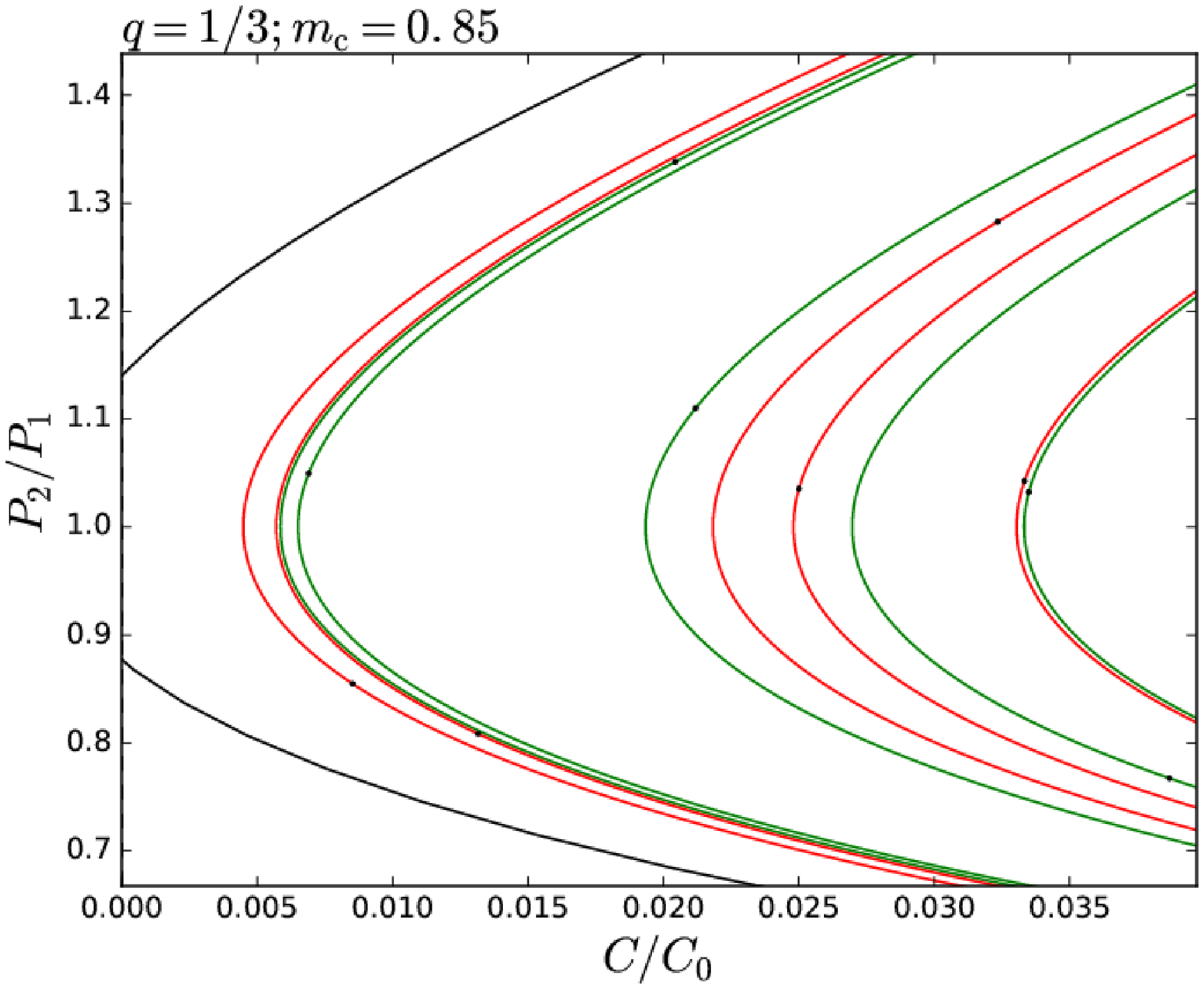}
\includegraphics[width=8.0cm]{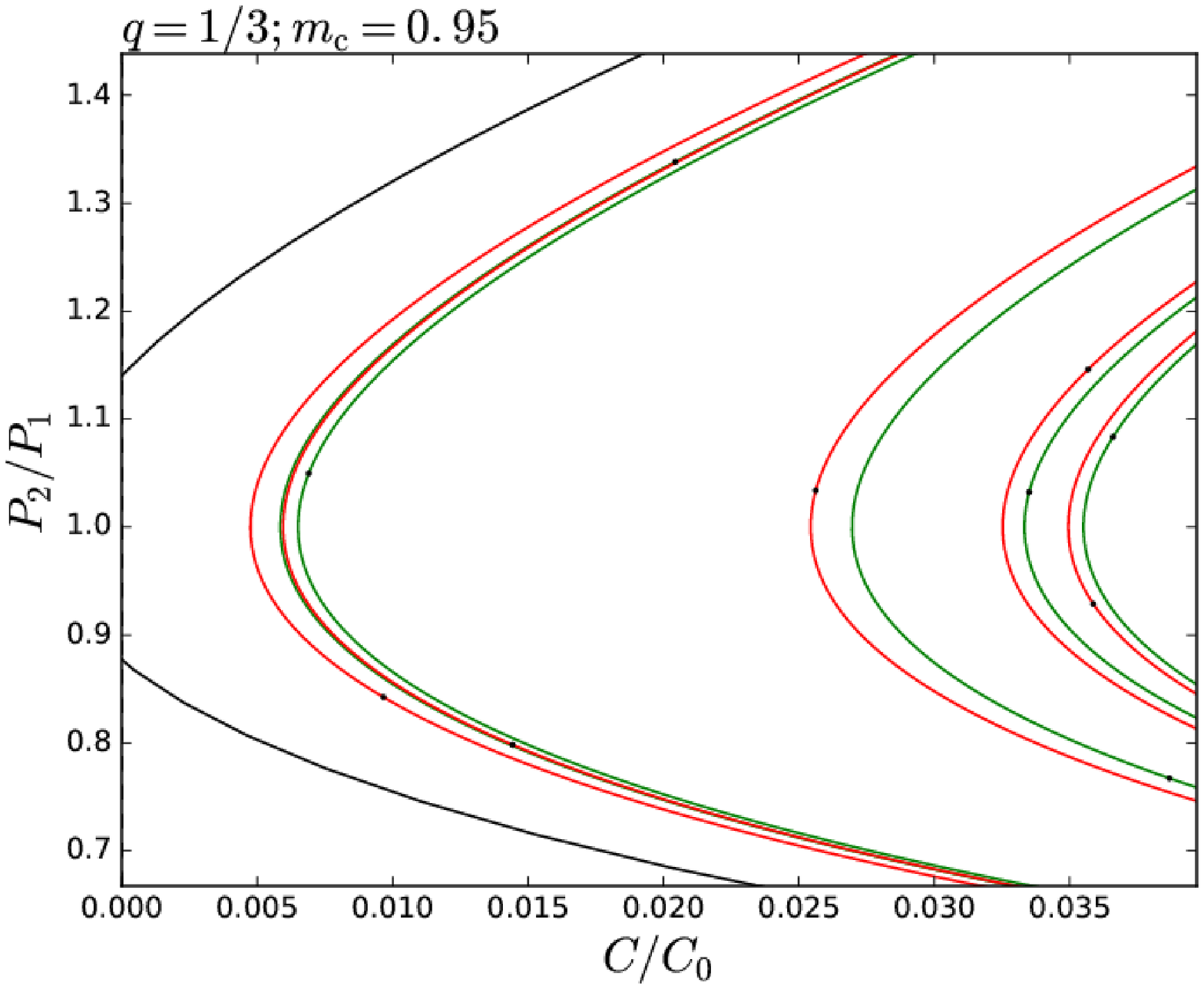}
\end{tabular}
\end{center}
\caption{Solutions of period ratio, $P_2/P_1$, and angular momentum deficit, $C$, scaled by $C_0 = \mathrm{M}_\mathrm{E}(\mathrm{GM}_\odot\ \mathrm{AU})^{1/2},$ before the collision (green) and after the collision (red), along with the coordinates of the first and last snapshots in the SPH calculation (black points). The solution for a Hill-stable orbit is shown for comparison (solid black line) for the pre-collision masses of each calculation, and we see that, in general, systems become more stable after the collision. Angular momentum deficits below zero are unphysical, and orbits with solutions where $C<0$ have period-ratio boundaries at the circular solutions.}
\label{Fig:Scattering_Plots1}
\end{figure*}

To quantify the effect the collisions have on the orbits, we calculate the angular momentum deficit, following \citet{1997A&A...317L..75L}, \begin{equation}C=\sum_k^nm_k\sqrt{GM_*a_k}\left(1-\sqrt{1-e_k^2}\cos(i_k)\right).\end{equation}
Enforcing conservation of energy and angular momentum, we calculate the angular momentum deficit as a function of the period ratio, and for each set of initial and post-collision orbits, the minimum angular momentum deficit, $C_\mathrm{min},$ where $C_\mathrm{min}\le0$ defines a set of orbits with a circular solution.
Figure~\ref{Fig:Scattering_Plots1} shows the angular momentum, period-ratio solutions for the initial and final orbits of each calculation, with the Hill-stable solution \citep{1993Icar..106..247G} as a reference.
In general, the minimum angular momentum deficits of the post-collision orbits are less than the minimum angular momentum deficits of the initial orbits, and closer to the Hill-stable solution.
We use the final position and velocities from the hydrodynamic calculations as initial conditions for followup dynamical integrations to determine the long-term stability of the system, and while each scattering outcome eventually results in a subsequent collision, further collisions may eventually stabilize the system.

During the collision, the lower mass planets ($m\sim4.0\ M_\mathrm{E}$), tend to lose some fraction of their gas envelopes, and only the largest planets ($m=12.0\ M_\mathrm{E}$) gain mass during the collision.
We develop a model for predicting the change in mass during a collision in \S\ref{Sec:Predictions}.
Using the same method as {\it Paper 1}, we find that the density ratios tend to become more extreme for systems with mass ratios, $q\ne1$, with the lower-mass planets losing more mass than the higher-mass companions.
These results suggest that two very tightly-packed planets with a large density ratio (e.g. Kepler-36; \citealt{2012Sci...337..556C}) may be evidence of a previous planet-planet collision (\citealt{2016ApJ...823..162L}, \citealt{2016ApJ...817L..13I}).

\subsection{Mergers}
Collisions that eventually result in the merger of the two cores generally undergo multiple episodes of physical collisions at the periapsis of each planet-planet orbit, losing some relative orbital energy at each periapsis passage.
While our treatment of the cores does not allow us to fully integrate through the merger, we are able to follow the changes in mass and orbital energy in each orbit until the cores physically collide.
In previous versions of these calculations, where the equations of state of both the core and gas are approximated simply as polytropes, we found several distinct outcomes, including fragmentation of the less-massive core, where some mass is accreted onto the more massive core and the remainder becomes bound, forming a planet-planet binary with a more extreme mass-ratio.
Preferential shedding of the mantle may lead to remnants where the less-massive planet survives with a much higher iron-core mass fraction, and the more-massive planet has an excess of mantle material.
Follow-up studies using equations of state that more accurately model direct core-core collisions are necessary to fully understand the details of these outcomes.

\subsection{Bound Planet-Pair}
\begin{figure}[htp]
\begin{center}
\begin{tabular}{cc}
\includegraphics[width=8.0cm]{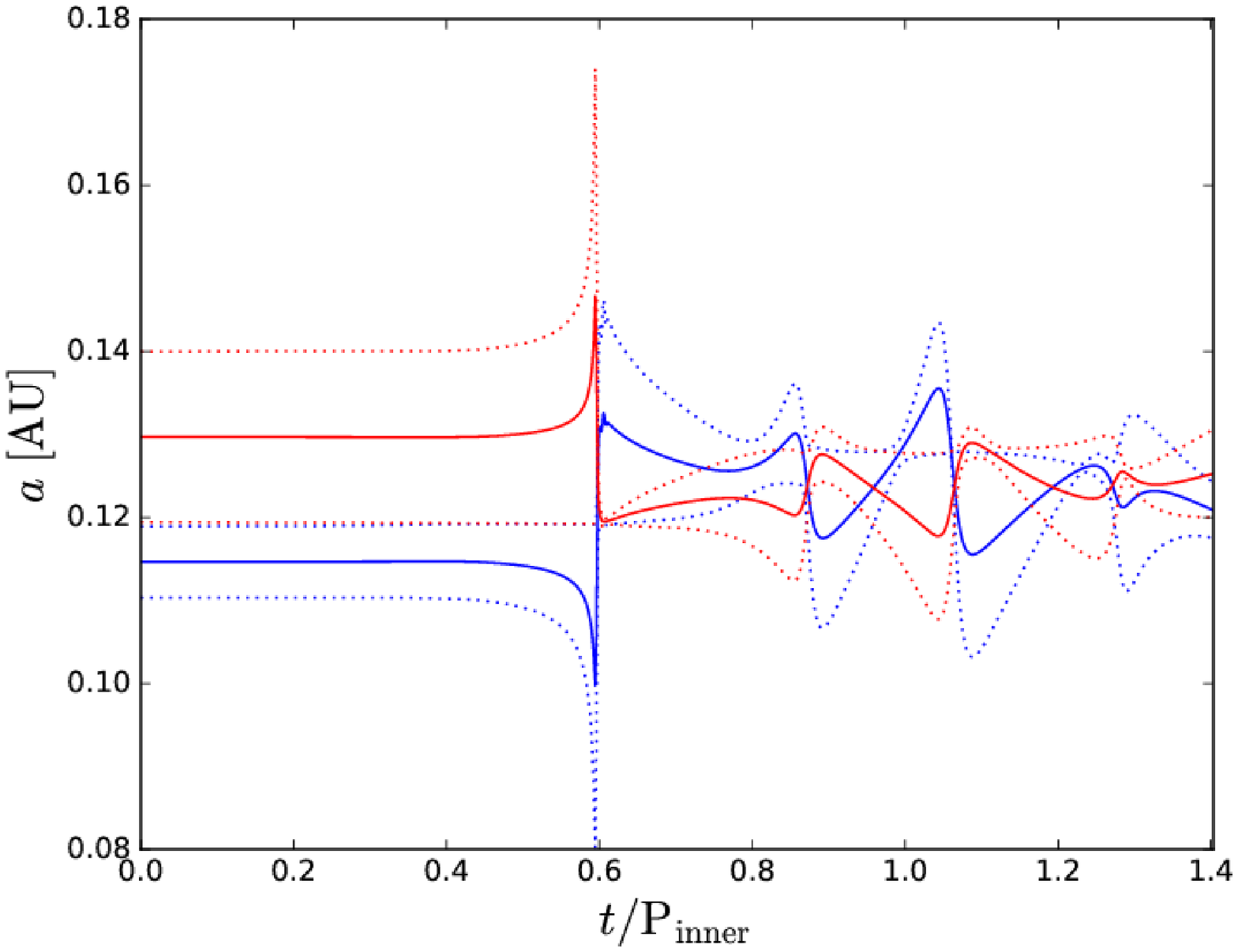} \\
\includegraphics[width=8.0cm]{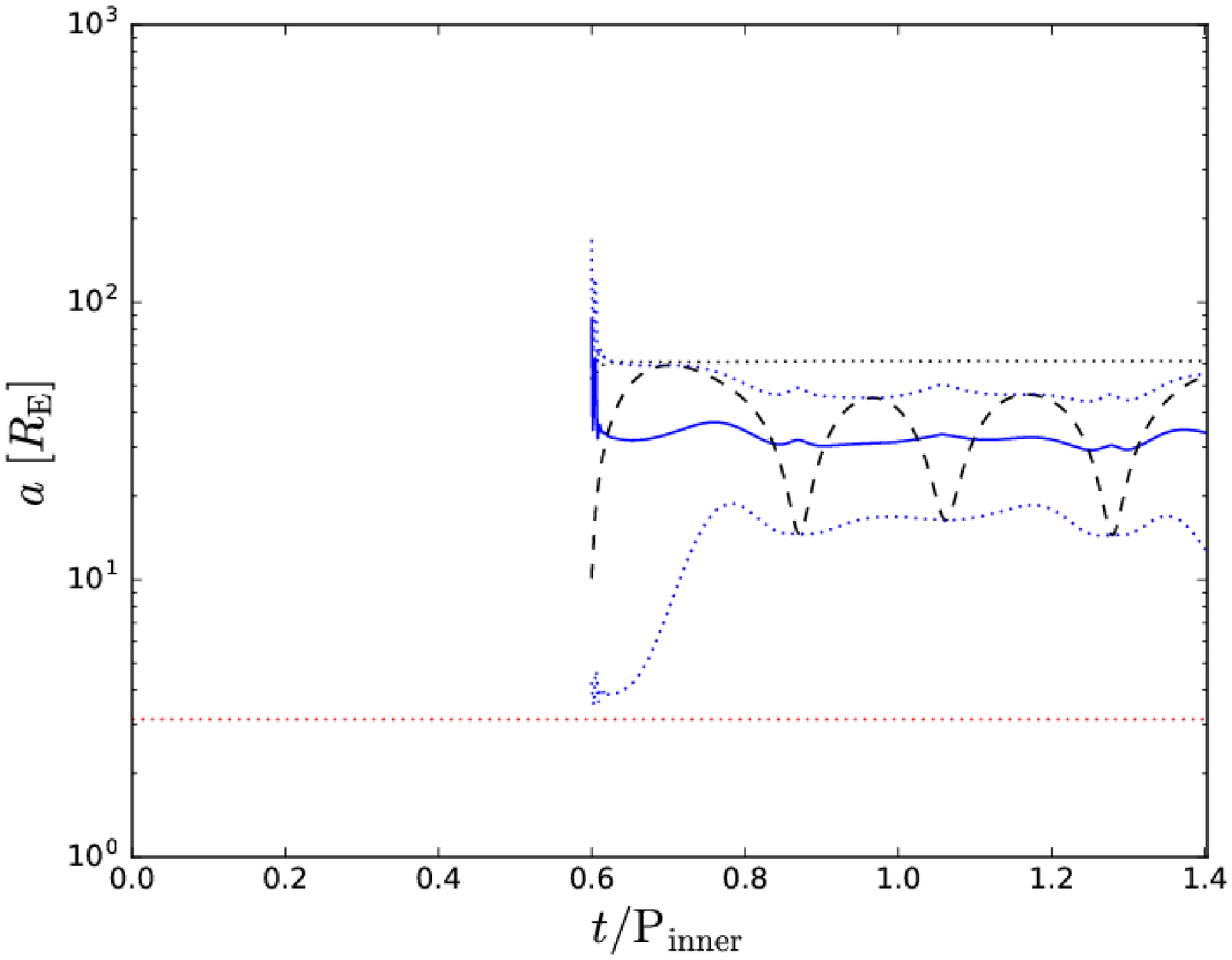}
\end{tabular}
\end{center}
\caption{Orbital evolution of a collison resulting in a potential planet-planet binary. The planet-star orbits (top), with the semi-major axes (solid line), and periastron and apoastron (dotted lines).
The planet-planet orbit (bottom) after the collision shows the physical separation (dashed black line), semi-major axes (blue), and periastron and apoastron (dotted lines) of the planet-planet binary.
The sum of the physical size of the cores (dotted red line) and the maximum periapsis (dotted black line) from \eqref{EQ:max_periapsis} for comparison.}
\label{Fig:Bound_Orbits}
\end{figure}

While most collisions resulting in the two planets becoming a bound pair have an orbit with a periapsis that results in a core-core collision, and a likely merger, one calculation shows a potentially long-lived planet-planet binary.
Following \citet{2010arXiv1007.1418P}, we define the inner binary as the two bound planets and the outer binary as the bound planet-pair and the host star, and compare the maximum apoapsis of the inner binary to the mutual Hill radius at the periapsis of the outer binary, \begin{equation}\label{EQ:max_apoapsis}r_\mathrm{1,apo} = a_\mathrm{in}(1+e_\mathrm{in})\left(\frac{m_2}{m_1+m_2}\right),\end{equation} \begin{equation}\label{EQ:R_Hill_periapsis}R_\mathrm{H,peri}=a_\mathrm{out}(1-e_\mathrm{out})\left(\frac{\mu}{3M_*}\right)^{1/3},\end{equation} where $m_2 > m_1$, and $a_\mathrm{in}$, $e_\mathrm{in}$, $a_\mathrm{out}$, and $e_\mathrm{out}$ are the semi-major axes and eccentricities of the inner and outer binary.
We estimate the inner binary's maximum apoapsis by enforcing the maximum distance of a planet from the center of mass of the inner binary be less than the Hill radius at the outer binary's periapsis, \begin{equation}\label{EQ:max_periapsis}r_\mathrm{max, apo} = a_\mathrm{out}(1-e_\mathrm{out})\left(\frac{\mu}{3M_*}\right)^{1/3}\left(\frac{m_2}{m_1+m_2}\right).\end{equation}
Figure~\ref{Fig:Bound_Orbits} shows the orbital evolution of the inner and outer binaries of this calculation, and we see that the orbit of the planet-planet binary has a periapsis large enough to prevent the cores from colliding and an apoapsis small enough that the planets remain inside the mutual Hill sphere.
Determing the long-term stability of this planet-planet binary is complex and requires including tides and decay of the inner binary's orbits due to the gas envelope.

\section{Recipes to Predict Collisional Outcomes}
\label{Sec:Predictions}
The details of the collision are important in determining both the outcomes and the structures of the remnant planet(s).
The maximum relative energy of a bound planet-planet pair (resulting in either a merger or a long-lived planet-planet binary) can be calculated, as the maximum orbital separation must be less than the mutual Hill Radius, \begin{equation}\label{EQ:E_max}E_\mathrm{max} = \frac{-Gm_1m_2}{2a_\mathrm{max}}.\end{equation} 
Using \eqref{EQ:max_apoapsis} and \eqref{EQ:R_Hill_periapsis}, \begin{equation}\label{EQ:a_max}a_\mathrm{max} = a_\mathrm{out}\left(\frac{1-e_\mathrm{out}}{1+e_\mathrm{in}}\right)\left(\frac{m_2}{m_1+m_2}\right)\left(\frac{\mu}{3M_*}\right)^{1/3},\end{equation} where $a_\mathrm{out}$, $e_\mathrm{out}$, $a_\mathrm{in}$, and $e_\mathrm{in}$ are the semi-major axes and eccentricities of the outer (planets' center of mass and host star) and inner (planet-planet) binaries, and $m_1$, $m_2$, and $\mu$ are the planet masses and reduced mass after the collision, where $m_1 < m_2$.
In each planet-planet collision, some amount of energy, $E_\mathrm{d},$ is lost from the planets' orbit, and we predict that the remnant will result in a scattering event, where both planets survive, if \begin{equation}E_\mathrm{c} - E_\mathrm{d} > E_\mathrm{max},\end{equation} and may otherwise become a capture, either a merger or bound planet-pair.

\subsection{Fitting Energy Dissipated and Change in Mass}
\label{SSec:Fits}
\begin{figure*}[htp]
\begin{center}
\begin{tabular}{cc}
\includegraphics[width=8.0cm]{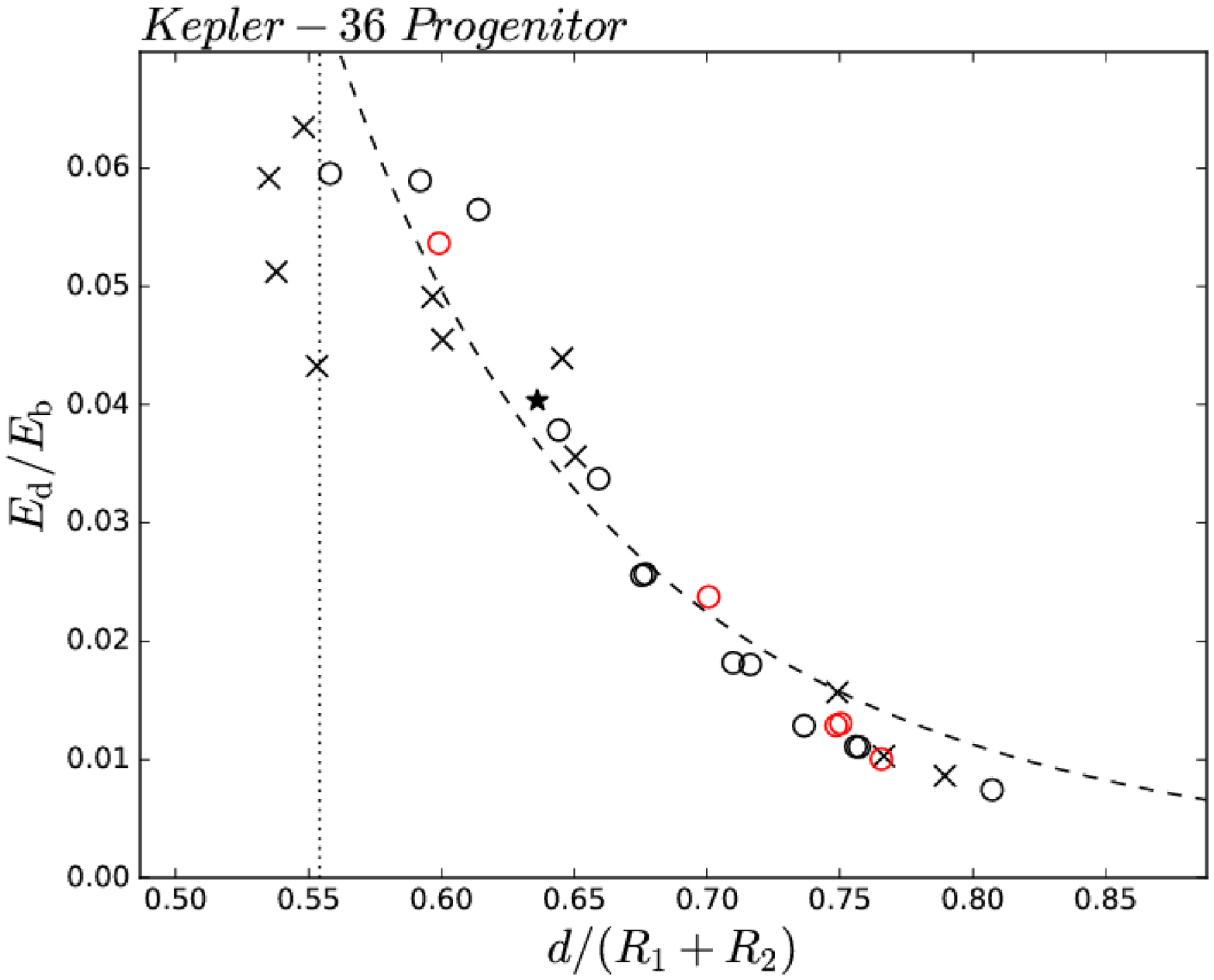}
\color{white}
\rule{8.0cm}{5.0cm}
\color{black} \\
\includegraphics[width=8.0cm]{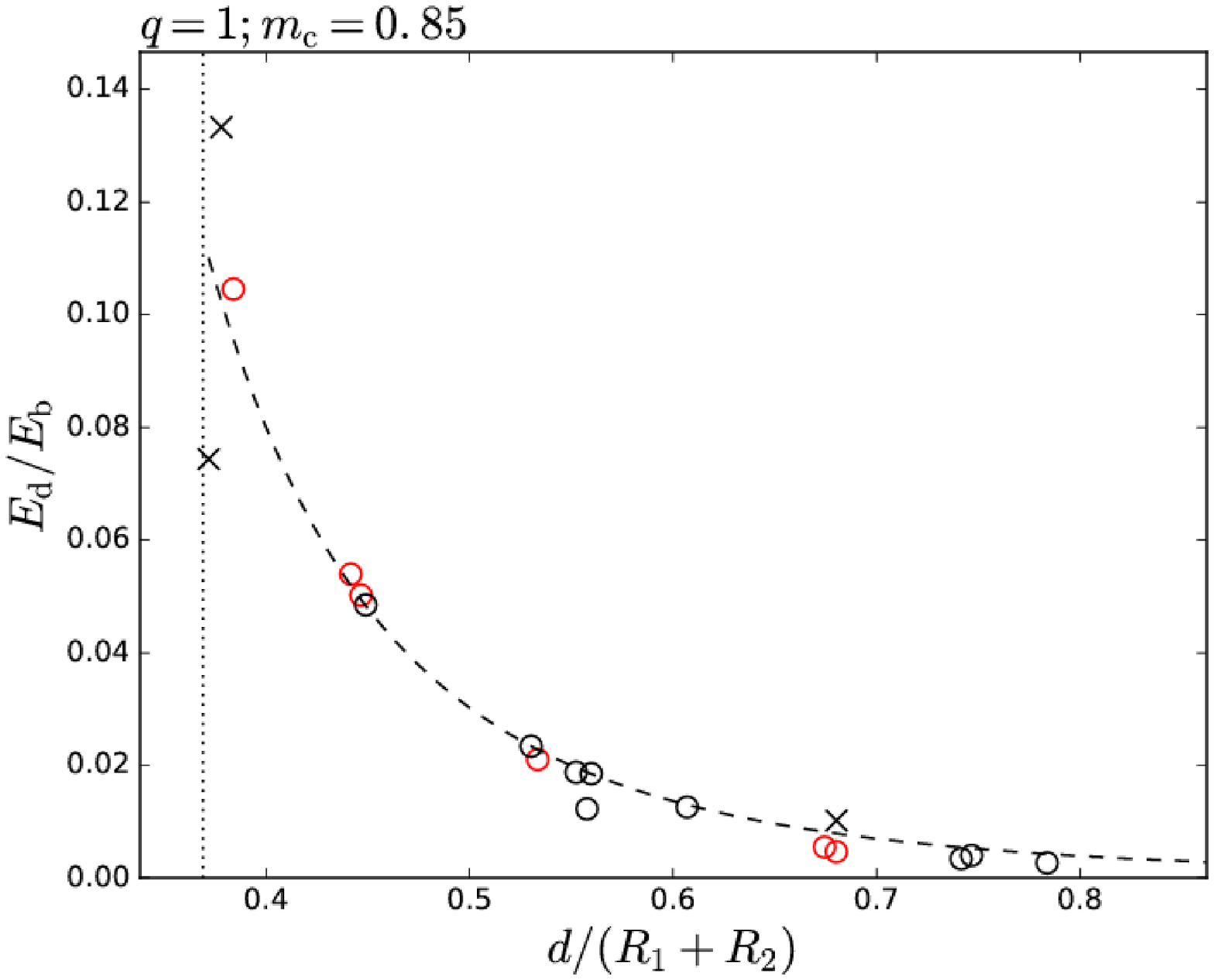}
\includegraphics[width=8.0cm]{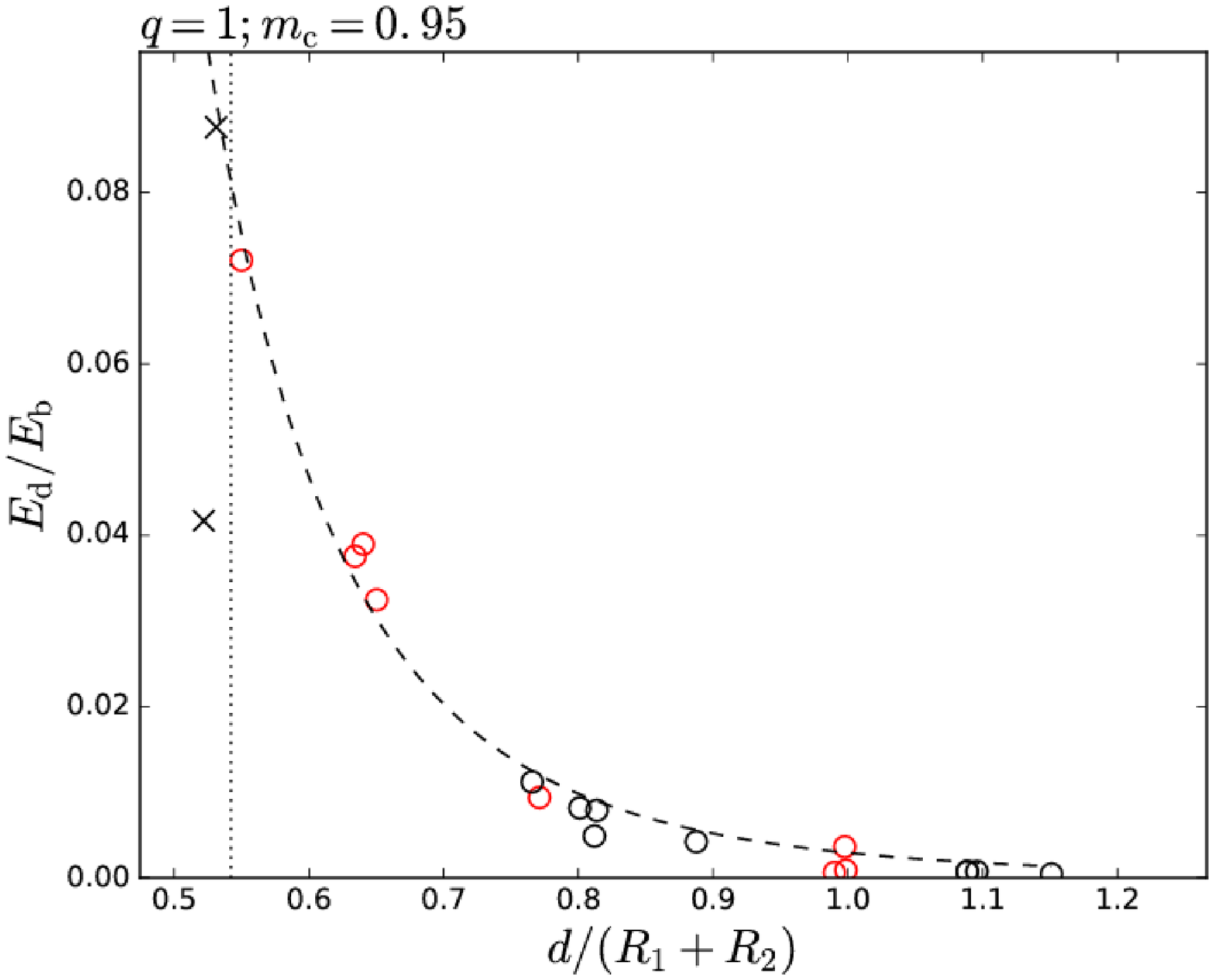} \\
\includegraphics[width=8.0cm]{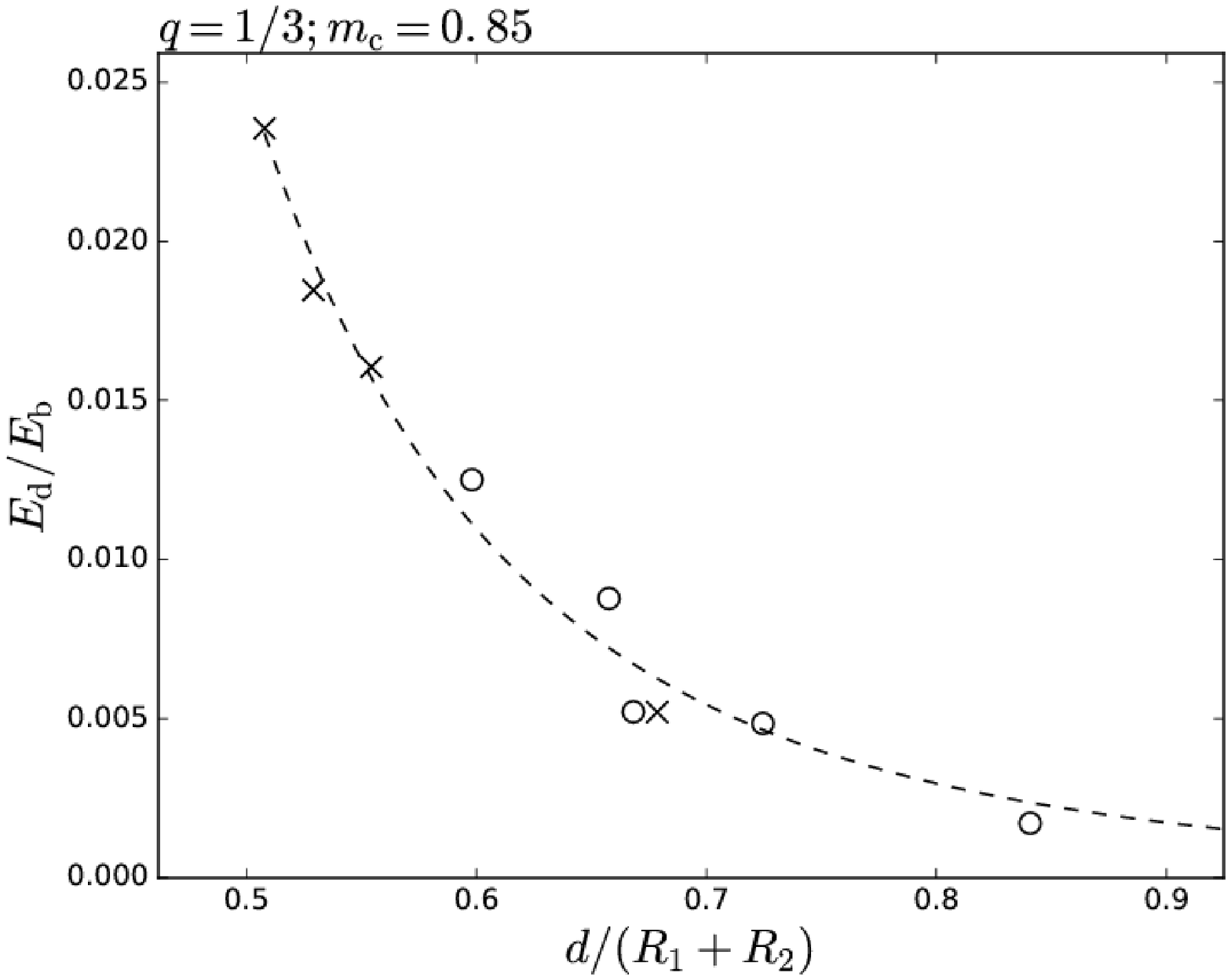}
\includegraphics[width=8.0cm]{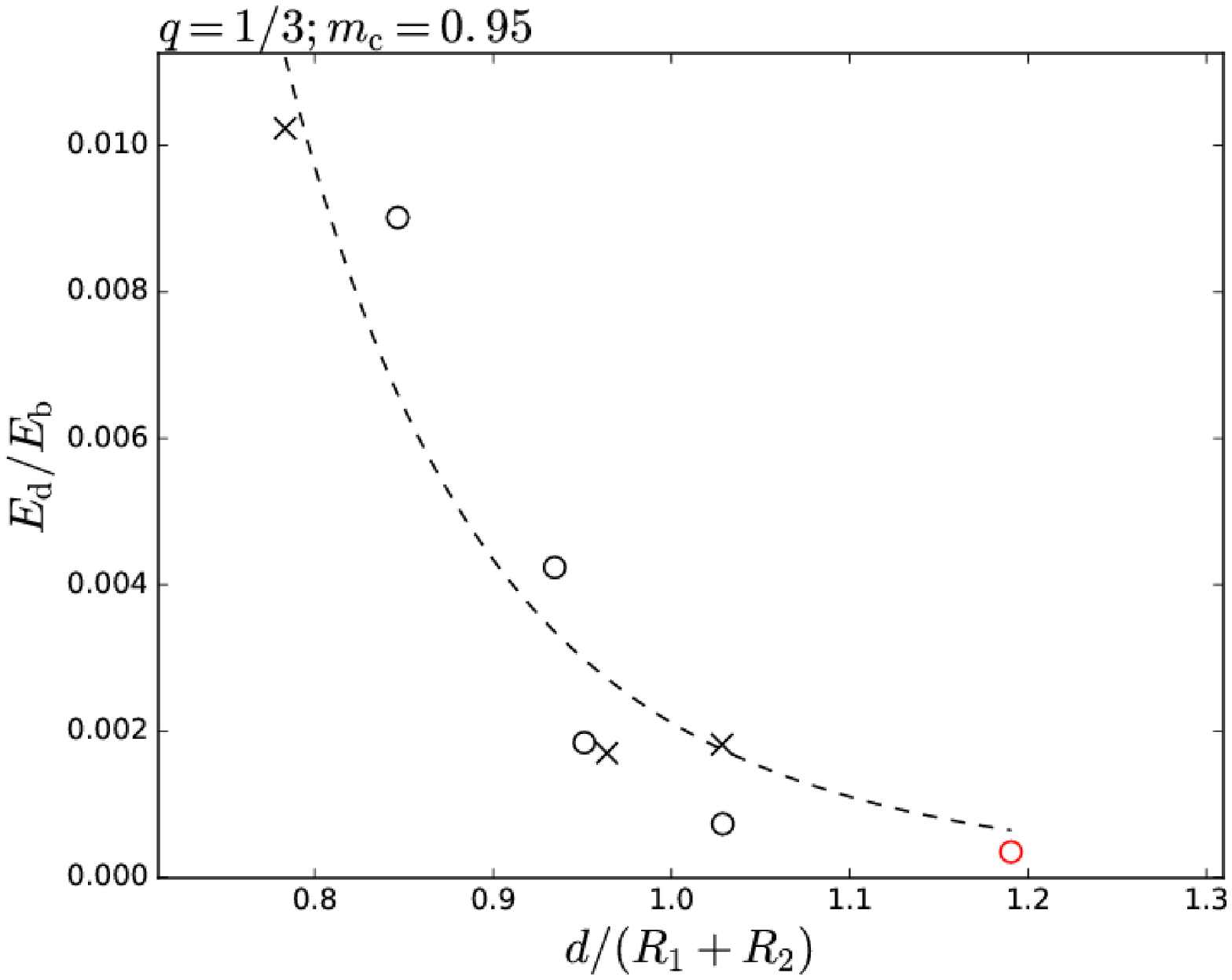}
\end{tabular}
\end{center}
\caption{Change of orbital energy of the planets as a function of the degree of contact, $\eta = d_\mathrm{min}/(R_1+R_2)$, where the symbols designate the outcomes as described in Figure~\ref{Fig:Collision_Results}.
We show the fits, $E_\mathrm{d}/E_\mathrm{b} = A_\mathrm{E}\eta^{k_\mathrm{E}},$ (dashed black lines) where the best-fit values are reported in Table~\ref{TBL:Best_fit_values}.}
\label{Fig:Energy_Dissipated}
\end{figure*}

\begin{figure*}[htp]
\begin{center}
\begin{tabular}{cc}
\includegraphics[width=8.0cm]{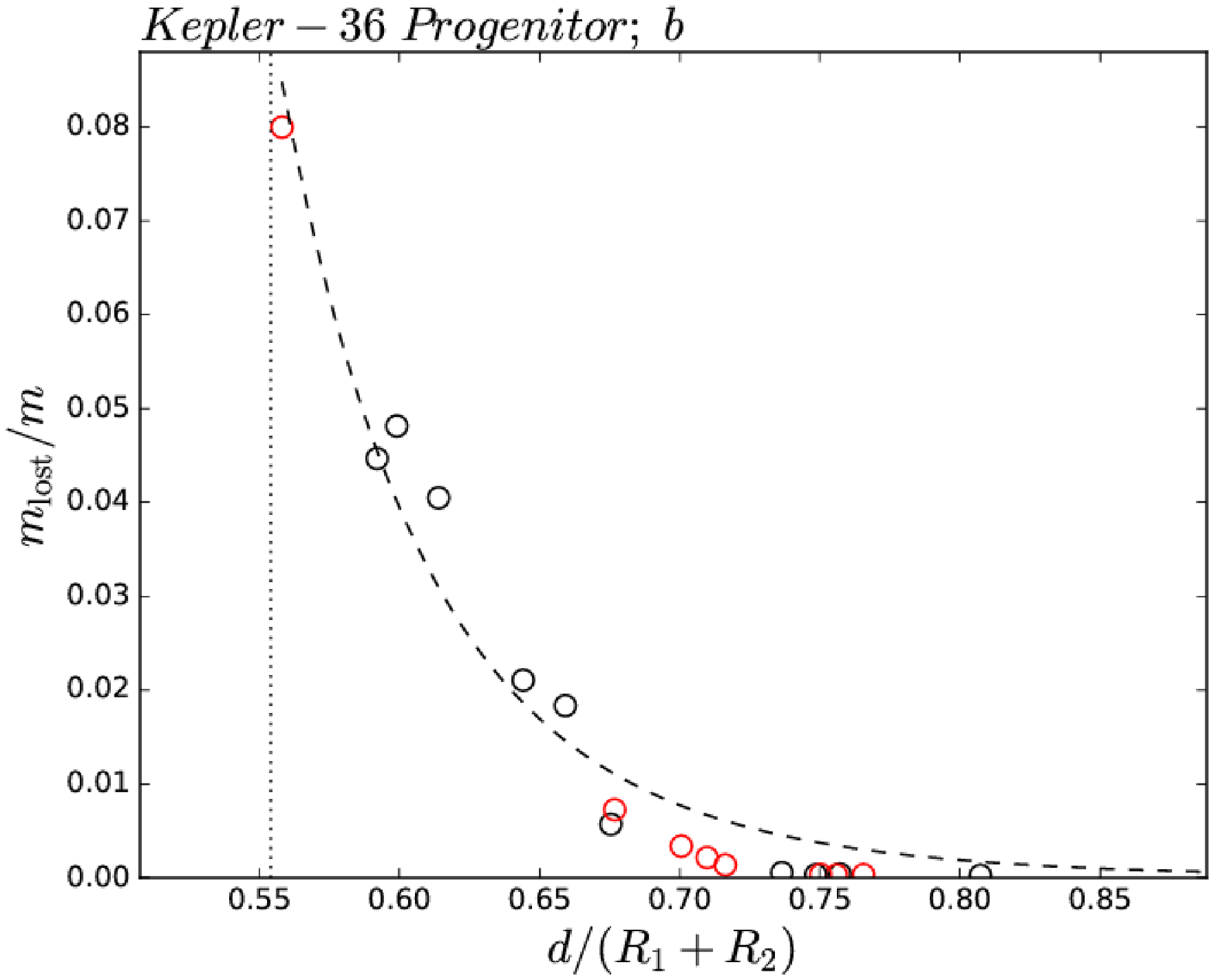}
\includegraphics[width=8.0cm]{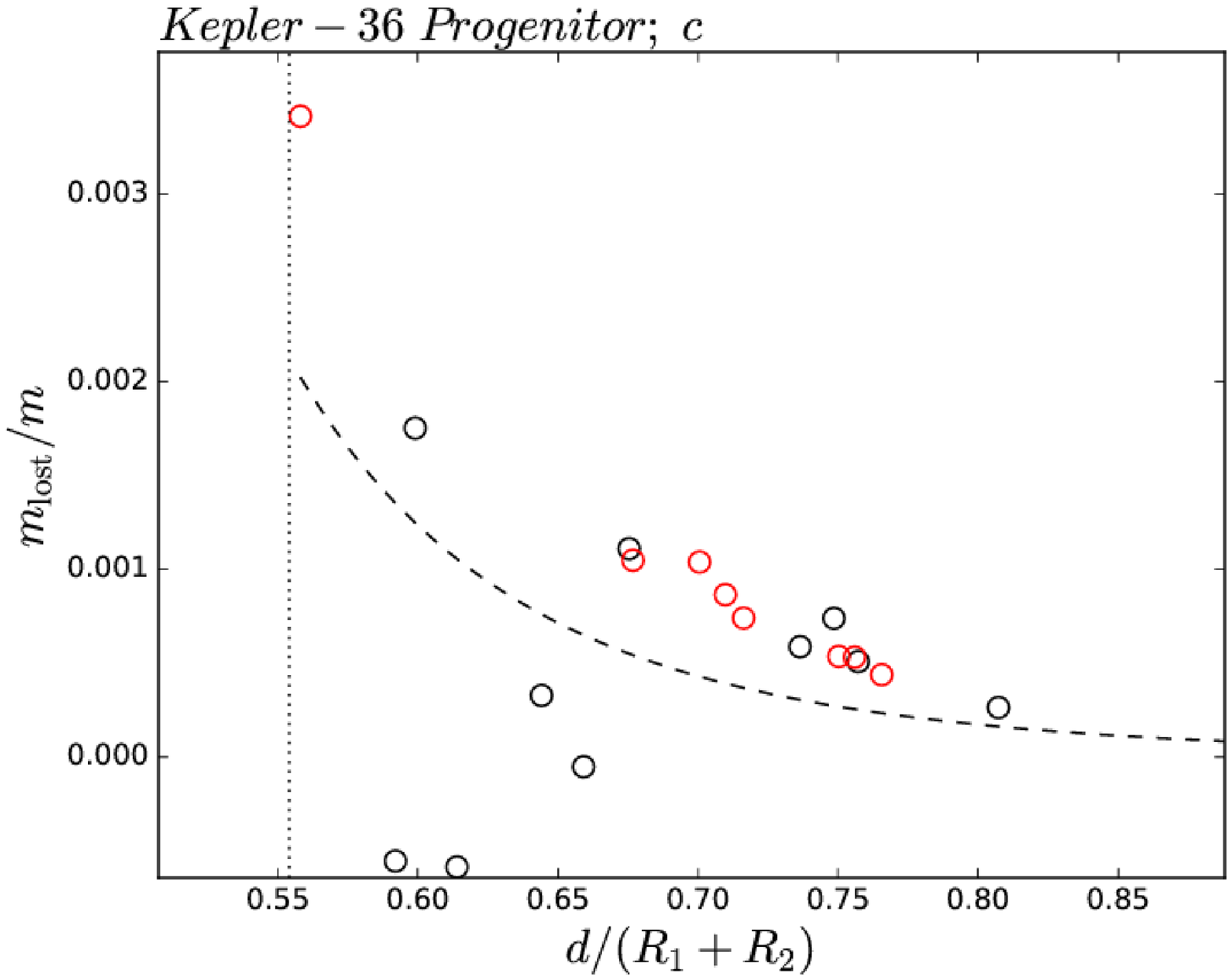} \\
\includegraphics[width=8.0cm]{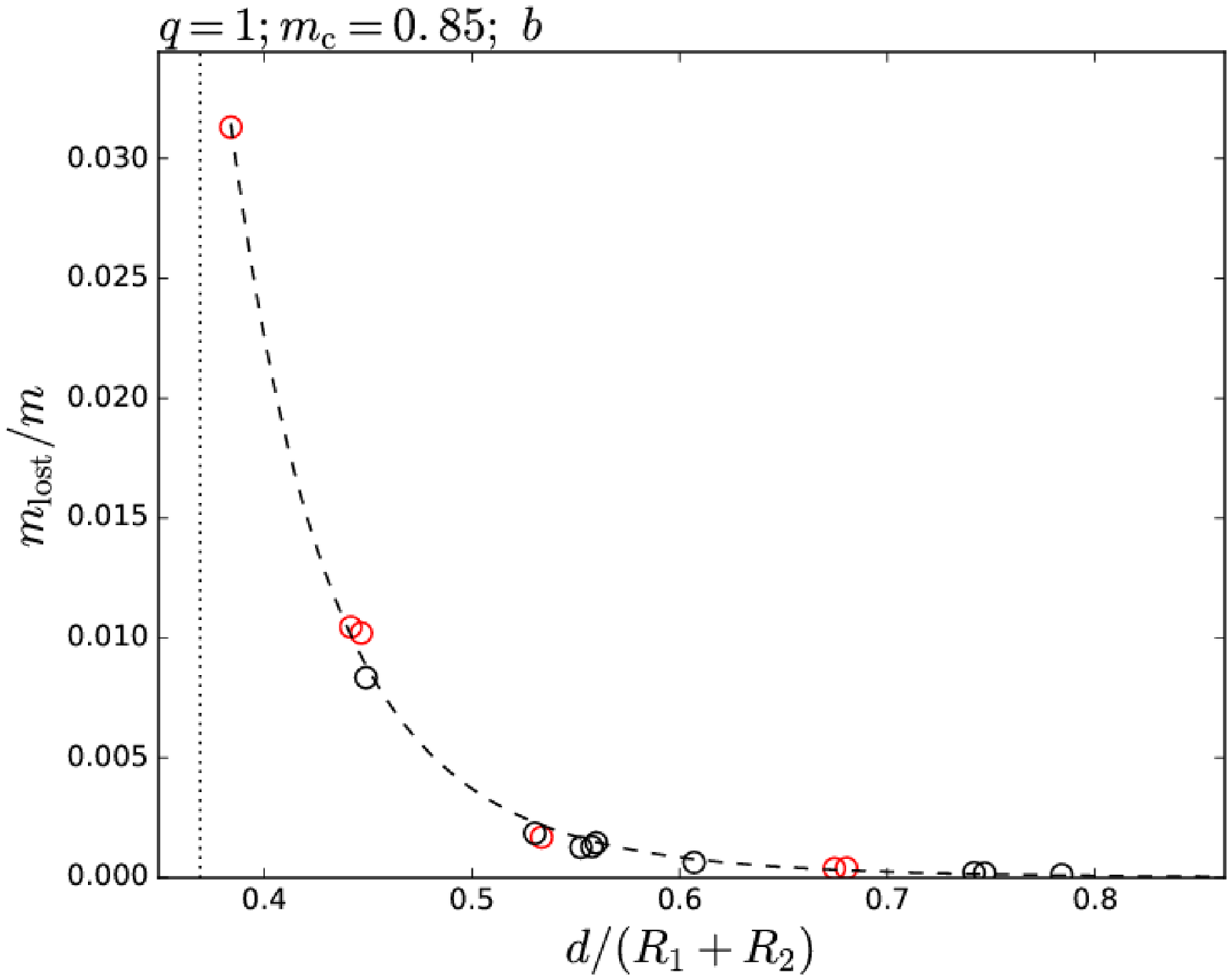}
\includegraphics[width=8.0cm]{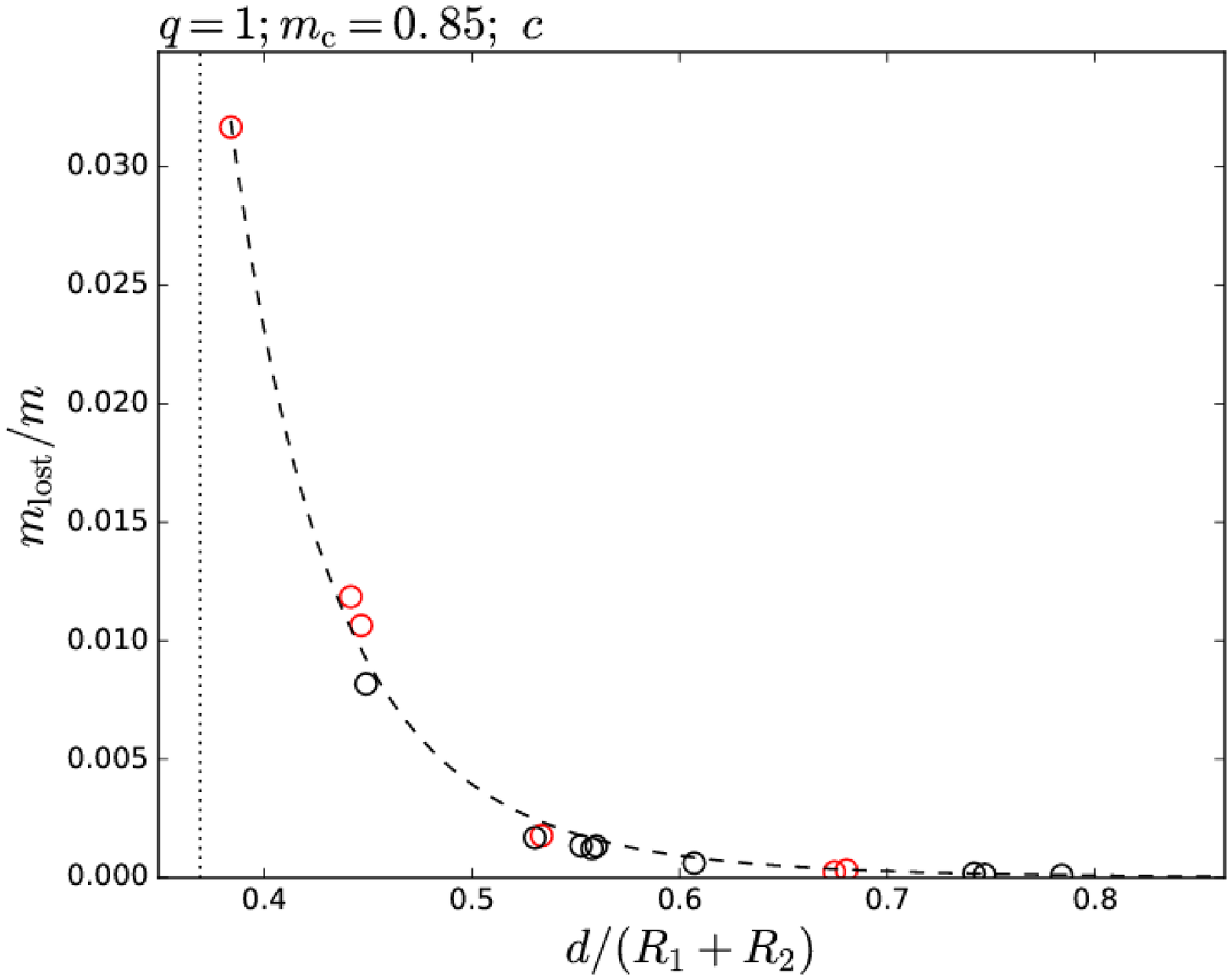} \\
\includegraphics[width=8.0cm]{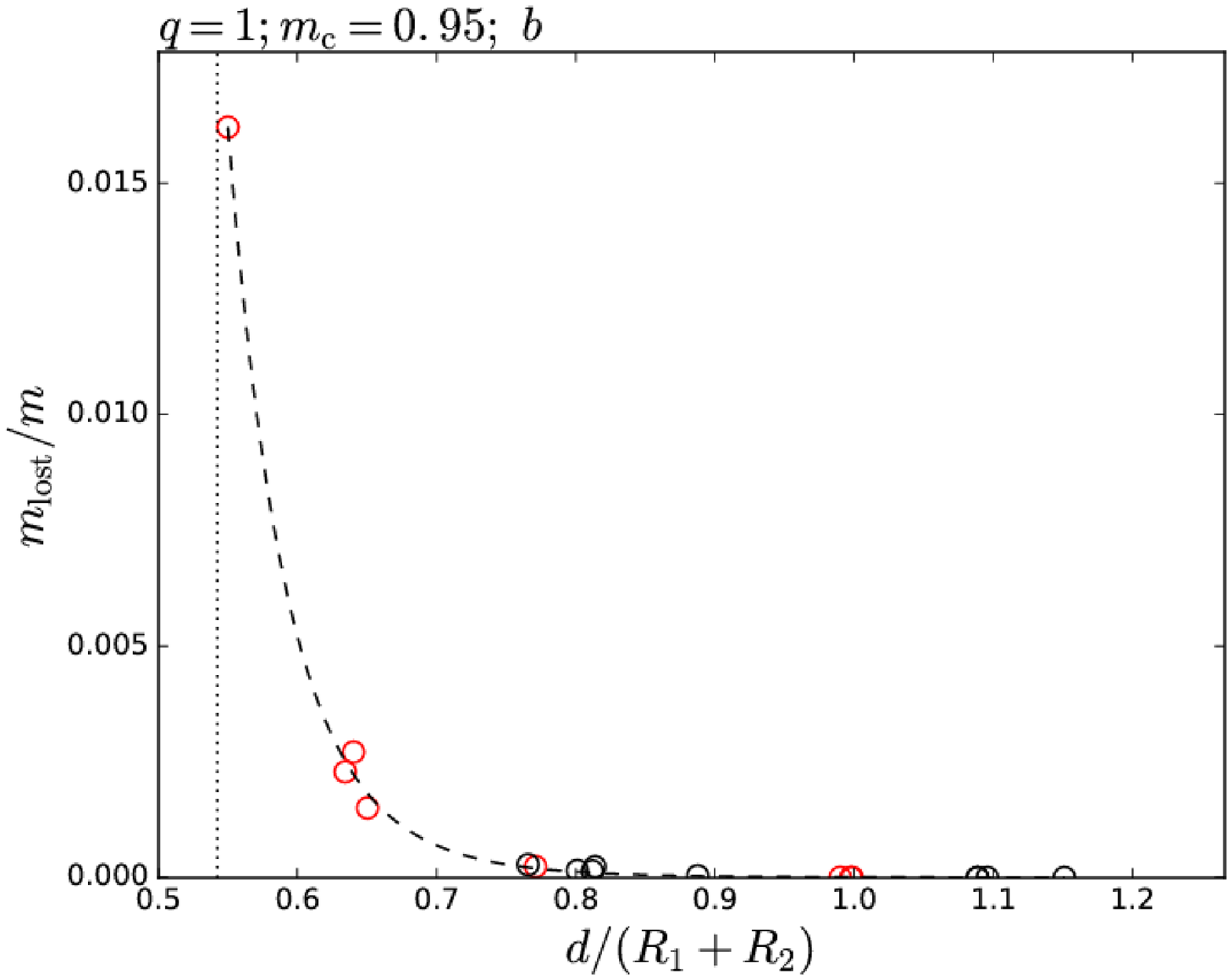}
\includegraphics[width=8.0cm]{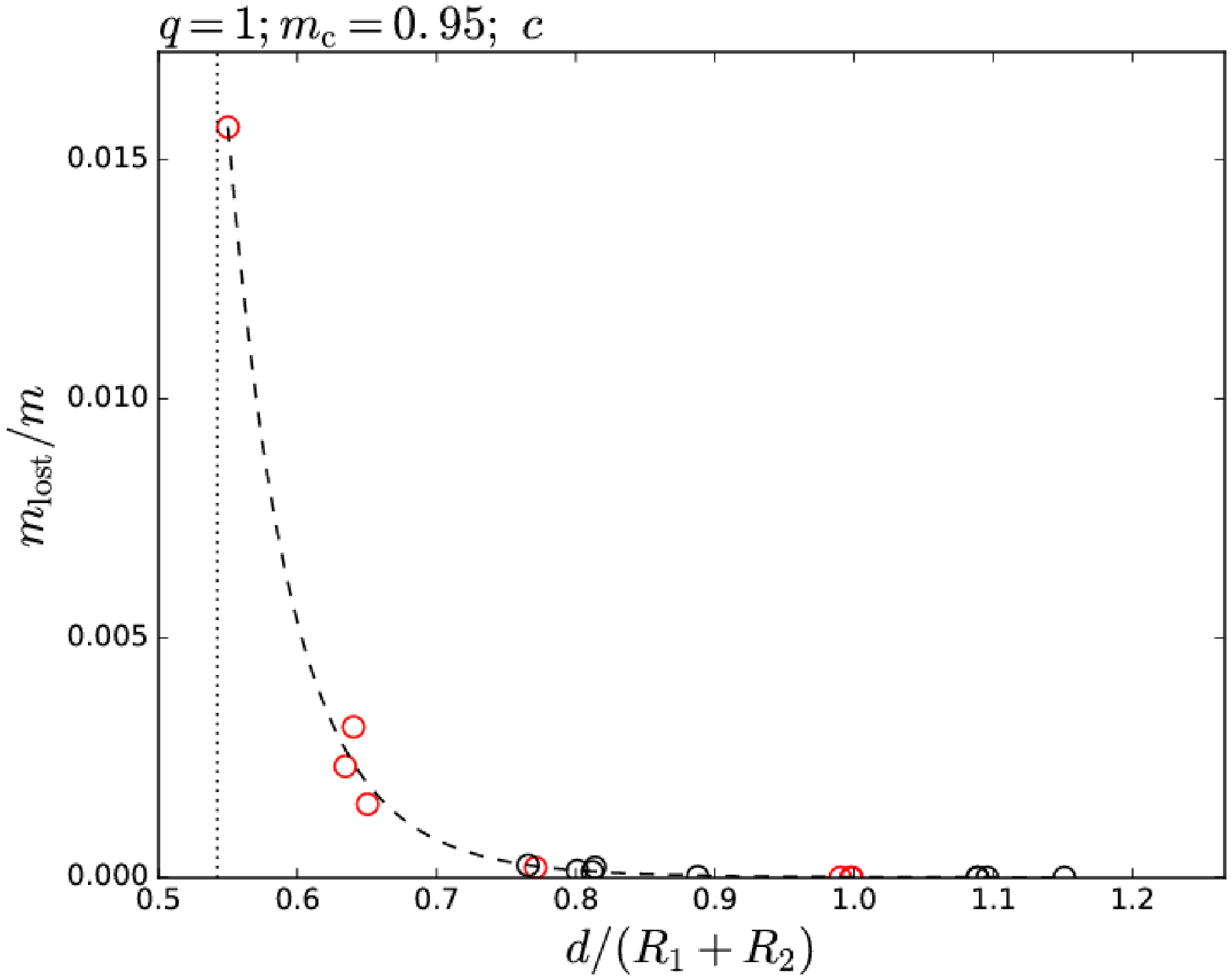}
\end{tabular}
\end{center}
\caption{Fractional change in mass after the first close encounter, of the inner (left) and outer (right) planet as a function of the degree of contact, $\eta = d_\mathrm{min}/(R_1+R_2)$, for the Kepler-36 progenitor (top), $q=1;\ m_\mathrm{c}=0.85$ (middle), and $q=1;\ m_\mathrm{c}=0.95$ (bottom) calculations, where the symbols designate the outcomes as described in Figure~\ref{Fig:Collision_Results}.
We show the fits, $m_{\mathrm{lost},i}/m_i = A_{\mathrm{m},i}\eta^{k_{\mathrm{m},i}},$ (dashed black lines) where the best-fit values are reported in Table~\ref{TBL:Best_fit_values}.}
\label{Fig:mass_loss1}
\end{figure*}

\begin{figure*}[htp]
\begin{center}
\begin{tabular}{cc}
\includegraphics[width=8.0cm]{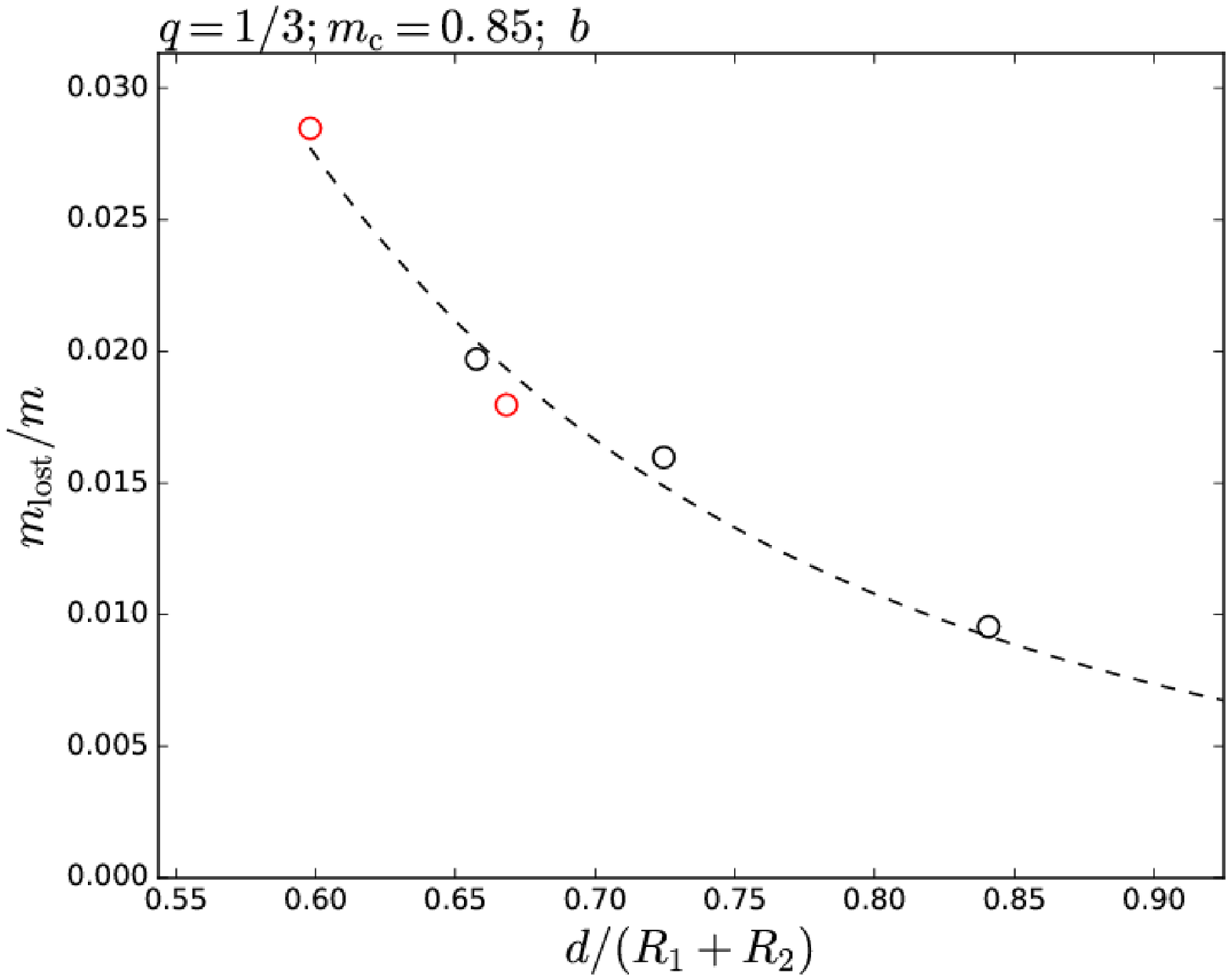}
\includegraphics[width=8.0cm]{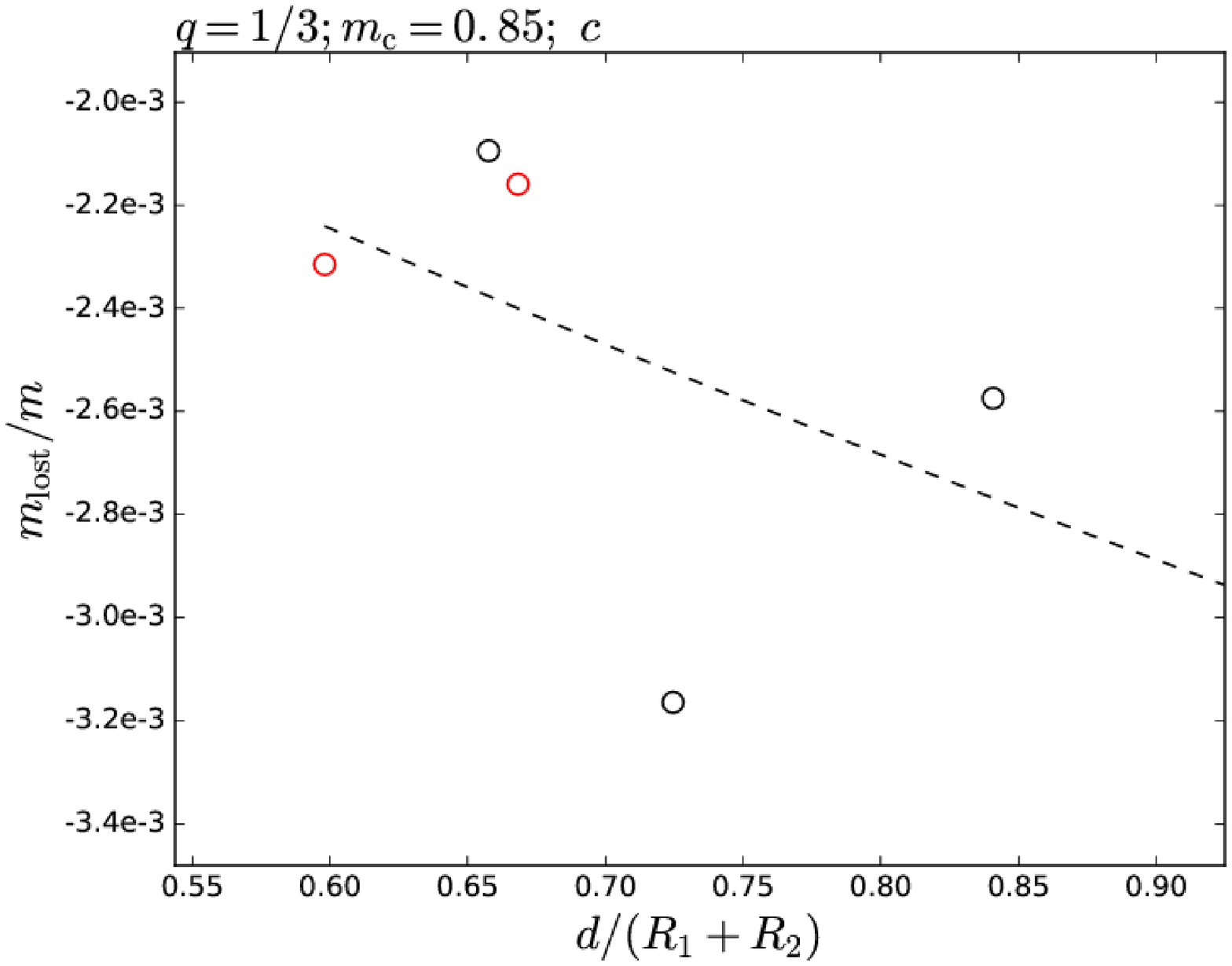} \\
\includegraphics[width=8.0cm]{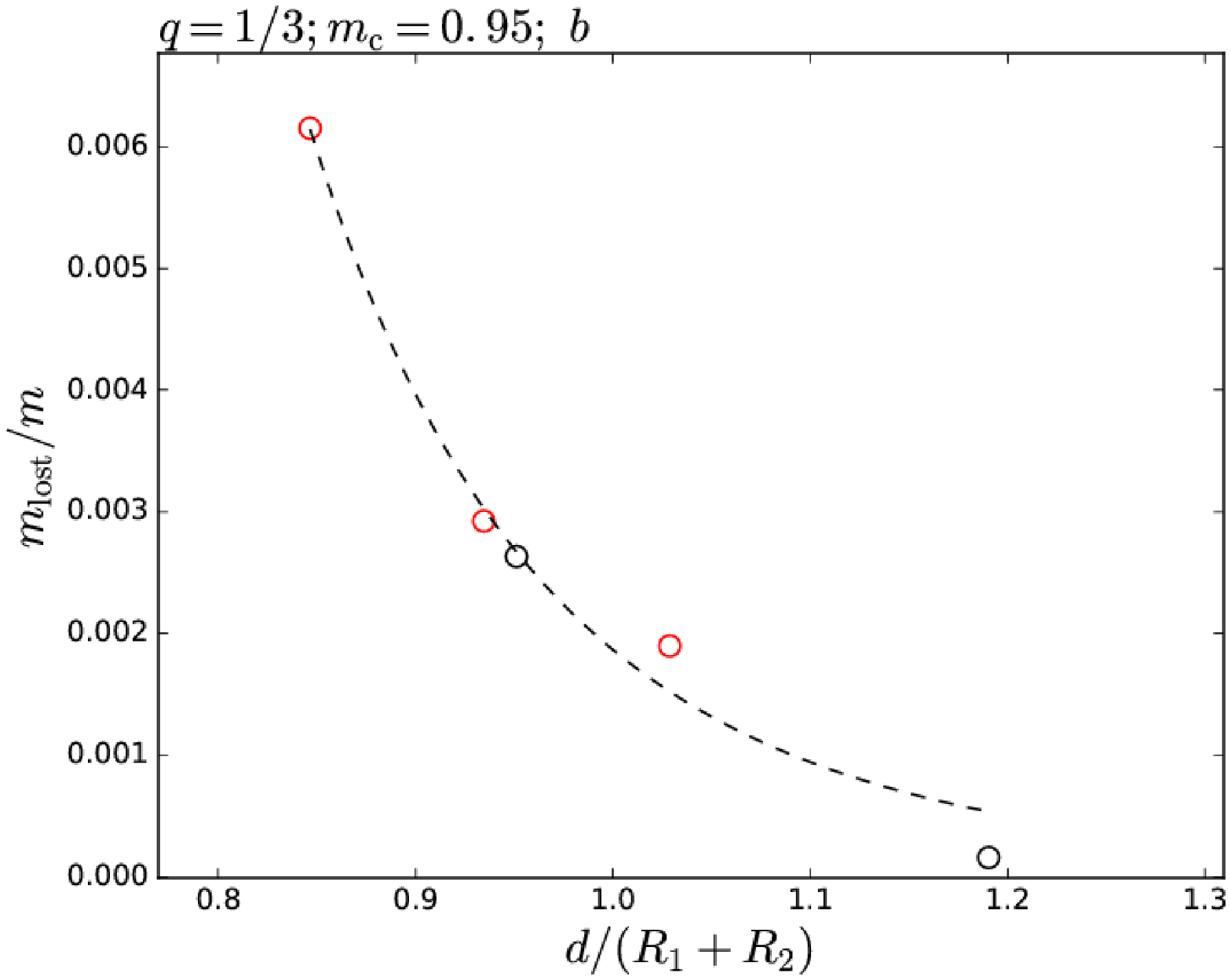}
\includegraphics[width=8.0cm]{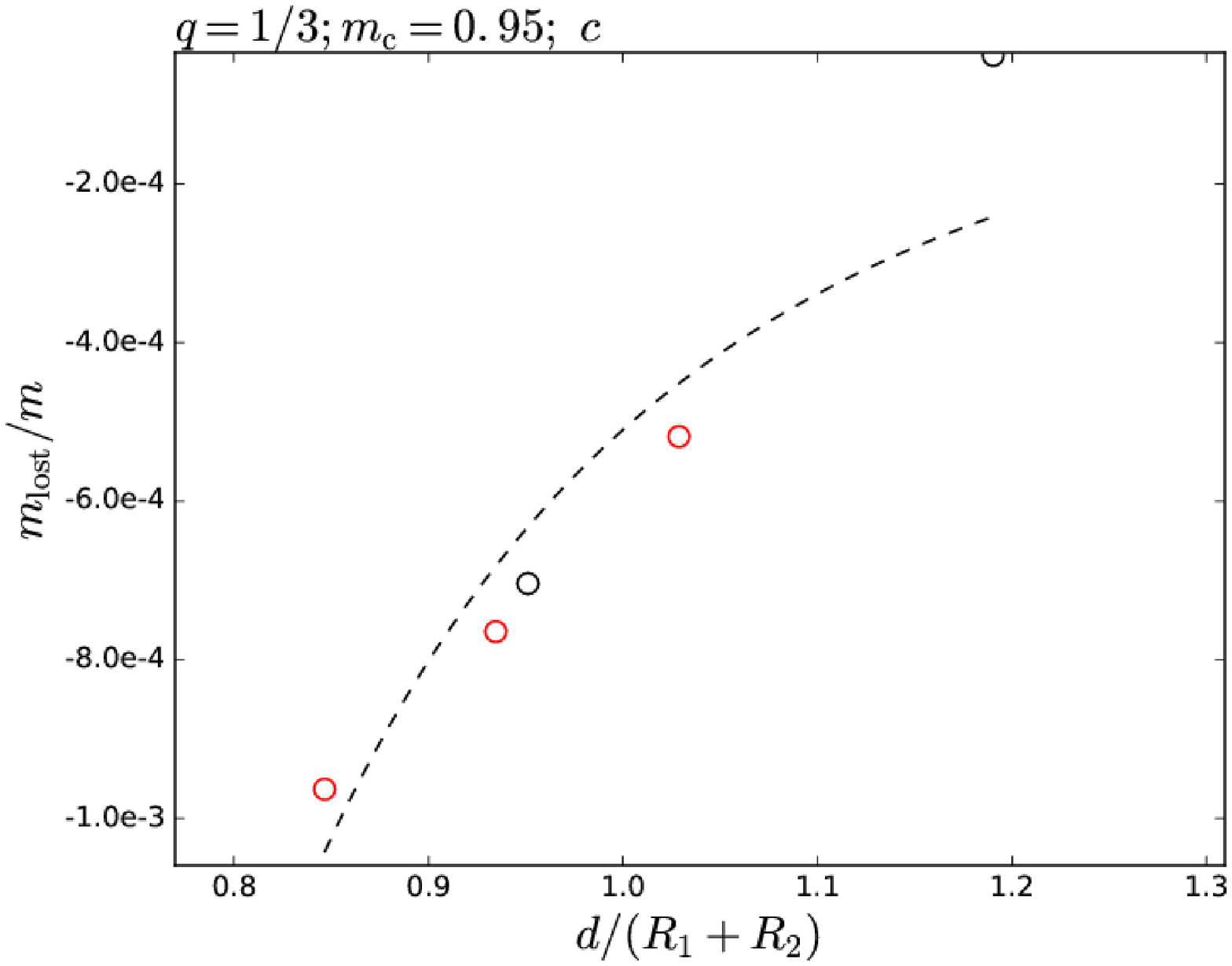}
\end{tabular}
\end{center}
\caption{Fractional change in mass after the first close encounter, of the inner (left) and outer (right) planet as a function of the degree of contact, $\eta = d_\mathrm{min}/(R_1+R_2)$, for the $q=1/3;\ m_\mathrm{c}=0.85$ (top), and $q=1/3;\ m_\mathrm{c}=0.95$ (bottom) calculations.
Symbols and fits are as in Fig.~\ref{Fig:mass_loss1}.}
\label{Fig:mass_loss2}
\end{figure*}

To predict the outcomes of a generic collision, we examine the energy dissipated and changes in mass during the first periastron passage of the two planets.
Figures~\ref{Fig:Energy_Dissipated}-\ref{Fig:mass_loss2} show the energy dissipated and changes in mass for each planet as a function of the distance of closest approach.
We fit the energy dissipated and change in mass of each planet with a power law  as a function of distance of closest approach, $f=Ad^k$, discarding collisions where we do not fully resolve the first passage, or with a distance of closest approach close enough to cause contact between the physical cores, as this introduces additional physics and requires a different fit.
We use only scattering collisions to fit the changes in mass, as we are interested in the masses of each planet after leaving their mutual Hill sphere.
Table~\ref{TBL:Best_fit_values} summarizes the best fit values for each set of calculations.

We find that higher mass ratios and core mass fractions exhibit a steeper exponent in the power law for both the energy dissipated and change in mass.
The change in mass of the less-massive planet is predicted well by the power-law fit, but we find that the change in mass of the more-massive planet in $q\ne1$ calculations is not as well modeled by a single power-law and likely depends on additional factors, for example, the angle of impact relative to the host star.

\begin{deluxetable}{lccccc}
\tabletypesize{\footnotesize}
\tablewidth{8.0cm}
\tablecolumns{3}
\tablecaption{Best Fits for Energy Dissipated and Change in Mass\label{TBL:Best_fit_values}}
\tablehead{
    \colhead{Set} & \colhead{$A$} & \colhead{$k$}}
\startdata
$E_\mathrm{d}/E_\mathrm{b} = A_\mathrm{E}\eta^{k_\mathrm{E}}$ \\
$ Kepler-36\ Progenitor	$ & $	3.57\times10^{-3}	$ & $	-5.15	$ & \\
$ q=1;\ m_\mathrm{c}=0.85	$ & $	1.47\times10^{-3}	$ & $	-4.36	$ & \\
$ q=1;\ m_\mathrm{c}=0.95	$ & $	2.92\times10^{-3}	$ & $	-5.43	$ & \\
$ q=1/3;\ m_\mathrm{c}=0.85	$ & $	1.07\times10^{-3}	$ & $	-4.55	$ & \\
$ q=1/3;\ m_\mathrm{c}=0.95	$ & $	2.11\times10^{-3}	$ & $	-6.83	$ & \\
\hline \\
$m_{\mathrm{lost},1}/m_1 = A_{\mathrm{m},1}\eta^{k_{\mathrm{m},1}}$ \\
$ Kepler-36\ Progenitor	$ & $	1.79\times10^{-4}	$ & $	-10.56	$ & \\
$ q=1;\ m_\mathrm{c}=0.85	$ & $	1.36\times10^{-5}	$ & $	-8.09	$ & \\
$ q=1;\ m_\mathrm{c}=0.95	$ & $	6.57\times10^{-6}	$ & $	-13.06	$ & \\
$ q=1/3;\ m_\mathrm{c}=0.85	$ & $	5.23\times10^{-3}	$ & $	-3.24	$ & \\
$ q=1/3;\ m_\mathrm{c}=0.95	$ & $	1.87\times10^{-3}	$ & $	-7.15	$ & \\
\hline \\
$m_{\mathrm{lost},2}/m_2 = A_{\mathrm{m},2}\eta^{\gamma_{\mathrm{m},2}}$ \\
$ Kepler-36\ Progenitor	$ & $	3.80\times10^{-5}	$ & $	-6.81	$ & \\
$ q=1;\ m_\mathrm{c}=0.85	$ & $	1.62\times10^{-5}	$ & $	-7.92	$ & \\
$ q=1;\ m_\mathrm{c}=0.95	$ & $	9.45\times10^{-6}	$ & $	-12.40	$ & \\
$ q=1/3;\ m_\mathrm{c}=0.85	$ & $	-3.08\times10^{-3}	$ & $	0.62	$ & \\
$ q=1/3;\ m_\mathrm{c}=0.95	$ & $	-5.10\times10^{-4}	$ & $	-4.29	$ & \\
\enddata
\tablenotetext{}{$Set$ designates the set name, and $A$ and $k$ are dimensionless variables used in models predicting the energy dissipated and changes in mass as a function of the distance of closest approach.}
\end{deluxetable}

\subsection{Prescription}
For each set of collisions, we develop models using the initial energy, distance of closest approach, masses, and orbits to predict the outcome and final masses and orbits of the remnant planet(s).
Using \eqref{EQ:E_max} and \eqref{EQ:a_max}, the critical initial collision energy required for a scattering may be expressed, \begin{equation}\label{EQ:E}E_\mathrm{c} = E_\mathrm{esc} + E_\mathrm{d},\end{equation} where \begin{equation}\label{EQ:E_esc}E_\mathrm{esc}=\frac{-Gm_1(m_1+m_2)}{2a_\mathrm{out}}\frac{1+e_\mathrm{in}}{1-e_\mathrm{out}}\left(\frac{\mu}{3M_*}\right)^{-1/3},\end{equation} and $m_1$, $m_2$, and $E_\mathrm{d}$ may be estimated as a function of $d_\mathrm{min}$ using the fits to discussed in the previous subsection.

\begin{figure*}[htp]
\begin{center}
\begin{tabular}{cc}
\includegraphics[width=8.0cm]{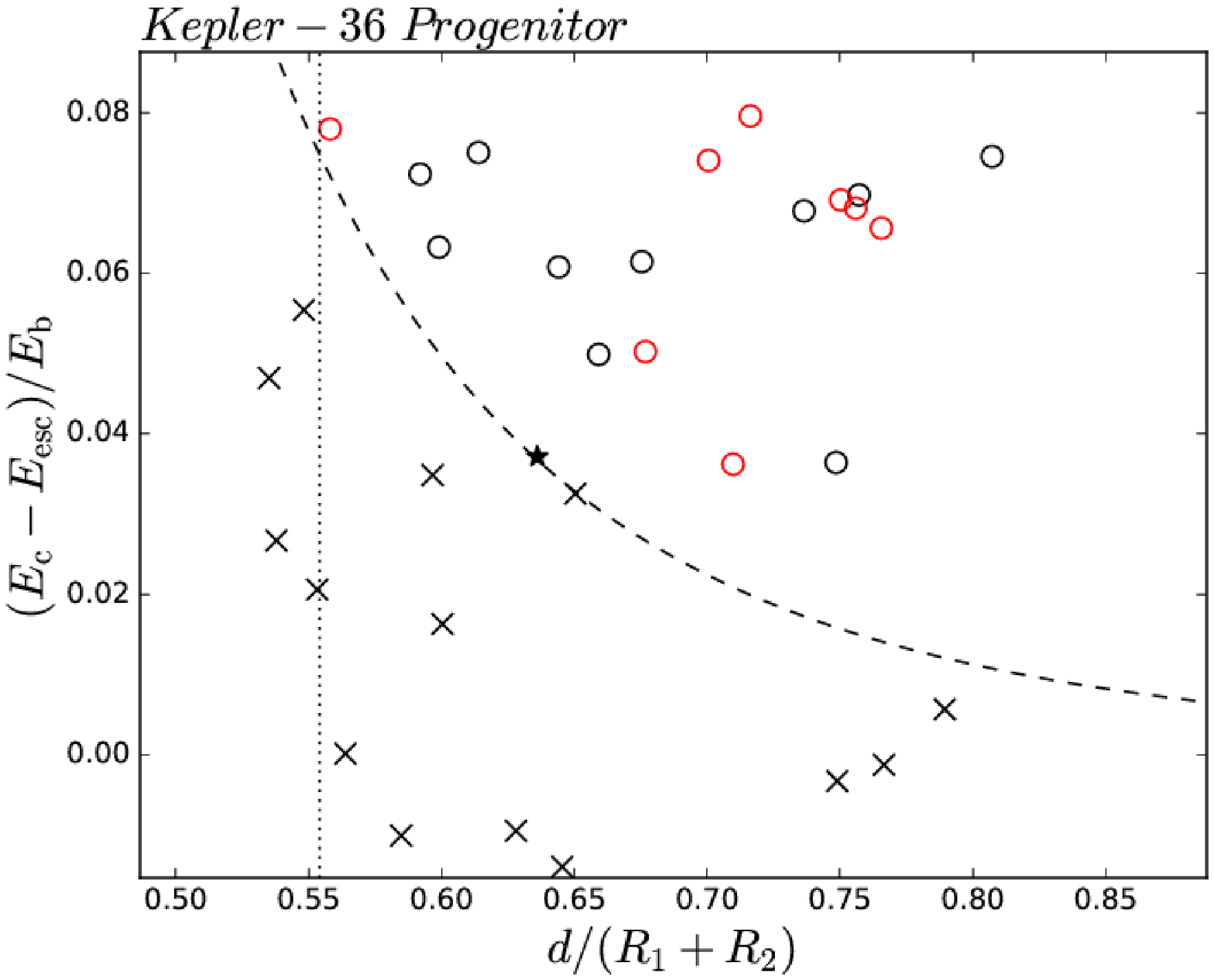}
\color{white}
\rule{8.0cm}{5.0cm}
\color{black} \\
\includegraphics[width=8.0cm]{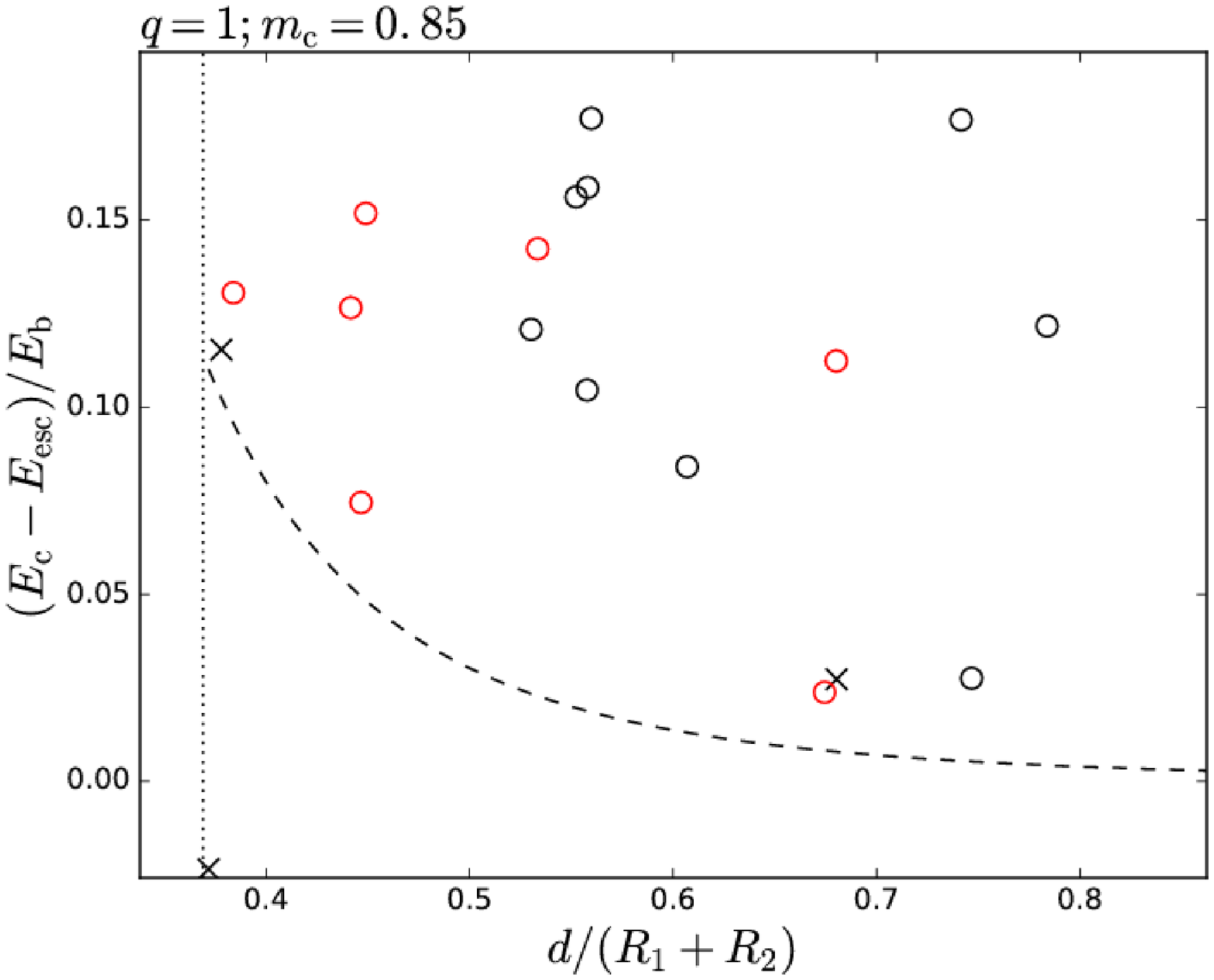}
\includegraphics[width=8.0cm]{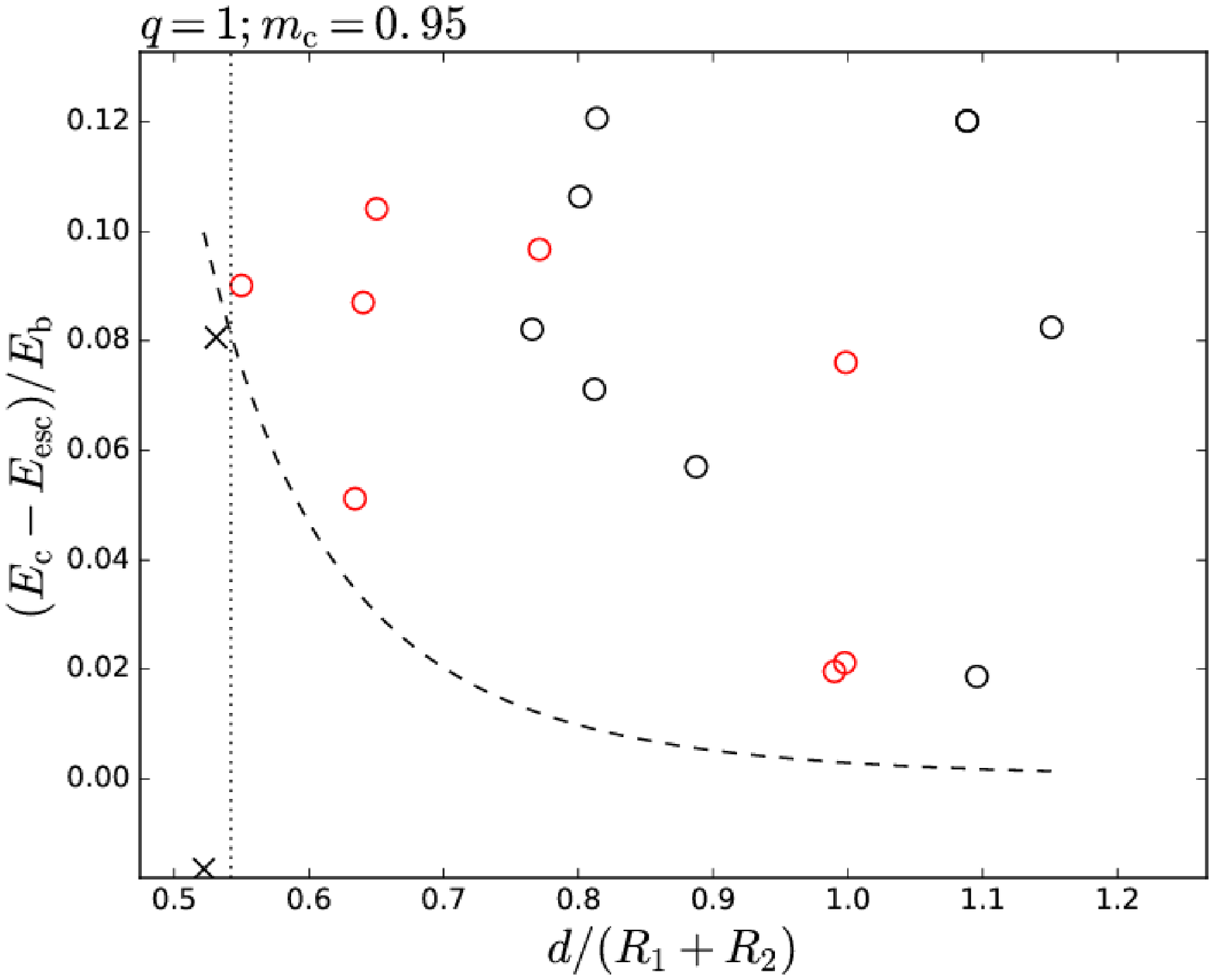} \\
\includegraphics[width=8.0cm]{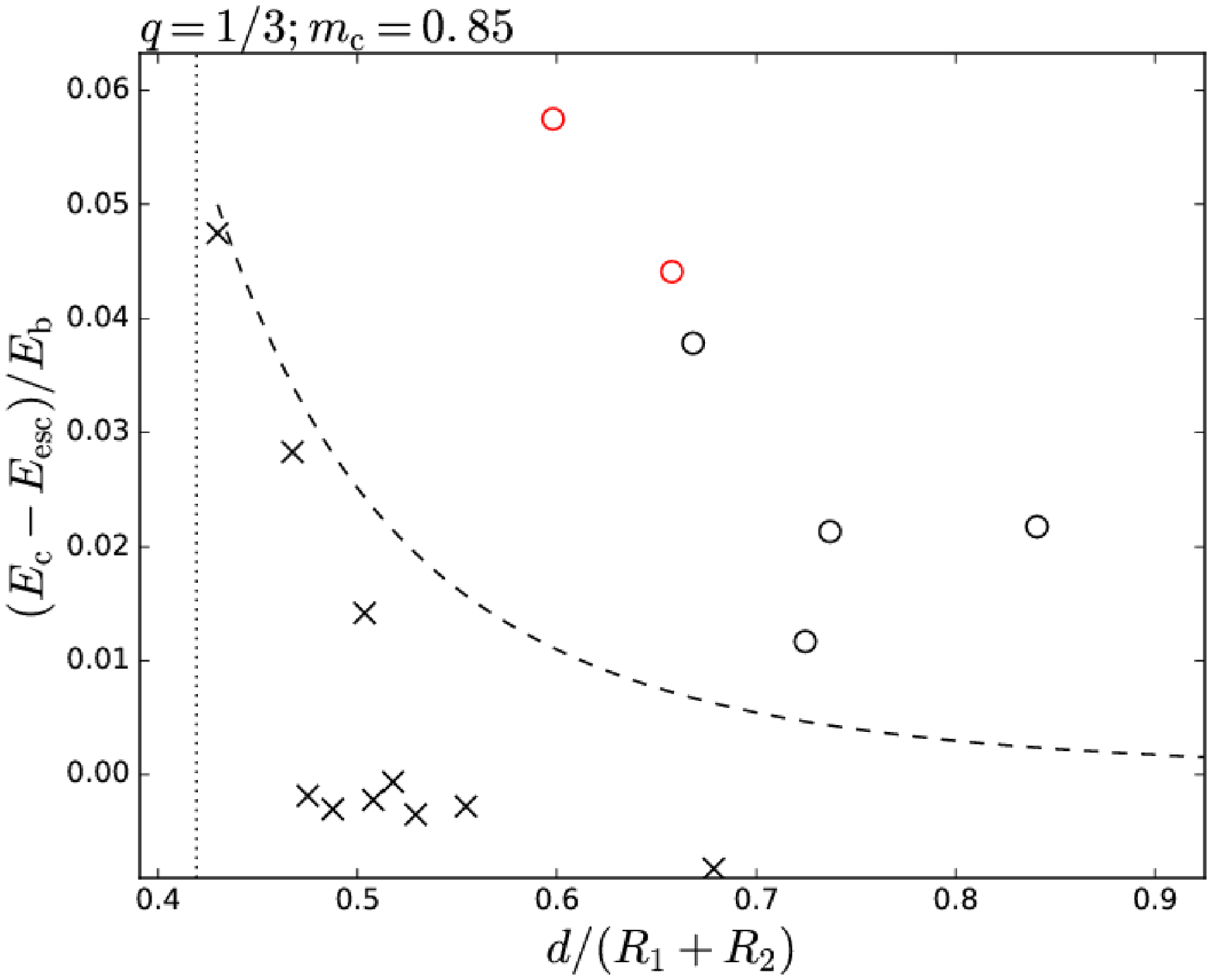}
\includegraphics[width=8.0cm]{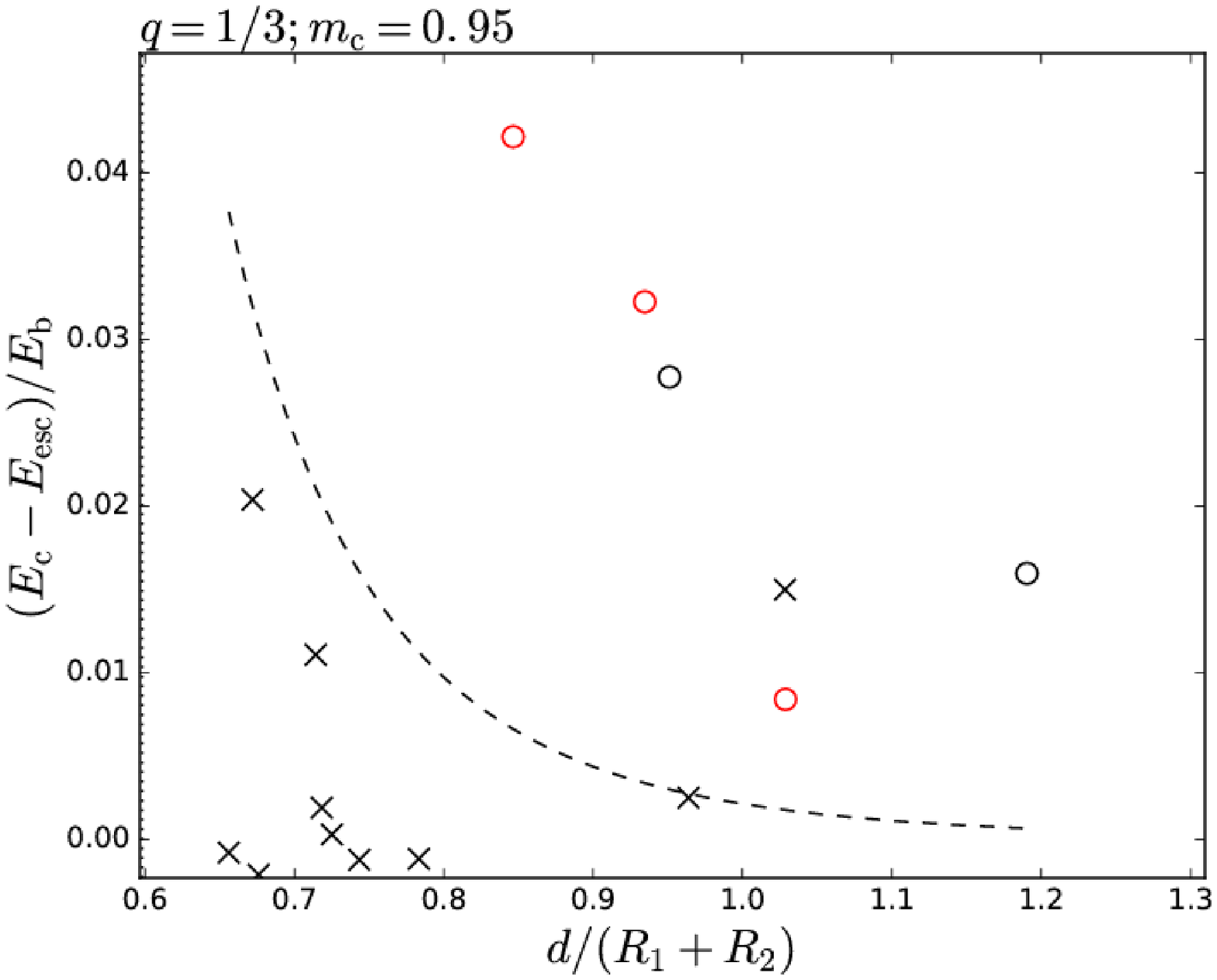}
\end{tabular}
\end{center}
\caption{The outcomes of each collision, designated as described in \ref{Fig:Collision_Results}, as a function of the degree of contact, $\eta = d_\mathrm{min}/(R_1+R_2),$ and the collision energy in units of the binding energy, as described by eq.~\eqref{EQ:binding_energy}, from the SPH calculations.
The collision energy is normalized by the escape energy as described by eq.~\eqref{EQ:E_esc}.
We show the predicted critical collision energy separating scattering collisions from captures (dashed black lines), and we see that the model accurately predicts the outcome in 98 out of 102 collisions, where 2 of the misclassified outcomes are the result of a second, separate close encounter, and 1 results in the only bound-planet pair found in our set of calculations.}
\label{Fig:Outcome_Predictions}
\end{figure*}

\begin{figure*}[htp]
\begin{center}
\begin{tabular}{cc}
\includegraphics[width=8.0cm]{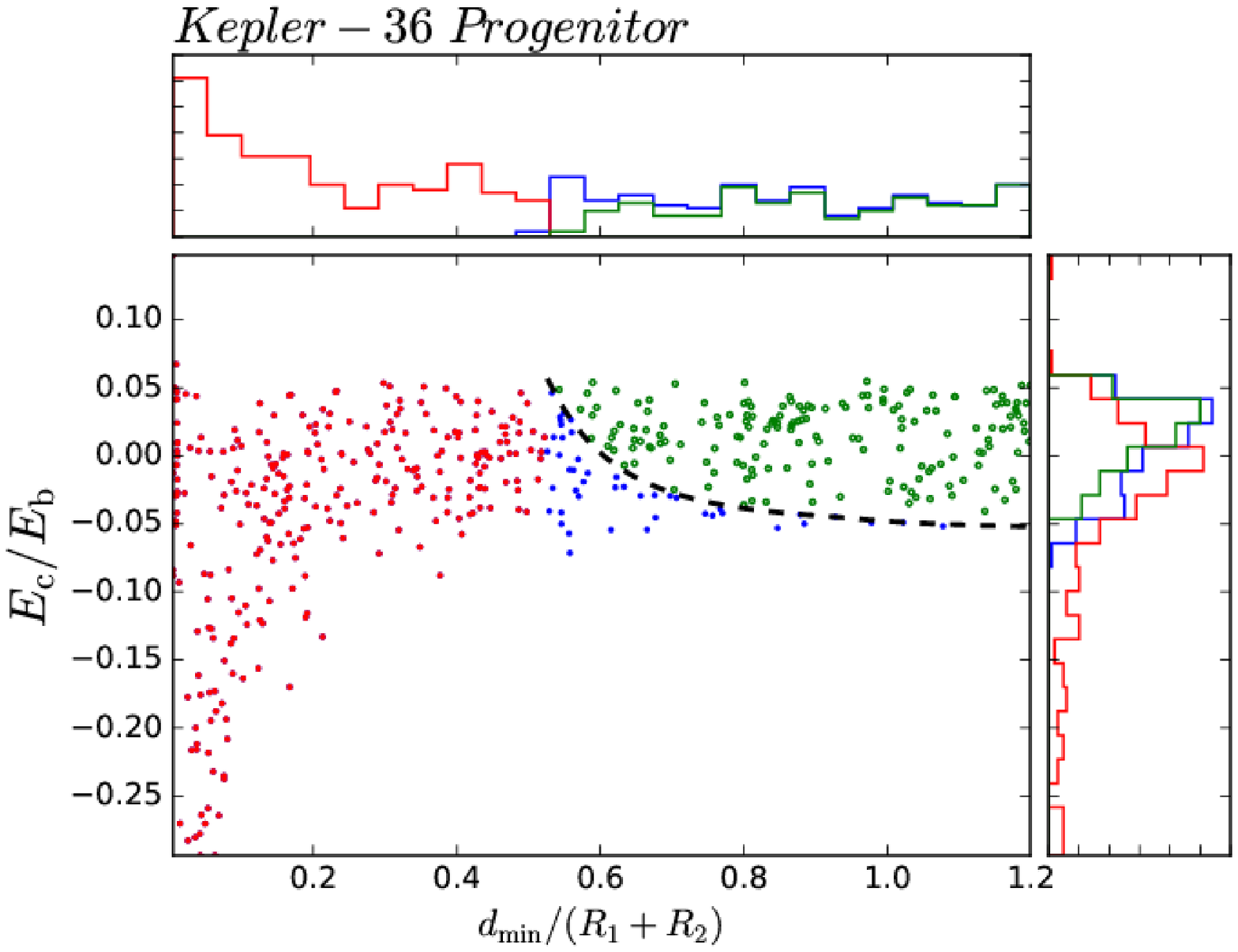}
\color{white}
\rule{8.0cm}{5.0cm}
\color{black} \\
\includegraphics[width=8.0cm]{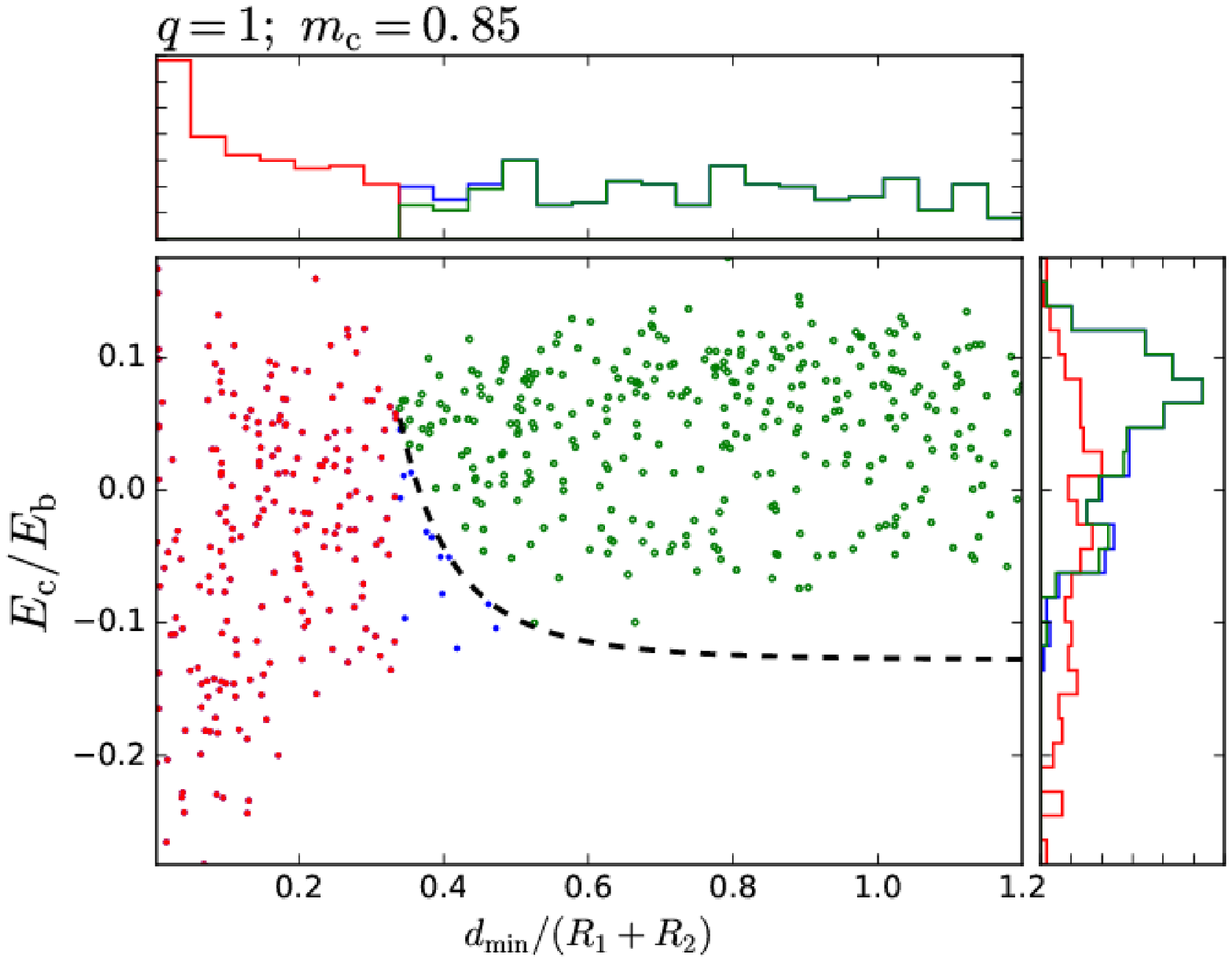}
\includegraphics[width=8.0cm]{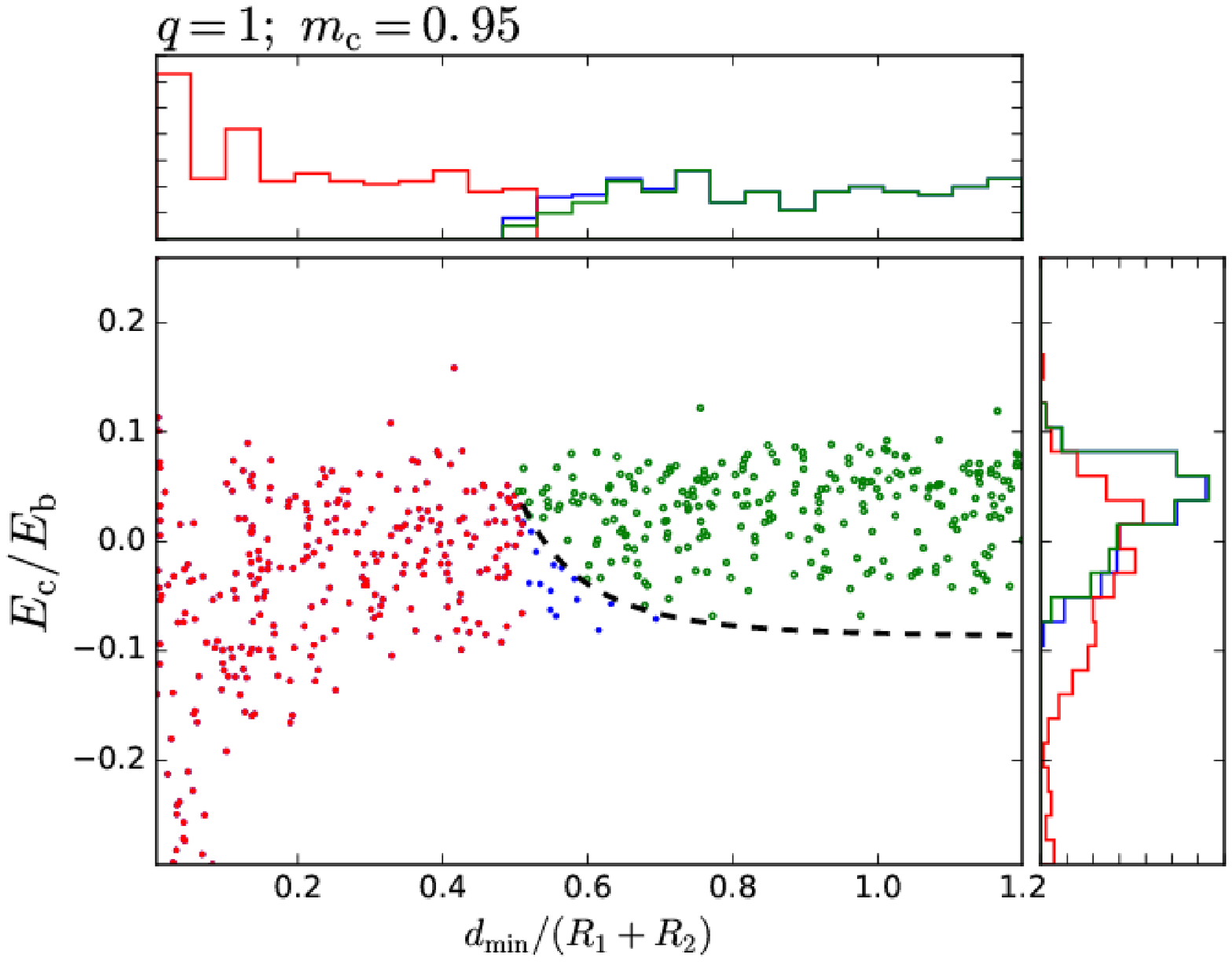} \\
\includegraphics[width=8.0cm]{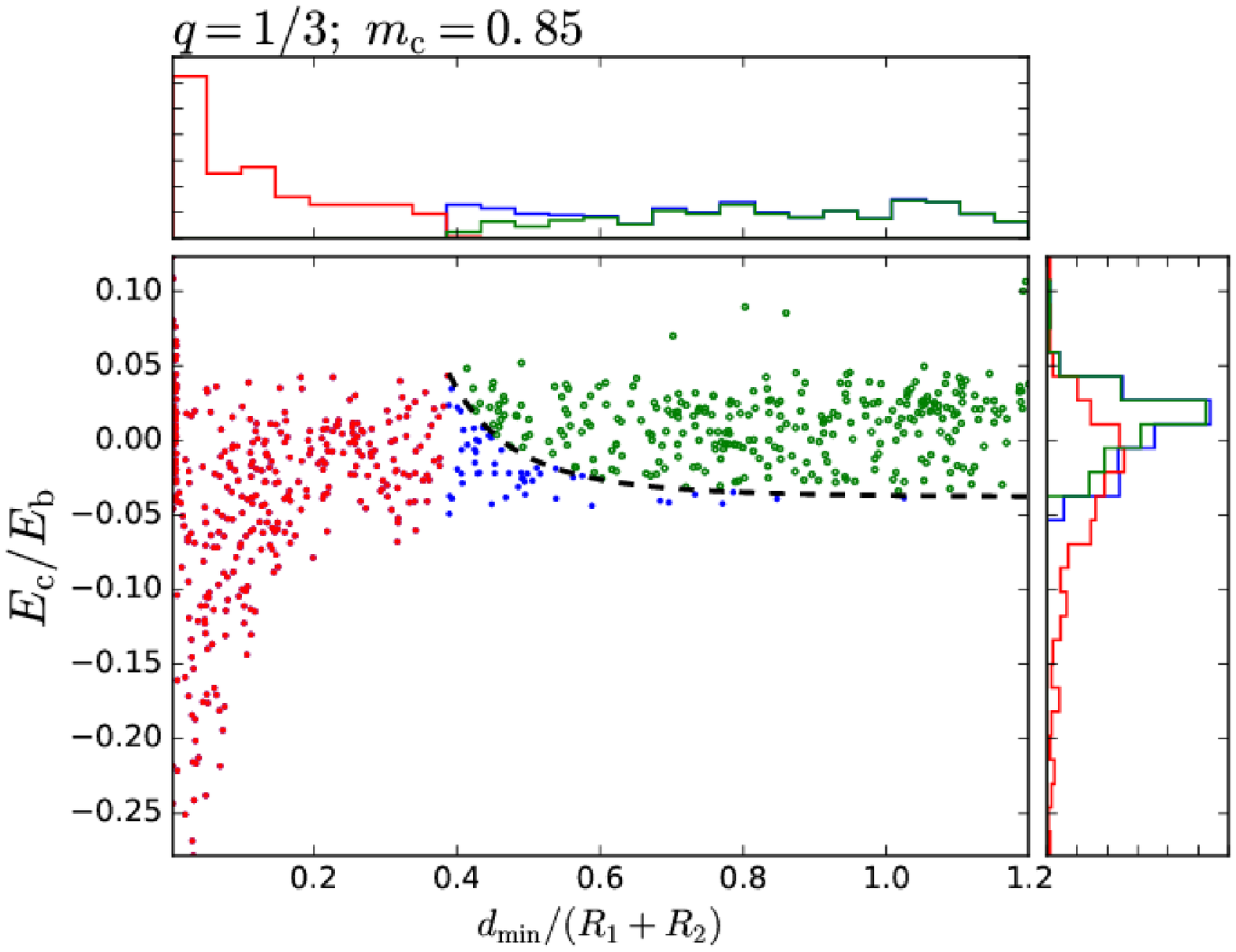}
\includegraphics[width=8.0cm]{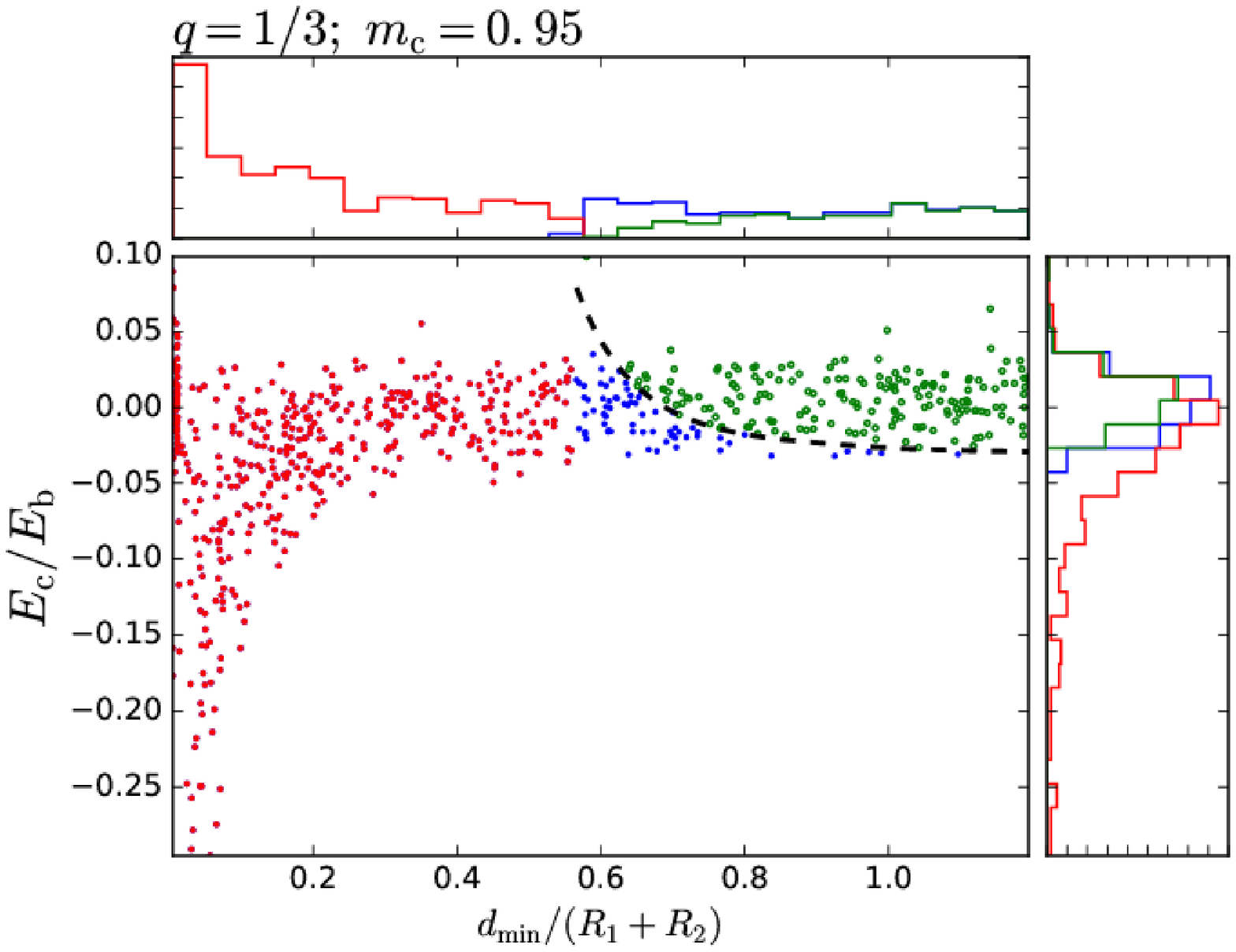}
\end{tabular}
\end{center}
\caption{Distribution of collisions characterized as in Figure~\ref{Fig:Collision_Parameters}, with the predicted critical initial collision energy described by eq.~\eqref{EQ:E} (dashed black line) separating grazing collisions resulting in mergers (blue) and scatterings (green).
Table~\ref{TBL:Predicted_Nbody} summarizes the predicted outcomes, where $72\%$ to $96\%$ of the grazing collisions result in a scattering, and are poorly modeled by the sticky-sphere approximation.}
\label{Fig:Collision_Parameters_Predictions}
\end{figure*}

Figure~\ref{Fig:Outcome_Predictions} shows, for each set of calculations, the critical initial collision energy in excess of the energy required for a planet to leave the mutual Hill sphere (to normalize the distance from the host star), as a function of the distance of closest approach, separating the predicted captures and scatterings.
We see that the potential planet-planet binary is very close to the critical energy separating captures and scatterings.
The model accurately predicts 98 out of 102 outcomes, and we examine closely the outcomes that are misclassified.
Of the four misclassified outcomes, two are mergers that occur after initially leaving the mutual Hill sphere, and can be attributed to a second, separate close encounter, and one is a collision very close to the critical collision energy, resulting in a potentially long-lived planet-planet binary.
This model may be adapted for a generic collision between sub-Neptunes using the fits from the calculations most similar to the target system.
Figure~\ref{Fig:Collision_Parameters_Predictions} shows the distribution of collisions from our dynamical integrations discussed in \S\ref{Sec:Nbody} with the predicted critical collision energy for separating grazing collisions resulting in mergers from scatterings.
Table~\ref{TBL:Predicted_Nbody} summarizes the outcomes of grazing collisions and near misses, where $72\%$ - $96\%$ of grazing collisions result in a scattering, motivating an improvement on the sticky-sphere approximation.
Based on the predicted outcome, the final properties of the collision remnant may be estimated, specifically, the amount of gas retained by a merger remnant, and the energy and gas lost during a scattering.

\begin{deluxetable*}{lcccccccc}
\tabletypesize{\footnotesize}
\tablewidth{6.0cm}
\tablecolumns{3}
\tablecaption{Predicted Outcomes of Grazing Collisions\label{TBL:Predicted_Nbody}}
\tablehead{
    \colhead{Set} & \colhead{$N_\mathrm{grazing}$} & \colhead{$N_\mathrm{scattering}$}}
\startdata
$ Kepler-36\ progenitor		$ & $	212	$ & $	167	$ & \\
$ q=1;\ m_\mathrm{c} = 0.85	$ & $	333	$ & $	320	$ & \\
$ q=1;\ m_\mathrm{c} = 0.95	$ & $	269	$ & $	255	$ & \\
$ q=1/3;\ m_\mathrm{c} = 0.85	$ & $	348	$ & $	295	$ & \\
$ q=1/3;\ m_\mathrm{c} = 0.95	$ & $	254	$ & $	183	$ & \\
\enddata
\tablenotetext{}{$Set$ designates the set name, $N_\mathrm{grazing}$ is the number of grazing collisions (including near misses described in Table~\ref{TBL:NBody_Stats}) that occur within $1000$ years and $N_\mathrm{scattering}$ is the number of grazing collisions that result in a scattering, where the planets leave their mutual Hill radii without merging.}
\end{deluxetable*}

\subsubsection{Scatterings}
We find that a majority of grazing collisions where the cores do not physically come into contact result in a scattering with both planets leaving their mutual Hill sphere.
During the scattering, both planets lose mass and orbital energy, which is deposited into unbinding gas and also converted into internal energy.

The change in mass and orbital energy during a scattering collision can significantly affect the dynamical evolution of the system.
The prescription for treating scattering collisions is as follows, using as inputs the degree of contact, masses, mass ratio, and core mass fractions of the planets:
\begin{itemize}\setlength{\itemsep}{-4pt}
 \item [1)] Integrate until reaching the projected minimum separation.
 \item [2)] Use the fits presented in Table~\ref{TBL:Best_fit_values} to estimate the mass fraction, $m_{\mathrm{lost},i}$, and orbital energy, $E_\mathrm{d}$, lost from both planets. The best fit values may be chosen either from the set of collisions most similar to the two colliding planets {\bf or using a linear interpolation.}
 \item [3)] Calculate the remnant masses for both planets \begin{equation}m^\prime_i = m_i(1-m_{\mathrm{lost},i}(\eta)).\end{equation} For cases where the mass ratio of the two planets is large ($q<2/3$), the change in mass of the larger planet is not as well characterized by our fits, likely depending on details such as the exact trajectory of the collision; the observed change in mass for such large planets is relatively small, and can be ignored.
 \item [4)] Decrease the magnitude of the relative velocity between the two planets to account for the decrease in orbital energy, conserving energy (accounting for the energy lost), \begin{equation}\frac{v^\prime}{v}=\sqrt{\frac{1}{\mu^\prime}\left[\frac{2G}{d_\mathrm{min}v^2}\left(m_1m_2-m^\prime_1m^\prime_2\right)+\mu-\frac{E_\mathrm{d}(\eta)}{v^2}\right]},\end{equation} where $\mu^\prime$ and $v^\prime$ are the remnant reduced-mass and magnitude of relative velocity.
 \item [5)] Numerically solve for the remnant velocities of each planet, conserving specific angular momentum, \begin{equation}v^\prime=|{\bf v^\prime}_1-{\bf v^\prime}_2|,\end{equation}\begin{equation}\frac{m_1{\bf r}_1\times{\bf v}_1+m_2{\bf r}_2\times{\bf v}_2}{m_1+m_2}=\frac{m^\prime_1{\bf r}_1\times{\bf v^\prime}_1+m^\prime_2{\bf r}_2\times{\bf v^\prime}_2}{m^\prime_1+m^\prime_2},\end{equation} where ${\bf v^\prime}_i=v^\prime_i{\bf \hat{v}}_i$ is the remnant velocity vector parallel to the initial velocity vector.
 \item [6)] (Optional) Adjust the radii of both planets to account for the loss of gas. This step assumes that subsequent collisions occur after the planets have relaxed back into equilibrium.
\end{itemize}

Using this prescription should result in a material difference in dynamical integrations that exhibit scattering collisions, generally driving the planets into more stable orbits, and in cases with a subsequent merger, less-massive remnants.

\subsubsection{Mergers}
While we do not integrate collisions resulting in a physical collision between the cores through the merger due to the computational cost, we find that the short term change in mass is largely determined by the planets' core masses.
$4\ M_\mathrm{E}$ planets tend to lose a majority of their envelope in addition to some mantle, while $12\ M_\mathrm{E}$ planets do not exhibit significant changes in mass.
Our method for treating for merger events is the same as from {\it Paper 1}, ejecting the gas from both planets and using the sticky-sphere approximation assuming most of the core mass ($\sim95\%-100\%$) is retained.
\citet{2016ApJ...817L..13I} provide a model for estimating the final gas mass fraction of the remnant planet after a merger as a function of the impact velocity; because the collisions in this study generally have energies comparable to the escape energy, our prediction that most of the gas envelope is ejected in a merger is consistent with this study.

\section{Conclusions}
\label{Sec:Conclusions}
We conduct a detailed study on collisions between two planets in initially unstable orbits, varying the mass ratios and core mass fractions of the planets to sample a range of typical neighbors in Kepler Multis, and summarize our findings as follows:
\begin{itemize}\setlength{\itemsep}{-4pt}
  \item[1)]102 collision calculations result in 62 scatterings, 39 mergers, and 1 bound planet-pair.
  \item[2)]While every scattering remnant eventually resulted in an eventual crossing orbit during post-collision dynamical integrations, we find that collisions tend to stabilize the system. Further collisions may eventually result in a stable system with two surviving planets.
  \item[3)]The collisions that result in mergers tend to eject a majority of the gas from the less-massive planet.
  Future calculations with a better treatment of the core are required to fully resolve planet-planet mergers.
  \item[4)]One collision results in a potentially long-lived planet-planet binary with an eccentric orbit such that the periapsis avoids collision of the planets' cores and the apoapsis is small enough that the planets remain inside the mutual Hill sphere.
  \item[5)]The outcome of a collision depends very sensitively on the distance of closest approach and the relative energy between the two planets at the collision; specifically, high relative energies and large distances of closest approach lead to more scatterings and low relative energies and low distances of closest approach are more likely to result in a merger or bound planet-pair.
  \item[6)]After a collision, the density ratio of $q\ne1$ planets tend to become more extreme, with the higher-mass core retaining more of, or even accreting, the disrupted gas, and the lower-mass core losing a higher fraction of gas; we found a minimum density ratio of $\rho_2/\rho_1 = 0.4$ in our Kepler-36 progenitor calculations with a pre-collision density ratio, $\rho_2/\rho_1 = 1.15$.
  \item[7)]Many interesting outcomes provide motivation for further study; specifically the possibility of a long-lived planet-planet binary, and potential moon-forming collisions encourage development of models capable of accurately modeling violent collisions of the dense core material and resolving the gas envelope.
\end{itemize}

We aggregate the collision calculations to create fits for the energy dissipated and mass lost during a collision as a function of the distance of closest approach and planet properties.
We use these fits to develop a prescription for use in N-body integrators to predict and model the outcomes of such collisions between sub-Neptunes.
We find that collisions resulting in two surviving planets leaving their mutual Hill sphere tend to increase the dynamical stability of the system and reduce the overall gas mass fraction.
The amount of gas lost strongly depends on the core masses of the planets and a low-mass gas-poor planet neighboring a high-mass gas-rich planet, with a very small dynamical separation, may be evidence of one or more previous planet-planet collisions.
The collisions that result in direct contact between the two cores are not fully resolved through the likely eventual merger in our simulations, and motivate further study to understand the details of these outcomes.
Previous calculations of these collisions with simple polytropic equations of state showed many distinct outcomes including fragmentation and partial accretion of the less-massive core.
If the lighter mantle material were preferentially stripped off of the less-massive core and accreted by the more-massive core, the remnant planets would exhibit very different core compositions: the less-massive planet with an excess iron-core mass fraction and the more-massive planet with an higher silicate-mantle mass fraction.

\subsection{Future Work}
Given our choice of equations of state to model the core, which were chosen to provide a stable boundary condition, we focused on studying grazing collisions and were not able to integrate core-core impacts through the entire merger.
Improving the algorithms used to resolve interfaces between components and improving our core equations of state to better models shocks and mixing, will allow the study of the deeper impacts, for example the long term evolution of mergers.

In this work we focused on performing many calculations to better understand the outcomes of collisions after the first periastron passage and we do not resolve collisions past a few orbital periods of the innermost planet.
Calculations that integrate scattering collisions for long timescales are important in fully understanding the behavior of the disrupted gas, particularly if the gas eventually falls back onto one of the planets.
A study of the long-term evolution of the gas will require additional physics such as a model to handle stellar winds.

We studied in detail five sets of 2-planet systems, made up of two sub-Neptunes, where both planets have a tenuous gas envelope that dominates the volume, varying the mass ratios and gas mass fractions.
However, there still remains much work to be done to better understand the likely common collisions in high-multiplicity, tightly-packed systems.
{\bf Performing similar calculations at lower planet-ages will increase the size of the planet envelopes, resulting in collisions at larger distances of closest approach, and the change in structure will likely lead to qualitatively different outcomes.
Increasing the semi-major axes of the planets will lower the impact of the host star, and including terrestrial planets and gas giants will also expand the scope of this type of study.}

The existence of long-lived planet-planet binaries is an interesting problem and the absence of an observation allows us to place an upper limit on the occurence rate.
A detailed treatment of the stability of such a system could provide insight to the types of collisions that are able to create these phenomena.

\subsection{Acknowledgements}
This research was supported in part through the computational resources and staff contributions provided for the Quest high performance computing facility at Northwestern University, which is jointly supported by the Office of the Provost, the Office for Research, and Northwestern University Information Technology.
FAR and JAH were supported by NASA Grant NNX12AI86G.
JAH was also supported by an NSF GK-12 Fellowship funded through NSF Award DGE-0948017 to Northwestern University.
JCL was supported by NSF grant number AST-1313091.
JHS was supported by NASA grants NNX16AK08G and NNX16AK32G.
We thank Joshua Fixelle for useful discussions while developing the equations of state used in the SPH calculations and Francesca Valsecchi for help with using {\it MESA} to generate sub-Neptune envelopes.
We thank the anonymous referee for providing very thorough and helpful feedback leading to major improvements in this manuscript.
This work used the SPLASH visualization software \citep{2007PASA...24..159P}.

\pagebreak
\appendix

\section{Appendix}

\subsection{Kepler-36 Progenitor Properties}
\label{A:Analytic}
In this section we show that an initially unstable system that undergoes a planet-planet collision can result in two planets with the same angular momentum and orbital energy distribution as an observed and dynamically stable system, using Kepler-36 as our nominal system.
We calculate the orbital elements and planet masses of potential prognenitor systems to the currently observed Kepler-36 system, assuming conservative mass transfer, \begin{equation}m_1=m_{1,\mathrm{o}}-\Delta m,\end{equation} \begin{equation}m_2=m_{2,\mathrm{o}}+\Delta m,\end{equation} where $m_1$ and $m_2$ are the initial planet masses, $m_{1,\mathrm{o}}$ and $m_{2,\mathrm{o}}$ are the observed planet masses, and $\Delta m$ is the amount of mass transfer, conservation of angular momentum, \begin{equation}h=\sum_{k=1,2}\mu_{k,o}\sqrt{G(M_*+m_{k,\mathrm{o}})a_{k,\mathrm{o}}(1-e_{k,\mathrm{o}}^2)},\end{equation} where $M_*=1.071\ M_\odot$ is the mass of the host star, $a_{1,\mathrm{o}}=0.1153\ \mathrm{AU}$ and $a_{2,\mathrm{o}}=0.1283\ \mathrm{AU}$ are the observed semi-major axes, $e_{1,\mathrm{o}}=0.04$ and $e_{2,\mathrm{o}}=0.04$ are the observed upper limits to the eccentricities, and $\mu_{1,\mathrm{o}}$ and $\mu_{2,\mathrm{o}}$ are the reduced masses \citep{2012Sci...337..556C}.
We assume that the collision results in the loss of some orbital energy, \begin{equation}E=\frac{-1}{\epsilon}\sum_{k=1,2}\frac{GM_*m_{k,\mathrm{o}}}{2a_{k,\mathrm{o}}},\end{equation}  where $\epsilon$ is the fraction of orbital energy conserved in the collision.

\subsubsection{Mass Transfer}
\label{SSec:MassTransfer}
\begin{figure}[htp]
\begin{center}
\begin{tabular}{cc}
\includegraphics[width=8.0cm]{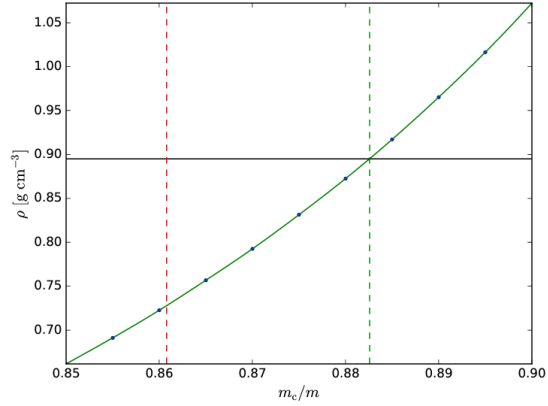}
\end{tabular}
\end{center}
\caption{Density as a function of core mass, using a series of {\it MESA} models (blue circles) to generate a fit (solid green). We find that the observed density of $\rho=0.895\ \mathrm{g}\ \mathrm{cm}^{-3}$ (solid black) corresponds to a fit core mass fraction of $m_\mathrm{c}/M=0.882$ (dashed green), which agrees reasonably well with the core mass fraction prediction of $m_\mathrm{c}/M=0.861$ from \citet{2014ApJ...787..173H}.}
\label{Fig:CoreMass}
\end{figure}

We estimate the distribution of gas in the progenitor Kepler-36 system, assuming the currently observed distribution of gas is a result of conservative mass transfer in a prior planet-planet collision.
\citet{2012Sci...337..556C} find the mass and radius measurement of {\it b} is consistent with a planet with little or no gas envelope and \citet{2014ApJ...787..173H} predict a core mass fraction of $m_\mathrm{c}/M=0.861$ for {\it c}.
We generate a series of models for {\it c} using {\it MESA} (\citealt{2011ApJS..192....3P}; \citealt{2013ApJS..208....4P}; \citealt{2015ApJS..220...15P}), varying the core mass fraction and using the core mass and core-density relationship from \citet{2014ApJ...787..173H} assuming a $32.5\%$ iron and $67.5\%$ perovskite differentiated core.
Figure~\ref{Fig:CoreMass} shows the density of our models as a function of the core mass fraction, where $m_\mathrm{2,c}/m_{2,\mathrm{o}}=0.882$ is consistent with the observed density, which is reasonably close to the value reported by \citet{2014ApJ...787..173H}, with differences potentially arising from our model including irradiation from the host star leading to a larger radius.

We use the result from \citet{2015ApJ...811...41L} to estimate the initial mass of the gas envelopes, \begin{equation}m_\mathrm{g}=m_{1,\mathrm{g}}+m_{2,\mathrm{g}},\end{equation} \begin{equation}m_\mathrm{g}=\Delta m\left(1+\left(\frac{m_{2,\mathrm{c}}}{m_{1,\mathrm{c}}}\right)^\alpha\right),\end{equation} where $m_\mathrm{g}$ is the gas mass observed on {\it c}, $m_{1,\mathrm{g}}$ and $m_{2,\mathrm{g}}$ are the gas mass of the progenitor planets, and $\alpha=2.6$, and estimate that the mass lost by {\it b} is $\Delta m=0.215M_\oplus.$

\subsubsection{Orbital Elements}
We estimate the orbital elements of a dynamically unstable progenitor system that, after a collision, results in a dynamically stable system, specifically a system resembling Kepler-36, assuming conservative mass transfer and conservation of angular momentum.
Because the progenitor system must allow for a crossing orbit, \begin{equation}a_1(1+e_1)=a_2(1-e_2),\end{equation} and we assume angular momentum is conserved, we can solve for the range of inner semi-major axes, \begin{equation}a_1=\left(\frac{h^2}{G(1+e_1)\left(\mu_1\sqrt{(M_*+m_{1,\mathrm{o}}-\Delta m)(1-e_1)}+\mu_2\sqrt{(M_*+m_{2,\mathrm{o}}+\Delta m)(2-(1+e_1)T^{-2/3})}\right)^2}\right),\end{equation} where $e_1$ and $e_2$ are the eccentricities of {\it b} and {\it c}, and $T$ is the period ratio.
For each solution we find the fractional loss in orbital energy required to move from our proposed progenitor system to the observed Kepler-36 system, \begin{equation}\epsilon_\mathrm{E}=\frac{-M_*(m_1+m_2T^{-2/3})}{2E_\mathrm{o}a_1}.\end{equation}
We use the constraints that the progenitor system does not initially have a crossing orbit, \begin{equation}\label{EQ:f}f=a_2(1-e_2)-a_1(1+e_1)\ge0,\end{equation} and \begin{equation}\frac{\partial f}{\partial e_2}=a_2\left(\frac{h_1e_2m_2}{h_2e_1m_1}-1\right),\label{EQ:dfde2}\end{equation} to determine the boundaries on $e_1$ and $e_2$, as a function of both $\Delta m$ and $T$, where $h_1$ and $h_2$ are the angular momentum of the planets at the crossing orbit solution.
Using $\Delta m=0.215M_\oplus$ from \S\ref{SSec:MassTransfer}, $T=1.3$, and maximizing conservation of energy, $\epsilon_\mathrm{E}=0.998$, we find $a_1=0.111\ \mathrm{AU}$, $a_2=0.132\ \mathrm{AU}$, $e_1\ge0.11$, and $e_2\le0.07$.

\end{document}